\documentclass{aa}
\usepackage{txfonts}
\usepackage{graphicx}
\usepackage{setspace}
\usepackage{natbib}
\usepackage{sidecap}
\bibpunct{(}{)}{;}{a}{}{,}
\usepackage{color}
\usepackage{subfigure}
\usepackage{floatflt}
\usepackage{rotating}
\usepackage{aalongtable, lscape, afterpage}

\newcommand{\comment}[1]{}

\begin{document}

% makros.tex

\newcommand{\st}[1]{\mathrm{#1}} % st = straight %
\newcommand{\bld}[1]{\textbf{#1}}
\newcommand{\pow}[2]{$\st{#1}^{#2}$}
\newcommand{\grad}{\hspace{-0.15em}\r{}}
\newcommand{\lm}{\lambda}

\newcommand{\average}[1]{\left\langle #1 \right\rangle}

\newcommand{\h}{~h_{71}~}
\newcommand{\hinv}[1]{~h_{71}^{#1}~}
\newcommand{\eq}[1]{(\ref{eq-#1})}
\newcommand{\wrt}{with respect to\ }
\newcommand{\mms}{\frac{M_{\odot}}{M}}
\newcommand{\mpy}{\st{M}_{\odot}~\st{yr}^{-1}}
\newcommand{\ct}{t_{\st{cool}}}
\newcommand{\rc}{r_{\st{cool}}}
\newcommand{\tvir}{T_{\st{vir}}}
\newcommand{\rvir}{R_{\st{500}}}
\newcommand{\mvir}{M_{\st{500}}}
\newcommand{\mbh}{M_{\st{BH}}}
\newcommand{\ms}{\st M_{\odot}}
\newcommand{\ls}{L_{\odot}}
\newcommand{\lxb}{L_{\st {Xb}}}
\newcommand{\lx}{L_{\st {X}}}
\newcommand{\lr}{L_{\st {R}}}
\newcommand{\lbcg}{L_{\st{BCG}}}
\newcommand{\lt}{L_{\st{X}}{\st -}\tvir}
\newcommand{\Dl}{D_{\st{l}}}
\newcommand{\hiflux}{\textit{HIFLUGCS}}
\newcommand{\chandra}{\textit{Chandra}}
\newcommand{\vla}{\textit{VLA}}
\newcommand{\gmrt}{\textit{GMRT}}
\newcommand{\atca}{\textit{ATCA}}
\newcommand{\XMM}{\textit{XMM-Newton}}
\newcommand{\einstein}{\textit{Einstein}}
\newcommand{\asca}{\textit{ASCA}}
\newcommand{\rosat}{\textit{ROSAT}}
\newcommand{\mdr}{\dot{M}_{\st{classical}}}
\newcommand{\smdr}{\dot{M}_{\st{spec}}}
\newcommand{\imdr}{\dot{M}_{\st{i, classical}}}

\title{What is a Cool-Core Cluster? A Detailed Analysis of the Cores
  of the X-ray Flux-Limited {\it HIFLUGCS} Cluster Sample.}

\author{D. S. Hudson\inst{1} \and R. Mittal\inst{1,2} \and
  T. H. Reiprich\inst{1} \and P. E. J. Nulsen\inst{3} \and
  H. Andernach\inst{1,4} \and C. L. Sarazin\inst{5}
  \institute{Argelander-Institut f\"ur Astronomie der Universit\"at
    Bonn, Auf dem H\"ugel 71, D-53121 Bonn, Germany \and Rochester
    Institute of Technology, 84 Lamb Memorial Drive, Rochester, NY
    14623 \and Harvard-Smithsonian Center for Astrophysics, 60 Garden
    Street, Cambridge, MA 02138, USA \and on leave of absence from
    Departamento de Astronom\'{i}a, Universidad de Guanajuato, AP 144,
    Guanajuato CP 36000, Mexico \and Department of Astronomy,
    University of Virginia, P.O. Box 400325, Charlottesville, VA
    22904-4325, USA}}

\date{Received/Accepted}

\abstract{

  We use the largest complete sample of 64 galaxy clusters (HIghest
  X-ray FLUx Galaxy Cluster Sample) with available high-quality X-ray
  data from Chandra, and apply 16 cool-core diagnostics to them, some
  of them new.  In order to identify the best parameter for
  characterizing cool-core clusters and quantify its relation to other
  parameters, we mainly use very high spatial resolution profiles of
  central gas density and temperature, and quantities derived from
  them. We also correlate optical properties of brightest cluster
  galaxies (BCGs) with X-ray properties.

  \hspace*{0.5cm}To segregate cool core and non-cool-core clusters, we
  find that central cooling time, $t_{\rm cool}$, is the best
  parameter for low redshift clusters with high quality data, and that
  cuspiness is the best parameter for high redshift clusters. 72\% of
  clusters in our sample have a cool core ($t_{\rm cool} <
  7.7~h_{71}^{-1/2}$~Gyr) and 44\% have strong cool cores ($t_{\rm
    cool}< 1.0~h_{71}^{-1/2}$~Gyr).  We find strong cool-core clusters
  are characterized as having low central entropy and a systematic
  central temperature drop.  Weak cool-core clusters have enhanced
  central entropy and temperature profile that are flat or decrease
  slightly towards the center. Non-cool-core clusters have high
  central entropies.

  \hspace*{0.5cm}For the first time we show quantitatively that the
  discrepancy in classical and spectroscopic mass deposition rates can
  not be explained with a recent formation of the cool cores,
  demonstrating the need for a heating mechanism to explain the
  cooling flow problem.

  \hspace*{0.5cm}We find that strong cool-core clusters have a
  distribution of central temperature drops, centered on 0.4$T_{\rm
    vir}$. However, the radius at which the temperature begins to drop
  varies. This lack of a universal inner temperature profile probably
  reflects the complex physics in cluster cores not directly related
  to the cluster as a whole. Our results suggest that the central
  temperature does not correlate with the mass of the BCGs and weakly
  correlates with the expected radiative cooling only for strong
  cool-core clusters. Since 88\% of the clusters in our sample have a
  BCG within a projected distance of 50~$h_{71}^{-1}$~kpc from the
  X-ray peak, we argue that it is easier to heat the gas (e.g. with
  mergers or non-gravitational processes) than to separate the dense
  core from the brightest cluster galaxy.

  \hspace*{0.5cm}Diffuse, Mpc-scale radio emission, believed to be
  associated with major mergers, has not been unambiguously detected
  in any of the strong cool-core clusters in our sample. Of the weak
  cool-core clusters and non-cool-core clusters most of the
  clusters~(seven out of eight) that have diffuse, Mpc-scale radio
  emission have a large ($>$50~$h_{71}^{-1}$~kpc) projected separation
  between their BCG and X-ray peak. In contrast, only two of the 56
  clusters with a small separation between the BCG and X-ray peak
  ($<$50~$h_{71}^{-1}$~kpc) show large-scale radio emission. Based on
  result, we argue that a large projected separation between the BCG
  and the X-ray peak is a good indicator of a major merger. The
  properties of weak cool-core clusters as an intermediate class of
  objects are discussed. Finally we describe individual properties of
  all 64 clusters in the sample.
}

\authorrunning{Daniel~Hudson~et~al.}  \titlerunning{What is a
  Cool-Core Cluster?}
\maketitle

\section{Introduction}
\label{Intro}
Early X-ray observations of galaxy clusters revealed that the
intracluster medium (ICM) in the centers of many clusters was so dense
that the cooling time of the gas was much shorter than the Hubble time
\citep[e.g][]{lea73,cowie77,fabian77,mathews78}.  These observations
led to the development of the cooling flow (CF) model.  In this model
the ICM at the centers of clusters with dense cores hydrostatically
cools, so that the cool gas is compressed by the weight of the
overlying gas.  Hot gas from the outer regions of the ICM flows in to
replace the compressed gas, generating a CF.  Although early X-ray
observations seemed to corroborate this model and there was some
evidence of expected H$\alpha$ emission and UV emission, optical
observations failed to detect the expected star formation rates, CO
and molecular gas \citep{1989AJ.....98.2018M,2001MNRAS.328..762E}.
Later, discrepancies between spectrally determined mass deposition
rates from {\it ASCA} and the classical determination from density
further brought the classical CF model into question \citep[][ and
references therein]{Makishima2001}. More recently, grating spectra
from {\it XMM}
\citep{2001A&A...365L.104P,2001A&A...365L..87T,2001A&A...379..107T,kaastra01,2002ApJ...579..600X,2002A&A...391..903S,peterson03,2008MNRAS.385.1186S}
have revealed that the gas in central regions is not cooling at the
rates predicted by the traditional cooling flow model. The classical
CF model assumes that no mechanism exists to significantly heat the
gas.  Therefore, this discrepancy has inspired a search for heating
models that can explain current observations.  Among those mechanisms
proposed are: conduction \citep[e.g.][]{zakamska03}, central AGN
heating via direct cosmic ray-ICM interaction+conduction
\citep{guo08}, AGN heating by bubble induced weak shocks
\citep[e.g.][]{mathews06}, soundwaves+conduction
\citep[e.g.][]{ruszkowski04} and turbulence+conduction
\citep[e.g.][]{dennis05}.  On the observational side recent studies
\citep[e.g.][]{birzan04,rafferty06,dunn06,dunn08,birzan08} have looked
at the mechanical energy released by AGN driven bubbles and compared
it to the energy needed to quench the CF.

The failure of the classical CF model has changed the nomenclature of
these centrally dense clusters from CF clusters to cool-core (CC)
clusters as suggested by \citet{molendi01}.  One problem with this
nomenclature is that it is unclear what distinguishes a CC cluster
from a non-CC (NCC) cluster.  The name implies that gas in the center
of the cluster is cool, but does that always imply a short cooling
time?  In fact authors define CC clusters differently often based on
central temperature drop \citep[e.g][]{sanderson06,burns07}, short
central cooling time \citep[e.g.][]{bauer05,ohara06,donahue07}, or
significant classical mass deposition rate \citep{chen07}.  There also
is a question as to whether there is a distinct difference between NCC
and CC clusters.  That is, is there a parameter that unambiguously
distinguishes NCC clusters from CC clusters?  This is a nontrivial
question since, when used as cosmological probes, clusters are often
segregated into CC/NCC subsamples.  Frequently CC clusters are chosen
for mass determination studies since they are believed to be
dynamically relaxed.  On the other hand, it requires a large amount of
energy to quench a CF, which may strongly affect the entire ICM.  For
example, \citet{ohara06} found that CC clusters (as defined as having
a central cooling time (CCT) more than 3~$\sigma$ below
7.1~$h_{70}^{-1/2}$~Gyr) have more scatter about scaling relations
than NCC clusters.  Therefore it becomes important to unambiguously
differentiate between CC and NCC clusters before proceeding with
determining their effects on scaling relations.

It is worth noting that a discrepancy between mass deposition rates
determined spectroscopically and those from images in some clusters
does not indicate in itself any breakdown of the simple cooling flow
picture, since it may be that these clusters just did not have enough
time to deposit the predicted level of mass. Here, we will show for
the first time that this recent formation solution to the cooling flow
discrepancy is very unlikely, based on statistical arguments.

Recently, \citet{chen07} investigated the cores of a sample of 106
galaxy clusters based on the extended HIghest X-ray FLUx Galaxy
Cluster Sample ({\it HIFLUGCS}) sample \citep{reiprich02} using {\it
  ROSAT} and {\it ASCA}, with their primary goal to study the effect
of CC versus NCC clusters on scaling relations.  They defined a CC
cluster as one that has significant classically defined mass
deposition rate ($\dot{M}_{\rm classical}$) and found 49\% of their
sample were CC clusters.  Additionally they determined that the
fraction of CC clusters in their sample decreased with increasing
mass.  \citet{sanderson06} investigated the cores of a statistically
selected sample of the 20 brightest\footnote{The Coma, Fornax, and
  Centaurus clusters were excluded because of their large angular
  size.} {\it HIFLUGCS} clusters using {\it Chandra}.  They find that
nine ($\sim$41\%) of their clusters are CC clusters, where they define
CC clusters as having a significant drop in temperature in the central
region.  Additionally they find that the slope of the inner
temperature profiles has a bimodal distribution with NCC clusters
having a flat temperature profile and CC clusters having a slope
(${\rm d}\log (T)/{\rm d}\log (r)$) of 0.4.  \citet{burns07}
investigated the properties of CC and NCC clusters from a cosmological
simulation which produced both types of clusters.  As with
\citet{sanderson06}, they define CC clusters based on the central
temperature decrease (by 20\% of T$_{\rm vir}$ in their case).  Their
simulations show that an early merger determines whether a cluster is
a CC cluster or not and suggest that CC clusters are not necessarily
more relaxed than NCC clusters.

Additionally, studies have begun on samples of distant (high-$z$)
clusters.  The goal of these studies is to determine the physical
evolution of CC clusters and whether the CC fraction changes with
redshift.  The major obstacle for these high-$z$ cluster studies is
that the traditional indications of CCs are difficult to measure for
distant clusters.  Therefore authors offer proxies, based on studies
of low-$z$ clusters, that can be measured for distant clusters with
limited signal.  \citet{vikhlinin07} suggested cuspiness as a proxy
for determining whether a distant cluster is a CC cluster or not.
Cuspiness is defined as:
\begin{equation}
  \alpha \equiv -\frac{\mathrm{d} \log(n)}{\mathrm{d} \log(r)} \;{\rm at}\; r = 0.04 r_{500},
\label{cuspiness}
\end{equation}
where $n$ is the electron density and $r$ is the distance from the
cluster center.  For their sample of 20 clusters with $z$ $>$ 0.5,
they find a lack of CC clusters.  That is, they find no {\it strong}
CC clusters ($\alpha$ $>$ 0.75 by their definition) and few weak CC
clusters ($\alpha$ = 0.5--0.75).  Approximately 67\% of their nearby
sample (a subsample of 48 {\it HIFLUGCS} clusters) has $\alpha$ $>$
0.5 versus 15\% for their $z$ $>$ 0.5 sample.  On the other hand,
using simulations, \citet{burns07} find that the CC fraction stays
constant back to a redshift of $z$ $\sim$1. Recently \citet{santos08}
compared a nearby sample of 11 clusters ($z = 0.15-0.3$) to a distant
sample of 15 clusters ($z = 0.7-1.4$). They suggest a concentration
parameter as a proxy for identifying CC clusters:
 \begin{equation}
C_{\rm SB}~\equiv~\frac{\Sigma(r<40~{\rm kpc})}{\Sigma(r<400~{\rm kpc})},
\label{concen}
\end{equation}
where $\Sigma(r<40\;{\rm kpc})$ and $\Sigma(r<400\;{\rm kpc})$ are the
integrated surface brightnesses within 40~kpc and 400~kpc
respectively.  As with \citet{vikhlinin07}, they find no strong CC
clusters in their high redshift sample.  On the other hand they find
roughly the same fraction of weak (or moderate) CC clusters in the
distant as in the local sample, in apparent conflict with
\citet{vikhlinin07}.  However, since both the nearby and distant
samples of \citet{santos08} are not complete, their fractions may be
biased by selection factors, so a straightforward comparison is not
possible.

In this paper we investigate a statistically complete, flux-limited
sample of 64 X-ray selected clusters and analyze their cores with the
{\it Chandra} {\it ACIS} instrument.  This is the first detailed core
analysis of a nearby, large, complete sample with a high resolution
instrument.  With its $\sim$0$\farcs$5 point spread function, {\it
  Chandra} is the ideal instrument for such a study.  Likewise, the
{\it HIFLUGCS} sample, which comprises the 64 X-ray brightest clusters
outside the Galactic plane \citep{reiprich02}, is an ideal sample.
Since the clusters are bright, they are nearby ($\langle
z\rangle=0.053$) making it possible to probe the very central regions
of the clusters.  Moreover they have high signal to noise allowing
precise measurements to be made.  It is worth noting that complete
flux-limited samples are not necessarily unbiased or representative
with respect to morphology \citep[e.g.][]{2006A&A...453L..39R}. The
point is that, for flux-limited samples, the bias can, in principle,
be calculated (e.g., Ikebe et al. 2002, Stanek et al. 2006). Also,
such samples are representative in the sense that their statistical
properties, like cooling core frequency, are directly comparable to
the same properties of simulated flux-limited samples. Samples
constructed based on availability of data in public archives are, in
general, not representative in this sense (``archive bias'').

Our goal in this paper is to determine if there is a physical property
that can unambiguously segregate CC from NCC clusters.  We use this
property to examine other parameters to see how well they may be used
as proxies when determining whether a cluster is a CC cluster or not.
The article is organized as follows. We outline our methods of data
reduction in Sect.~\ref{OaM}. We present our results in
Sect.~\ref{RaA}, wherein we describe the various parameters
investigated for determining whether a cluster is a CC or an NCC
cluster in Sect.~\ref{BaH}, determine the best parameter to segregate
distant CC and NCC clusters in Sect.~\ref{define_param}, compare this
parameter to others in Sect.~\ref{comparison} and determine the best
diagnostic for cool cores in distant clusters in
Sect.~\ref{bestproxy}. We discuss our results in Sect.~\ref{disc},
wherein we describe the basic cooling flow problem in
Sect.~\ref{CC_PROB} and discuss the inner temperature profiles in
Sect.~\ref{kTProfiles}, the central temperature drop seen in some
clusters in Sect.~\ref{kTProfiles}, the relation between cluster
mergers and the projected separation between the BCG and X-ray peak in
Sect.~\ref{bcgsepmerger} and the WCC clusters in
Sect.~\ref{wccclusters}. Finally we give our conclusions in
Sect.~\ref{concl}. To aid clarity, we give in Table~\ref{Nomenclature}
some of the abbreviations used throughout the paper.

\begin{table}
  \centering
  \caption{Nomenclature}
  \label{Nomenclature}
  \begin{tabular}{c c}
    \hline\hline
    Abbreviation &  \\
    \hline
    ICM               & IntraCluster Medium   \\                      
    CCT              & Central Cooling Time at 0.4\%$\rvir$ \\
    SCC              & Strong Cool Core \\
    WCC             & Weak Cool Core \\
    NCC             & Non Cool Core \\
    CC                & Cool Core (SCCs+WCCs)\\
    CF                & Cooling Flow \\
    BCG              & Brightest Cluster Galaxy\\
    EP                 & Emission Peak (X-ray) \\
    BCG-EP         & Projected separation between BCG and EP \\
    \hline
  \end{tabular}
    \label{abb}
\end{table}

In this work, we assume a flat $\Lambda$CDM Universe with $\Omega_{\rm
  M}$ = 0.3, $\Omega_{\rm vac}$ = 0.7, and $H_{0}$ =
$h_{71}$~71~km~s$^{-1}$~Mpc$^{-1}$.  Unless otherwise noted, $k$ is
the Boltzmann constant and we use the following nomenclature for our
coordinates: $x$ is the projected distance from the cluster center,
$r$ is the physical 3-D distance from the cluster center and $l$ is
the distance along the line of sight.  All errors are quoted at the
1~$\sigma$ level unless otherwise noted.

\section{Observations and Methods}
\label{OaM}
All 64 {\it HIFLUGCS} clusters have been observed with {\it Chandra},
representing 4.552 Ms of cleaned data.  For our analysis we used all
unflared data taken after the CCD focal plane temperature was reduced
to -120 $^\circ$C (2001-Jan-29) and was publicly available as of
2007-May-07, with a few exceptions.  (1)~In cases where more than one
data set was publicly available and one set was heavily flared, the
entire flared data set was discarded, (2)~for A754\footnote{There are
  later observations of A754, but they are all far offset and not
  useful for core studies and therefore not included.} and
A401\footnote{There is a post-2001-Jan-29 observation of A401, however
  it is offset such that the cluster center is in the corner of a CCD.
  We therefore included the centered pre-2001-Jan-29 observation as
  well.}, (3)~in the cases of A2597 and A2589 the original
observations were heavily flared and the PIs (T. Clarke and D. Buote,
respectively) of the newer proprietary data sets graciously provided
them to us before they became publicly available (4)~the Coma Cluster
(A1656), has a recently released calibration observation (2008-Mar-23)
which we used instead of the pre-2001-Jan-29 observations.  Several
other PI's kindly provided data sets before they were publicly
available, although they since have become publicly available (see
acknowledgments).  Details on the number and length of the
observations can be found in Table~\ref{obs}.  Note that we use
redshifts compiled from NASA/IPAC Extragalactic Database
(NED)\footnote{{\tt http://nedwww.ipac.caltech.edu}} and the values
did not differ significantly ($\Delta$z$<$ 0.001) from the values used
by \citet{reiprich02}.

\addtocounter{table}{1}

\subsection{Data Reduction}
\begin{figure}
\includegraphics[width=90.0mm,angle=0]{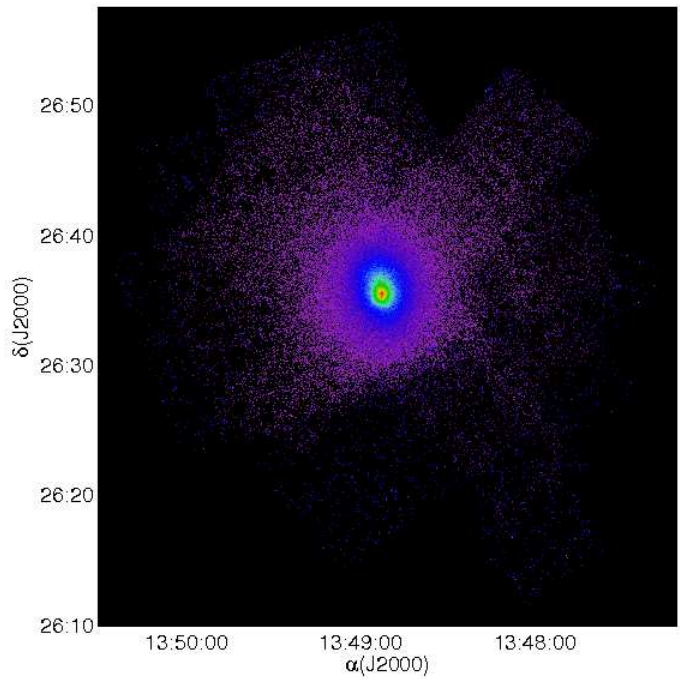}
\caption[A1795 Image] {An example of a mosaiced image.  This is the
  background subtracted, exposure corrected image created from the 12
  {\it Chandra} exposures of A1795.  \label{A1795_Image}}
\end{figure}

The basic data reduction was done using CIAO 3.2.2 and CALDB 3.0
following the methods outlined in \citet{hudson06}.  Since we are
interested in the cores of clusters, we took the cluster center to be
the emission peak~(EP) from the background subtracted, exposure
corrected image.  See \citet{hudson06} for details of our image
creation technique\footnote{For this analysis we did not apply
  adaptive smoothing.}.  To avoid statistical fluctuations in
determining the cluster peak, we smoothed all images with a Gaussian.
In general we used a $\sim$4$\arcsec$ kernel, however in a few
particularly noisy cases (noted in Appendix~\ref{noic}) we used a
larger smoothing kernel.  Fig.~\ref{A1795_Image} shows an example of a
mosaiced image.

\subsection{Temperature Profiles}
\label{tp}
Using the EP as the cluster center we created a temperature profile
for each cluster using the following method.  In order to get good
statistics for our temperature profiles, we estimated that 10\,000
source counts per annulus would give us a $\sim$5\% constraint on $T$.
However, in several cases we had a small number of total source counts
($\sim$30\,000), so that using 10\,000 source counts per annulus was
not practical.  We therefore extracted annuli with at least 5000
source counts for any cluster with fewer than 100\,000 total source
counts and 10\,000 source counts for any cluster with more than
100\,000 total source counts.  We created {\it pie} shaped annuli, not
extending beyond the boundary of the chips.  Although this decreased
our efficiency, it allows for unbiased density measurements when using
the normalization of the thermal model (a full circle extending beyond
the chip edge would generally cover more of the inner part of the
annulus than the outer).  In our subsequent studies it will also allow
us to use the same spectra for deprojection models as well as helping
to constrain the residual cosmic X-ray background (CXB) in the outer
cluster regions\footnote{The normalizations for the cluster emission
  for partial annuli of different observations can be scaled and tied
  when trying to determine the residual CXB.}.  We then created
spectra and fit the spectrum to an absorbed thermal model
(WABS*APEC*EDGE\footnote{The edge component is a correction to
  underestimation of {\it Chandra's} efficiency at $\sim$2 keV
  \citep[see][]{vikhlinin05}.}).  For the majority of observations we
used the radio measured hydrogen absorption columns from the
Leiden/Argentine/Bonn HI-Survey \citep{kalberla05} or, when available,
pointed observations from E. Murphy (private communication).  In a few
cases, noted in Appendix~\ref{noic}, the X-ray measured value was
significantly higher than the radio measured value and so we left the
hydrogen absorption column as a free parameter.  See
\citet{baumgartner06} for an explanation of the discrepancy between
radio and X-ray absorption column density measurements.  As noted in
Table~\ref{par} and Appendix~\ref{noic}, there were 14 cases
($\sim$22\%) in which we added a second thermal component to one or
more of the central annuli in order to improve the fit.  In these
cases, the lower of the two temperatures was used for our annular
analysis.  There were also a few cases in which using non-solar
metalicity ratios significantly improved the fit, however since the
best-fit temperature was unaffected, solar ratios were used for
simplicity.  See Appendix~\ref{noic} for notes on the clusters where
this was the case.

\subsection{Cluster Virial Temperature, Radius and Mass}
\label{kTvir}
The cluster virial temperature ($T_{\rm vir}$) is the temperature of
the X-ray emitting gas that is in hydrostatic equilibrium with the
cluster potential.  Although $T_{\rm vir}$ is not an imperative
quantity for core studies, it is useful for scaling clusters for
comparison purposes (e.g. relative temperature drop) or estimating the
cluster virial radius and mass.  In this section we describe our
method for determining $T_{\rm vir}$ and how we use it to estimate the
virial mass and radius.

%%%%% Check "is _this_ largest source of bias..."  

%%%%% Check "Additionally we removed the outer annuli...clusters _may
%%%%% a_ have"

\begin{figure}
\includegraphics[width=90.0mm,angle=0]{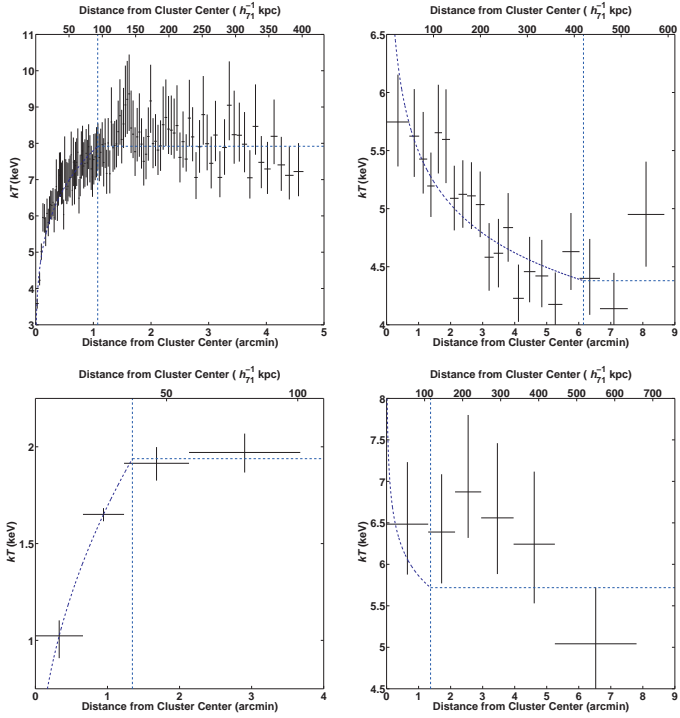}
\caption[A2029 $T$ Profile] {These figures show the broken powerlaw
  fit to four typical temperature profiles representing long and short
  exposures of clusters with a central temperature drop and clusters
  without a central temperature drop. The dashed blue lines indicate
  the fit and the core radius. {\it Top left}: A2029 - a long
  exposure cluster with a central temperature drop. {\it Top right}:
  A3158 - a long exposure of a cluster without a central temperature
  drop. {\it Bottom left}: A3581 - a short exposure of a cluster with
  a central temperature drop. {\it Bottom right}: ZwCl1215 - a short
  exposure of a cluster without a central temperature drop.
  \label{exprof}}
\end{figure}

The temperature decline in the central regions of some clusters, if
included, is the largest source of bias in the determination of
$T_{\rm vir}$. The reason for this temperature drop is a current topic
of debate and discussed in more detail in Sect.~\ref{kTProfiles}. Here
we simply discuss the removal of this central region, so it does not
bias the fit to $T_{\rm vir}$. In order to determine the size of the
central region to be excluded, we fit our temperature profiles to a
broken powerlaw. The {\it core} radius in the powerlaw was free and
the index of the outer component was fixed to be zero. Additionally we
removed the outer annuli where accurate subtraction of the blank sky
backgrounds~(BSB) becomes critical and where clusters may have a
decreasing temperature profile
\citep{1998ApJ...503...77M,2002ApJ...567..163D,vikhlinin05,burns07}.
The core radius in the powerlaw was taken to be the radius of the
excluded central region. Table~\ref{obs} gives this {\it core} radius
as well as $kT_{\rm vir}$ and overall cluster
metalicity. Fig.~\ref{exprof} shows examples of the broken powerlaw
fit to the temperature profiles of four representative clusters: (1)~a
long exposure of a cluster that has a central temperature drop, (2)~a
long exposure of a cluster without a central temperature drop, (3)~a
short exposure of a cluster with a central temperature drop and (4)~a
short exposure of a cluster without a central temperature drop. If the
core radius in the powerlaw was consistent with zero and/or the
temperature gradient~($\Gamma$ in column 5 of Table~\ref{obs}) was
positive, no central region was removed when determining $T_{\rm
  vir}$. This method allowed us to take full advantage of the largest
possible signal, while being certain that $T_{\rm vir}$ was not biased
by cool central gas. Our results suggest that using a fixed fraction
of the virial radius is not efficient and can be dangerous since the
{\it core} radius, as defined above, does not scale with the cluster
virial radius.

Since we include all data outside of the {\it core} radius when
determining $T_{\rm vir}$, there are regions where residual background
becomes important.  This was especially true for some of our more
distant clusters (e.g. RXJ1504, $z$ = 0.2153).  The residual
background results from three factors: (1)~since the BSB is scaled to
remove the particle background by matching the BSB 9.5 - 12 keV rate
to that of the observation, if the scaling factor is much different
from unity, the cosmic X-ray background (CXB) in the BSB will be
significantly increased or decreased before it is subtracted, (2)
likewise, the soft CXB component varies over the sky
\citep[e.g.][]{kuntz01} so the CXB in the observation and BSB will be
different and (3)~in back-illuminated (BI) CCDs, there are sometimes
residual soft flares.  To compensate for these effects a second soft
thermal component was included as well as, in the BI CCDs, an
un-folded powerlaw (i.e. a powerlaw that is not folded through the
instrument response). The temperature of the soft thermal component
was not allowed to exceed $1$~keV with a frozen solar abundance and
zero redshift. The normalization of the soft component was also
allowed to be negative in the case of oversubtraction of the CXB
(e.g. for a large scaling factor).  For all CCDs of the same type
(i.e. front or back illuminated) in a given observation, we tied the
residual background components with the normalization scaled by area.
We emphasize that for most clusters this step was taken as a
precaution since the cluster emission dominated over the residual CXB.
Only in the most distant clusters of our sample, where the outer
regions are beyond $\sim$0.5 $r_{\rm vir}$, is this correction
essential.  As an example the best-fit temperature for RXJ1504 without
this correction is 7.1 - 7.9 keV versus 8.4 - 10.9 keV with the
residual BSB correction.  On the other hand for A0085 ($z$ = 0.055) no
significant difference in temperature is found with or without the
correction.

Once we measured $T_{\rm vir}$, we used it to determine a
characteristic cluster radius and mass.  In the case of radius we used
the scaling relation determined by \citet{evrard96}:
\begin{equation}
  \left ( \frac{r_{j}}{h_{71}^{-1}~{\rm Mpc}} \right ) = N_{j}  \left ( \frac{kT_{\rm vir}}{10\;{\rm keV}} \right )^{\frac{1}{2}} ,
\label{r500}
\end{equation}
where $j$ is the average overdensity, within $r_{j}$, above the
critical density ($\rho_{c}$ $\equiv$ 3$H_{0}^2$/$8\pi G$) desired and
$N_{j}$ is the normalization.  We use normalizations $N_{500}$ = 2.00
and $N_{180}$ = 2.75, and define $r_{\rm vir}$ $\equiv$ $r_{180}$.
%\citep{evrard96}.
We note that these values may be overestimates of $r_{500}$ and
$r_{180}$ since they are based on early simulations
\citep[e.g.][]{finoguenov01}.  For the purposes of core studies,
however, this approximation should provide an adequate approximation
of a characteristic cluster radius.

To estimate a characteristic cluster mass ($M_{500}$), we used the
formulation of \citet{finoguenov01} derived from the {\it ROSAT} and
{\it ASCA} observations of the {\it HIFLUGCS} clusters,
\begin{equation}
  % M_{500} = 3.53^{+0.33}_{-0.30}~h_{50}^{-1}~\times 10^{13}~kT_{\rm
  %   vir}^{1.676 \pm 0.0054}~M_{\odot},
\left( \frac{M_{500}}{10^{13}~h_{71}^{-1}~M_{\odot}} \right ) = ( 2.5\pm0.2) ~ \left( \frac{kT_{\rm vir}}{1 ~{\rm keV}} \right )^{1.676 \pm 0.054}.
\label{M500}
\end{equation}

\subsection{Central Temperature Drops}
To quantify the central temperature drop we calculated two quantities
shown in columns 4 and 5 of Table~\ref{par}.  One is simply the
temperature of the central annulus ($T_{0}$) divided by the virial
temperature ($T_{\rm vir}$).  The second is the slope of the powerlaw
fit to the central temperature profile.  For clusters with a measured
central temperature drop (Sect.~\ref{kTvir}), we fit the slope out to
the break in the powerlaw.  For those clusters with no central
temperature drop, we fit the three innermost annuli\footnote{In a few
  cases there were clusters with central drops that did not have three
  bins out to the core radius.  In these cases we also fit the
  powerlaw to the inner three annuli.}.  The purposes of this
measurement was (1)~to check for a universal central temperature
profile shape and (2)~differentiate between a true systematic drop and
random temperature fluctuations seen in merging clusters (e.g. A0754).

\subsection{Density Profiles}
As with the temperature profile, the surface brightness profile
becomes uncertain at large radii due to uncertainties in the residual
background.  Since we are only interested in the central regions and
the surface brightness falls off rapidly at large radii, we focused on
fitting the central regions only.  We argue that any uncertainty in
the shape of the profile in the outer regions has a negligible effect
when deprojecting the central regions.  Since one of the parameters we
are interested in, cuspiness (see Sect.~\ref{cusp} and
Eqn.~\ref{cuspiness}), is defined in terms of the derivative of the
$\log$ of the density profile at 0.04~$r_{500}$,%:
%\begin{equation}
%\alpha \equiv -\frac{\mathrm{d} {\log}(n)}{{\mathrm{d} {\log}(r)} \;{\rm at}\; r = 0.04 r_{500}
%\label{cuspiness}
%\end{equation}
%\citep{vikhlinin07}, we defined the central region as being 20\%
% larger than the radius where $\alpha$ is measured ($r_{\rm
%   central}\equiv$0.048 $r_{500}$).
we decided to use a slightly larger (20\%) region for determining the
density profiles. This way we would have a constraint on the
derivative of the $\log$ of the density profile at 0.04~$r_{500}$.
Specifically, we extracted a spectrum from the projected central
region (0-0.048~$r_{500}$) and fit it with an absorbed thermal model
(WABS*APEC*EDGE).  We used the best-fit parameters of this model to
create a {\it spectrumfile}\footnote{See {\tt
    http://cxc.harvard.edu/ciao/ahelp/mkinstmap.html} for the
  definition of a {\it spectrumfile}.}, which we used in turn to
create a weighted exposure map\footnote{See {\tt
    http://space.mit.edu/CXC/docs/expmap\_intro.ps.gz} for more
  details on this procedure.}.

We then created a background subtracted, exposure corrected image, in
the 0.5 - 7.0 keV range, similar to the method described in
\citet{hudson06}.  The difference in our method here and that
described in \citet{hudson06} is that: (1)~we only used a single
weighted exposure map instead of many monochromatic exposure maps for
different energy bands, (2)~we created an {\it error image}
($\sigma_{\rm img}$) and (3)~we created a background subtracted image
with no exposure correction.  The {\it error image} was created
assuming the errors in the observation, background and {\it readout
  artifact} or out-of-time events (OOTs) were Poisson ($\sqrt{N}$) and
could be added in quadrature.  % Specifically the {\it error image} is
% defined as the square of the net uncertainty ($\sigma_{\rm img}$
% $\equiv$ $\sigma_{\rm net}^2$), where
% \begin{equation}
% \sigma_{\rm net} = \sqrt{n_{\rm obs} + n_{\rm bkg} + n_{\rm OOTs}},
% \end{equation}
% and $n_{\rm obs}$, $n_{\rm bkg}$, $n_{\rm OOTs}$ are the number of
% counts in a given pixel from the observation, background and OOTs,
% respectively.  For an extracted region of $k$ pixels, the
% net %observation, background and OOT
% source counts ($N$) are:
% \begin{equation}
%   N = \sum_{i=1}^{k} \left( n_{{\rm obs}_{i}} - n_{{\rm bkg}_{i}} - n_{{\rm OOTs}_{i}} \right ).
% \end{equation}
% Therefore the uncertainty in a given extraction region is:
% \begin{equation}
% \sigma_{\rm net} = \sqrt{\sum n_{\rm obs} + \sum n_{\rm bkg} + \sum n_{\rm OOTs} },
% \end{equation}
% and the sum of the values of the pixels in the error image are:
% \begin{equation}
%   \sum \sigma_{\rm img} = \sum (n_{\rm obs} + n_{\rm bkg} +  n_{\rm OOTs} ) = \sigma_{\rm net}^2.
% \end{equation}
% Therefore the uncertainty of any extraction region is the square root
% of the sum of the values of the {\it error image} in the given region.

We used the background subtracted image with no exposure correction to
determine annuli with at least 500 source counts\footnote{In a few
  cases, if the $\beta$ model fit was poor, this was reduced to $>$100
  counts.} per annulus and extracted a surface brightness profile and
errors from the background subtracted exposure corrected image and the
error image respectively.  The surface brightness profile was then fit
to a single $\beta$ model:
\begin{equation}
  \Sigma = \Sigma_{0} \left [ 1 + \left( \frac{x}{x_{c}} \right )^{2} \right ]^{-3\beta + 1/2}
\label{sbbeta}
\end{equation}
or a double $\beta$ model:
\begin{equation}
  \Sigma = \Sigma_{01} \left [ 1 + \left( \frac{x}{x_{c_{1}}} \right )^{2} \right ]^{-3\beta_{1} + 1/2} + \Sigma_{02} \left [ 1 + \left( \frac{x}{x_{c_{2}}} \right )^{2} \right ]^{-3\beta_{2} + 1/2}
\label{sbdbeta}
\end{equation}
\citep{cavaliere76}, where $x_{c}$ is the core radius, $\Sigma_{0}$
($\Sigma_{0}$ = $\Sigma_{01}$ + $\Sigma_{02}$ for a double $\beta$
model) is the central surface brightness and $\beta$ is constrained to
be less than 2.

From $\beta$ and $x_{c}$ for a fit to a $\beta$ model ($\beta_{1}$,
$\beta_{2}$, $x_{c_1}$ and $x_{c_2}$ for a fit to a double $\beta$
model), we extracted the shape of the density profile.
\begin{equation}
  n = n_{0} \left [ 1 + \left ( \frac{r}{r_{c}} \right )^{2} \right ]^{-\frac{3 \beta}{2}}
\label{sbm}
\end{equation}
and
\begin{equation}
  n = \left ( n_{01}^2 \left [ 1 + \left ( \frac{r}{r_{c_1}} \right )^{2} \right ]^{-3 \beta_1} + n_{02}^2 \left [ 1 + \left ( \frac{r}{r_{c_2}} \right )^{2} \right ]^{-3 \beta_2} \right )^{\frac{1}{2}}
\end{equation}
for a single and double $\beta$ model respectively, where $n_{0}$ is
the central electron density, $r_{c}$ is the physical core radius
associated with $x_{c}$ for our cosmology and $n_{0} = \sqrt{n_{01}^2
  + n_{02}^2}$.  In order to calculate the normalization ($n_{0}$) of
the model, we used the best-fit spectral parameters from the 0-0.048
$r_{500}$ regions, as described below.

For a single $\beta$ model the central electron density, $n_{0}$ is:
\begin{equation}
  % n_{0} = \left ( \frac{10^{14} \: 4\pi \: D_{A} \: D_{L} \; \zeta
  %     \; {\rm NORM}}{EI} \right )^{\frac{1}{2}},
  n_{0} = \left ( \frac{10^{14} \: 4\pi \: D_{A} \: D_{L} \; \zeta \; {\mathcal N}}{{\rm EI}} \right )^{\frac{1}{2}},
\label{n0-beta_model}
\end{equation}
where ${\mathcal N}$ is the normalization of the {\it APEC} model,
$D_{A}$ is the angular diameter distance, $D_{L}$ is the luminosity
distance, $\zeta$ is the ratio of electrons to protons ($\sim 1.2$)
and EI is the emission integral divided by the central density:
\begin{equation}
{\rm EI} \equiv \int{\left ( \frac{n}{n_{0}} \right )^2~\mathrm{d}V}.
\label{EI1}
\end{equation}
Inserting Eqn.~\ref{sbm} into Eqn.~\ref{EI1}, we get:
\begin{equation} {\rm EI} = 2\pi \int_{-\infty}^{\infty} \int_{0}^{R}
  x~\left( 1 + \frac{x^2 + l^2}{x_{c}^{2}} \right)^{-3\beta} \mathrm{d}x
  \mathrm{d}l
\label{EI}
\end{equation}
for the 0-0.048~$r_{500}$ region fitted with the {\it APEC} model
\citep{mewe85,mewe86,smith00,smith01b} and $R$ = 0.048~$r_{500}$.

For a double $\beta$ model the central electron density, $n_{0}$ is:
\begin{equation}
  n_{0} = \left [ \frac{10^{14} \: 4\pi\;  (\Sigma_{12}\:{\rm LI}_{2} + {\rm LI}_{1}) \: D_{A} \: D_{L} \; \zeta \:{\mathcal N}}{\Sigma_{12} \:{\rm LI}_{2}\:{\rm EI}_{1} + {\rm LI}_{1}\: {\rm EI}_{2}} \right ]^{\frac{1}{2}},
\label{n0-dbeta_model}
\end{equation}
with the same definition of variables as Eqn.~\ref{n0-beta_model}.
Additionally, $\Sigma_{12}$ is the ratio of the central surface
brightness of model$-{1}$ to model-${2}$, EI$_{i}$ is the emission
integral for model$-{i}$ as defined by Eqn.~\ref{EI}, and LI$_{i}$ is
the line emission measure for model$-{i}$, defined as,
\begin{equation} 
{\rm LI}_{i} \equiv \int_{-\infty}^{\infty} \left( 1
    + \frac{l^2}{x_{c_{i}}^{2}} \right)^{-3\beta_{i}} \mathrm{d}l
\label{LI}
\end{equation}
See Appendix~\ref{DBCalc} for details on this calculation.

\subsection{Cooling Times}
\label{cooltimes}
Once we created the density profiles, we used them, together with the
cooling function of the best-fit temperature, to estimate the average
cooling time of the gas using:
\begin{equation}
\label{tcooldefn}
t_{\rm cool} = \frac{3}{2} \frac{(n_{\rm e}+n_{\rm i})kT}{n_{\rm e} n_{\rm H} \Lambda(T,Z)},
\label{cteqn}
\end{equation}
where $n_{\rm i}$ and $n_{\rm e}$ are the number density of ions and
electrons respectively. The cooling function, $\Lambda(T,Z)$, was
calculated by spline interpolation on a table of values for the {\it
  APEC} model, for an optically thin thermal plasma, kindly provided
by R. K. Smith. We define the central cooling time (CCT) as:
\begin{equation}
  t_{\rm cool}(0) = \frac{3}{2} \zeta~\frac{(n_{{\rm e}0}+n_{{\rm i}0})kT_{48}}{n_{{\rm e}0}^2 \Lambda(T_{48})},
\label{ccteqn}
\end{equation}
where $n_{{\rm e}0}$ and $n_{{\rm i}0}$ are the central electron and
ion densities, respectively, and $T_{48}$ is the average temperature of
the \mbox{$0 - 0.048~r_{500}$} region.

We note here that in order to remove the bias introduced by different
physical resolutions due to different cluster distances, we took any
parameter calculated at $r$ = 0, (e.g. $n_{0}$ and CCT) to be the
value at $r_{0}$ $\equiv$ 0.004~$r_{500}$.  The exception to this is
$T_{0}$, which is defined as the average temperature of the central
annulus, which in all cases had a radius $<0.004$~$r_{500}$.  We
define $T_{\rm ctr}$ to be the central temperature of the cluster,
either $T_{48}$ or $T_{0}$ depending on the parameter considered.  In
general $T_{48}$ $\approx$ $T_{0}$ ($\langle T_{0}/T_{48} \rangle$ =
0.83).

\subsection{Entropy}
As is typically done in X-ray studies of galaxy clusters, we define
entropy as $K \equiv kTn^{-2/3}$ and central entropy as $K_{0}$ =
$kT_{0}$$n_{0}^{-2/3}$, where $n_{0}$ is the central electron density
and $T_{0}$ is the temperature of the innermost annulus from the
temperature profile. Additionally we define a purposely {\it biased}
entropy ($K_{\rm BIAS}$) that takes advantage of the inherent
difference in the density distribution of CC and NCC clusters to
further separate them.  Specifically, taking into account the fact
that CC clusters should have higher density cores, they will, in
general, have a smaller central annulus (in the temperature profile,
see Sect.~\ref{tp}) than NCC clusters (when fitting spectra).  That is,
the 10\,000 (or 5000) source count threshold for the central annulus
will be reached at a smaller radius for CC clusters than NCC clusters,
as the density is higher and typically the temperature is lower in
this region for CC clusters than for NCC clusters.  We then define the
density as the average density for the innermost annulus,
\begin{equation}
  n_{\rm annulus} = \left ( \frac{3 \times 10^{14} \: D_{A} \: D_{L} \; \zeta \; {\mathcal N}}{r_{\rm annulus}^3} \right )^{\frac{1}{2}},
\label{avgdens}
\end{equation}
where the variables are defined as before and $r_{\rm annulus}$ is the
physical radius of the central annulus.  Since the central annulus for
clusters with a central peak in density will be smaller, its value of
$n_{\rm annulus}$ will be closer to $n_{0}$ than for clusters without
a centrally peaked density.  Unfortunately, since we define $r_{\rm
  annulus}$ by a constant number of source counts there are additional
biases due to length of the observation and redshift of the cluster.
We argue, however that central density is a stronger effect.  As an
example we give the comparison of A2204 and A3158.  A2204 has a
redshift of $z = 0.1522$ with a good observation time of $\sim$18.6
ks.  A3158 - a cluster without a bright core, has a redshift of
$z=0.06$ and good observing time of $\sim$55.7 ks.  The innermost
annulus\footnote{For comparison purposes this is the radius for
  10\,000 counts. However when making the temperature profile for
  A2204 we used 5000 count annuli.} of A2204 has a radius of
7$\farcs$0 = 18.3~$h_{71}^{-1}$~kpc, smaller than the innermost
annulus of A3158 which has a radius of 42$\farcs$9 =
48.8~$h_{71}^{-1}$~kpc.  Even in the extreme case of A3667 ($z =
0.0566$) that has a good observing time ($\sim$485.3 ks) more than 20
times longer than A2204, the innermost annulus has a radius ($r_{\rm
  annulus}= 15\farcs$9 = 17.0~$h_{71}^{-1}$~kpc) comparable to A2204.
Based on this intentional bias, we define $K_{\rm BIAS} = kT_{0}
n_{\rm annulus}^{-2/3}$, where as with $K_{0}$, $T_{0}$ is the
temperature of the central annulus from the temperature profile.

\subsection{Mass Deposition Rates}
\label{MDR}
We define three mass deposition rates: the classically determined mass
deposition rate ($\dot{M}_{\rm classical}$), the spectrally determined
mass deposition rate ($\dot{M}_{\rm spec}$) and the modified
spectrally determined mass deposition rate ($\dot{M}_{\rm
  spec2}$). $\dot{M}_{\rm classical}$ is calculated from the gas
density and temperature assuming no energy input.  Using the
formulation of \citet{fabian79}, within radius $r$,
\begin{equation}
\label{mdotclasdef}
\dot{M}_{\rm classical}(<r) \simeq \frac{M_{\rm gas}(r)}{t_{\rm cool}(r) - t_{\rm cool}(0)}.
\end{equation}
Usually $\dot{M}_{\rm classical}$ is calculated at a certain radius
$r_{\rm cool}$, where we have chosen $r_{\rm cool}$ as the radius at
which $t_{\rm cool}$ = 7.7~$h_{71}^{-1/2}$~Gyr.  We chose 7.7~Gyr,
since it corresponds to the look back time of $z = 1$, a
representative period over which a cluster has time to relax and form
a cooling flow \citep{rafferty06}.  However, for uniformity in
comparison with our spectral mass deposition rates, we took $r$ to be
the smaller of 0.048~$r_{500}$ and $r_{\rm cool}$. Furthermore, to
check the effect of this cut, we extrapolated our density profiles to
calculate the classical mass deposition rates out to the cooling
radius, assuming the average gas characteristics remain the same and
the slope of the profile remains at our fit value. The mean cooling
radius was $\sim$0.07~$r_{500}$, and the mass deposition rates at the
cooling radius were indistinguishable, within the errors in
$\dot{M}_{\rm classical}$, from those calculated at 0.048~$r_{500}$.

\begin{figure}
\includegraphics[width=80.0mm,angle=0]{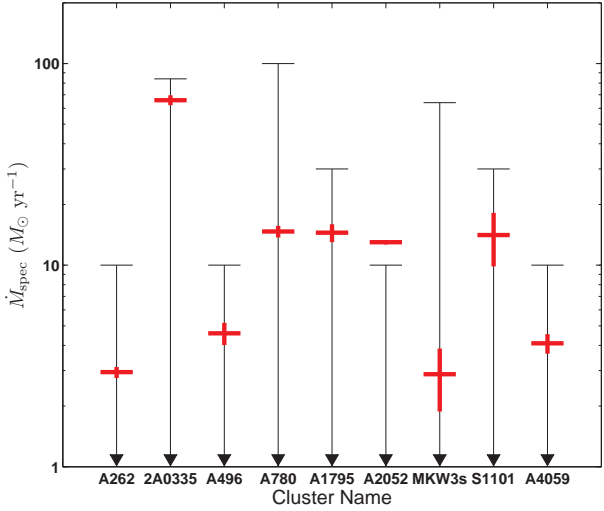}
\caption[{\it XMM RGS} - {\it Chandra ACIS} Spectroscopic Cooling
Rates] {{\it XMM RGS} - {\it Chandra ACIS} spectroscopic cooling
  rates. Here we plot the upper limits measured by \citet{peterson03}
  with the {\it XMM RGS} (black), compared to our {\it Chandra ACIS}
  measured values (thick red) for the nine clusters in common.  For
  eight of the nine clusters, the {\it Chandra ACIS} gives results
  consistent with the {\it XMM RGS}.  For the one cluster (A2052) for
  which they are not consistent, the value measured with the {\it
    Chandra ACIS} is $\sim$30\% higher (3 $M_{\odot}$~yr$^{-1}$) than
  the {\it XMM RGS} upper limit. This comparison corroborates the {\it
    Chandra ACIS} spectrally determined mass deposition
  rates.\label{RGS_acis}}
\end{figure}

$\dot{M}_{\rm spec}$ is a direct measurement of the amount of gas that
is cooling by fitting the expected line emission
\citep[e.g.][]{peterson03} from the multiphase gas.  In order to
determine $\dot{M}_{\rm spec}$, we fit the 0-0.048~$r_{500}$ region to
an absorbed thermal model with a cooling flow model
\citep{mushotzky88} (WABS*[APEC+MKCFLOW]*EDGE).  The higher
temperature component of the cooling flow model was tied to the
thermal model and the lower temperature component was frozen at 0.08
keV.  %zero\footnote{The actual value is 0.08 keV, the lowest
      %temperature value allowed by the model.}.
Given the moderate spectral resolution of the {\it Chandra} {\it
  ACIS}, we checked the reliability of the measured values.
Fig.~\ref{RGS_acis} shows the comparison between the values obtained
with the {\it XMM RGS} and the {\it Chandra ACIS} for the nine
clusters in common with our sample and the 14 clusters reported by
\citet{peterson03}.  Fig.~\ref{RGS_acis} shows that the values
obtained with the {\it Chandra} {\it ACIS} are, in all but one
case\footnote{In the case of A2052, the {\it Chandra} {\it ACIS} value
  is $\sim$30\% higher (although only a difference of 3
  M$_{\odot}$~yr$^{-1}$) than the upper limit obtained with the {\it
    XMM RGS}.}, consistent with the upper limits obtained by
\citet{peterson03}.  Based on this we argue that the values obtained
with the {\it Chandra ACIS} are reliable to well within an order of
magnitude for CC clusters.  As we discuss later, this value may not be
accurate for NCC clusters.

$\dot{M}_{\rm spec2}$ is determined similarly to $\dot{M}_{\rm spec}$,
except that the lower temperature component is left as a free
parameter.  $\dot{M}_{\rm spec2}$ represents the spectral fit to a
cooling flow model that is stopped at a lower-limit temperature
$T_{\rm low}$.  The interpretation of this model is discussed in more
detail in Sect.~\ref{CC_PROB}.

\subsection{Cuspiness}
\label{cusp}
\citet{vikhlinin07} suggested cuspiness as a proxy for identifying
cool-core clusters at large redshift.  Eqn.~\ref{cuspiness} gives the
definition of cuspiness.  For our low redshift sample, we can test the
correlation of cuspiness with the parameters used to define a cool
core.  Since our density profile is based on a model, we define it in
terms of the model parameters, with $r = 0.04~r_{500}$.  For a single
$\beta$ model
\begin{equation}
\alpha = \frac{3 \beta r^2}{ \left ( r_c^2 + r^2 \right )}.
\end{equation}
For a double $\beta$ model
\begin{equation}
  \alpha = 3 r^2 \frac{\Sigma_{12} {\rm LI}_{2} \beta_{1} r_{c_{1}}^{-2}b_{1}^{\prime} + LI_{1}\beta_{2}r_{c_2}^{-2} b_{2}^{\prime}}{\Sigma_{12} {\rm LI}_{2} b_{1} + {\rm LI}_{1} b_{2}},
\end{equation}
where $\Sigma_{12}$, LI$_{i}$ are defined as before and $b_{i} \equiv
\left ( 1 + \left ( \frac{r}{r_{c_i}} \right )^2 \right )^{-3
  \beta_{i}}$ and $b_{i}^{\prime} \equiv \left ( 1 + \left (
    \frac{r}{r_{c_i}} \right )^2 \right )^{-3 \beta_{i} - 1}$.

\addtocounter{table}{1}

\section{Results}
\label{RaA}

\subsection{Bimodality and Histograms}
\label{BaH}

In order to determine the best parameter to separate CC clusters from
NCC clusters, we tested 16 parameters for bimodality (and trimodality
in some cases) using the Kaye's Mixture Model (KMM) algorithm
\citep[e.g.][]{1994AJ....108.2348A}.  For each parameter we used a
single covariance for both (all three) subgroups when determining the
significance of the rejection of the single Gaussian hypothesis.  We
chose to do this because analytic errors are only statistically
meaningful when the same covariance is used for each subgroup.
However, since there is no reason to believe the bimodality in any
parameter is symmetric, we considered independent covariances for the
subgroups when determining the subgroup assignments.  Moreover the
analytic errors when using different covariances for each subgroup can
still give a good guideline on fit improvement
\citep[e.g.][]{1994AJ....108.2348A}, even if their true significance
is unknown.

Fig.~\ref{Histogram} shows histograms of the 16 parameters with
Gaussians (created from the KMM algorithm results) overplotted. For
each parameter the CC and NCC subgroups were independently determined.
We constructed Gaussians using the means and covariances returned by
the KMM algorithm and calculated the normalization of the Gaussians so
that the integral of the Gaussian was equal to the {\it area} of the
bins (sum of the number of clusters in each bin times the width of the
bin) within its relevant subgroup.  Bins of clusters from the CC
subgroup are colored blue and bins of clusters from the NCC subgroup
are colored red.  In the cases with a third subgroup between CC and
NCC clusters, the bins are colored black.

\subsubsection{Central Surface Brightness~(A): $\Sigma_{0}$}
The histogram for $\Sigma_{0}$ appears to have two peaks on either
side of $\sim$10$^{-6}$ photons cm$^{-2}$ s$^{-1}$ arcsec$^{-2}$ (0.5
- 7.0 keV). They are, however, not well separated and the
KMM-algorithm does not reject a single Gaussian hypothesis for this
data set ($<$85\% confidence).  If divided into two Gaussians, the KMM
algorithm finds a partition at $\sim$0.80 $\times$ 10$^{-6}$ photons
cm$^{-2}$ s$^{-1}$ arcsec$^{-2}$, which divides that sample into 37 CC
clusters and 27 NCC clusters.  The centers of the subgroups are 3.88
$\times$ 10$^{-6}$ photons cm$^{-2}$ s$^{-1}$ arcsec$^{-2}$ and 0.32
$\times$ 10$^{-6}$ photons cm$^{-2}$ s$^{-1}$ arcsec$^{-2}$ for the CC
and NCC clusters respectively.

\subsubsection{$\beta$ Model Core Radius (B): $r_{\rm c}$~(\%~$r_{500}$)}
In the case of a double $\beta$ model the inner core radius was used.
The histogram of $r_{\rm c}$ as a percentage of $r_{500}$ shows little
evidence of bimodality, with four distinct peaks, however the KMM
algorithm rejects the single Gaussian hypothesis at $\sim$99\%
confidence.  Using the KMM algorithm we find a split at $\sim$1.3
$\times$ 10$^{-2}$~$r_{500}$, which divides the sample into 30 CC
clusters and 34 NCC clusters. The centers of the subgroups are 0.35
$\times$ 10$^{-2}$~$r_{500}$ and 4.31 $\times$ 10$^{-2}$~$r_{500}$ for
the CC and NCC clusters respectively.

\subsubsection{Central Electron Density~(C): $n_{0}$}
The histogram for $n_{0}$ appears to have a single peak at $\sim2.5$
$\times$ 10$^{-2}$~$h_{71}^{1/2}$ cm$^{-3}$, with a lower, wider
distribution below $\sim$1 $\times$ 10$^{-2}$~$h_{71}^{1/2}$
cm$^{-3}$. The KMM algorithm rejects the single Gaussian hypothesis at
$>$99\% confidence.  We separate the two distributions at $\sim$1.5
$\times$ 10$^{-2}$~$h_{71}^{1/2}$ cm$^{-3}$ which divides the sample
into 37 CC clusters and 27 NCC clusters.  The centers of the subgroups
are 4.02 $\times$ 10$^{-2}$~$h_{71}^{1/2}$ cm$^{-3}$ and 0.46 $\times$
10$^{-2}$~$h_{71}^{1/2}$ cm$^{-3}$ for the CC and NCC clusters
respectively.

\subsubsection{Central Entropy~(D): $K_{0}$}
The histogram for $K_{0}$ looks like it may have a tri-modal
distribution with peaks at $\sim10$~$h_{71}^{-1/3}$ keV cm$^{2}$,
$\sim40$~$h_{71}^{-1/3}$ keV cm$^{2}$ and $\sim250$~$h_{71}^{-1/3}$
keV cm$^{2}$.  The KMM algorithm rejects the single Gaussian
hypothesis at more than 99\% confidence and the likelihood ratio test
statistic (LRTS) \citep{1994AJ....108.2348A} suggests $K_{0}$ is one
of the most significant bimodal distributions.  The KMM algorithm
finds a division between the two subgroups at $\sim$25~$h_{71}^{-1/3}$
keV cm$^{2}$, which divides the sample into 27 CC and 37 NCC clusters.
The division in $K_0$ was first reported by \citet{hudson07} and
\citet{reiprich07} and fits well with the results of \citet{voit08}.
Adding a third subgroup, improves the LRTS further and divides the
distribution at $\sim$22 and $\sim$150~$h_{71}^{-1/3}$ keV cm$^{2}$,
which gives 24 {\it strong} CC (SCC) clusters, 22 {\it weak} CC or
transition (WCC) clusters and 18 NCC clusters.  The centers of the
subgroups are 11~$h_{71}^{-1/3}$ keV cm$^{2}$, 59~$h_{71}^{-1/3}$ keV
cm$^{2}$ and 257~$h_{71}^{-1/3}$ keV cm$^{2}$ for the SCC, WCC and NCC
clusters respectively.

\subsubsection{Biased Central Entropy~(E): $K_{\rm BIAS}$}
Na{\"i}vely one would expect $n_{\rm annulus}$ $<$ $n_{0}$ (see
Eqn.~\ref{avgdens} for the definition of $n_{\rm annulus}$), however
since $n_{\rm annulus}$ is not deprojected and $n_{0}$ is deprojected,
it is possible that $n_{\rm annulus}$ will be larger than $n_{0}$,
implying an appreciable amount of emission along the line of sight.
This emission increases the apparent central density if it is not
subtracted before determining the central density. In general,
however, $n_{\rm annulus}$ will be smaller than $n_{0}$, especially
for NCC clusters which require a large region to obtain our count
criterion.  The effect of this intentional bias is that NCC clusters
end up with a large value of $K_{\rm BIAS}$ (as can be seen, the
maximum value of $K_{\rm BIAS}$ is larger than the maximum value of
$K_{0}$). On the other hand CC clusters will have a value of $K_{\rm
  BIAS}$ similar to $K_{0}$, separating the distribution.  The net
effect of the bias seems to be to shift the 15 transition clusters in
the $K_{0}$ distribution toward the NCC peak. The KMM algorithm
rejects the single Gaussian hypothesis at $>$99\% confidence.
Surprisingly the LRTS suggests the bimodal distribution for $K_{\rm
  BIAS}$ is less significant than for $K_{0}$.  The two CC/NCC
subgroups are divided at 40~$h_{71}^{-1/3}$ keV cm$^{2}$, which gives
29 CC clusters and 35 NCC clusters.  The centers of the subgroups are
11~$h_{71}^{-1/3}$ keV cm$^{2}$ and 131~$h_{71}^{-1/3}$ keV cm$^{2}$
for the CC and NCC clusters respectively.

\subsubsection{Cooling Radius (\%~$r_{\rm vir}$)~(F) }
We made two assumptions when calculating the cooling radius:
(1)~$t_{\rm cool}$($r_{\rm cool}$) = 7.7~$h_{71}^{-1/2}$~Gyr (see
Sect.~\ref{MDR}) and (2)~the gas at the cooling radius has the density
extrapolated from the 0-0.048~$r_{500}$ density profile.  Therefore,
any cluster with a CCT longer than 7.7~$h_{71}^{-1/2}$~Gyr, has a
cooling radius of zero.  Beyond that, there is a broad distribution
above about $0.02~r_{500}$.  The KMM algorithm rejects the single
Gaussian hypothesis at 98\% confidence, however the algorithm does not
converge if different covariances are used for each subgroup (possibly
because of the large peak at zero).  Using a common covariance for the
two subgroups, the KMM algorithm partitions the two subgroups at
$0.043~r_{500}$ which divides the sample into 37 CC clusters and 27
NCC clusters.  The centers of the subgroups are $0.081~r_{500}$ and
$0.012r_{500}$ for the CC and NCC clusters respectively. We note that
since a cluster with $t_{\rm cool}$(0) $>$ 7.7~$h_{71}^{-1/2}$ will,
by definition, have cooling radius of zero, it makes a poor parameter
for comparison studies.

\subsubsection{Scaled Spectral Mass Deposition Rate~(G):
  $\smdr$/$M_{500}$}
The histogram for $\dot{M}_{\rm spec}$/$M_{500}$ shows strong evidence
of a bimodal distribution.  There are 13 clusters with a best-fit
value of $\dot{M}_{\rm spec}$ = 0 and there is a second peak at
$\sim$10$^{-14}$~$h_{71}$~yr$^{-1}$.  Since the KMM algorithm was
fitted to the $\log$ value of the data, the clusters with
$\dot{M}_{\rm spec}$ = 0, were assigned to $\log$[$\dot{M}_{\rm
  spec}$/($M_{500}$ $10^{-14}$~$h_{71}$~yr$^{-1}$)] = -5.  The KMM
algorithm rejects the single Gaussian hypothesis at $>$99\%
confidence.  The LRTS confirms that the bimodality is one of the most
significant of the 16 parameters.  The KMM algorithm identifies 23
clusters with the first (NCC) subgroup.  However, due to the large
variance in the first subgroup, the two clusters with the largest
value of $\dot{M}_{\rm spec}$/$M_{500}$ are assigned to the first
subgroup (see Fig.~\ref{Histogram}-G).  Since this is physically
unreasonable, we take the partition value to be $\sim0.23$ $\times$
$10^{-14}$~$h_{71}$~yr$^{-1}$, which divides the sample into 43 CC
clusters and 21 NCC clusters.  The centers of the subgroups are 1.3
$\times$ $10^{-14}$~$h_{71}$~yr$^{-1}$ and 0.045 $\times$
$10^{-14}$~$h_{71}$~yr$^{-1}$ for the CC and NCC clusters
respectively.  As with cooling radius, the existence of zero values
makes $\dot{M}_{\rm spec}$/$M_{500}$ problematic as a parameter for
comparison studies.

\subsubsection{Scaled Classical Mass Deposition Rate~(H): $\mdr$/$M_{500}$}
As with cooling radius, the 18 clusters with CCT $>$
7.7~$h_{71}^{-1/2}$~Gyr, have $\dot{M}_{\rm classical}$ = 0.  Similar
to $\dot{M}_{\rm spec}$/$M_{500}$, those clusters were assigned
$\log$[$\dot{M}_{\rm classical}$/($M_{500}$$10^{-14}$~$h_{71}$
yr$^{-1}$)] = -5 when applying the KMM algorithm to the $\log$ of the
data.  Besides the peak at zero, there seems to be a second Gaussian
distribution at $\sim$20 $\times$ $10^{-14}$~$h_{71}$~yr$^{-1}$.  The
KMM algorithm rejects the single Gaussian hypothesis at more than 99\%
confidence.  The LRTS identifies $\dot{M}_{\rm classical}$/$M_{500}$
as having the most significant bimodality.  The partition between the
two subgroups is $\sim$0.5 $\times$ $10^{-14}$~$h_{71}$~yr$^{-1}$,
which divides the sample into 43 CC clusters and 21 NCC clusters.  The
centers of the subgroups are 20.5 $\times$ $10^{-14}$~$h_{71}$
yr$^{-1}$ and 0.0106 $\times$ $10^{-14}$~$h_{71}$~yr$^{-1}$ for the CC
and NCC clusters respectively.  We note, as with cooling radius, that
since a cluster with $t_{\rm cool}$(0) $>$ 7.7~$h_{71}^{-1/2}$ will,
by definition, have $\dot{M}_{\rm classical}$ = 0, it makes a poor
parameter for comparison studies.% since it is not continuous.

\subsubsection{Cuspiness~(I): $\alpha$}
The histogram for cuspiness (defined in Eqn.~\ref{cuspiness}) appears
to have a Gaussian distribution around a mean at $\sim$1.2.  There
seems to also be a broad distribution at $\sim$0.4 and a sharp peak at
$\sim$0.9.  The single Gaussian hypothesis is rejected by the KMM
algorithm at $\sim$99\% confidence.  The LRTS suggests that the
bimodal distribution is not as significant as for other parameters
(e.g. $\dot{M}_{\rm classical}$/$M_{500}$).  Using a trimodal
distribution is a significant improvement, but produces an unphysical
distribution.  Basically four of the CC clusters centered around
$\sim$1.2 are separated out of the CC subgroup (i.e. there are CC
clusters on both sides of the four clusters, as opposed to them being
between SCC and NCC clusters).  For the bimodal distribution, the
partition of the subgroups is $\sim$0.7 which divides the sample into
35 CC clusters and 29 NCC clusters.  The centers of the subgroups are
1.07 and 0.412 for the CC and NCC clusters respectively.
\citet{vikhlinin07} suggested a break of $\alpha$=0.5 for CC versus
NCC clusters and $\alpha>$0.7 for SCC clusters.  Using these cuts
there are 45 CC clusters (10 of which are weak CC clusters) and 19 NCC
clusters.

We note here that for many of the clusters that we have in common with
\citet{vikhlinin07}, we generally find larger values of $\alpha$ than
they do \citep[based on Fig.~2 in][]{vikhlinin07}.  There are several
possible explanations for this discrepancy.  The most obvious is the
different way in which we calculated $r_{500}$.  We calculated
$r_{500}$ from $T_{\rm vir}$ using the formula of \citet{evrard96},
which, as we noted earlier, may overestimate $r_{500}$.  On the other
hand \citet{vikhlinin07} estimate $r_{500}$ from their mass model.  If
our value of $r_{500}$ is larger than the value used by
\citet{vikhlinin07}, it makes sense that our values of $\alpha$ will
be larger, since the profile should steepen around 0.04~$r_{500}$ (see
Fig.~1 in \citet{vikhlinin07}).  The second major difference is that
our values of $\alpha$ are derived from surface brightness profile
models, whereas \citet{vikhlinin07} use direct deprojection and derive
$\alpha$ from density profile models.  Finally there are minor points
that can contribute to the discrepancies, such as the center used to
create the profile (we both use the X-ray peak, but for some NCC
clusters this is not well-defined), the energy band for the surface
brightness profile and the techniques used to create the profiles.  In
the end, we argue that these differences lead to an intrinsic scatter
in the values of $\alpha$ obtained which are dependent on the method
used to determine it.

\subsubsection{Scaled Core Luminosity~(J): $L_{X}$/[$M_{\rm gas}$ $kT_{\rm vir}$]}
We define a scaled version of central luminosity $L_{\rm X}$.  Scaled
$L_{\rm X}$ is taken from the projected spectral fit to the 0 -
0.048$r_{500}$ region.  This luminosity is then scaled by $kT_{\rm
  vir}$ and the gas mass of the region, calculated from the gas
density profile ($L_{\rm X}$/[$M_{\rm gas}\:kT_{\rm vir}$]).  The
histogram of the scaled $L_{\rm X}$, appears to be a single
distribution with a peak at $\sim0.6$ $\times$
$10^{30}$~$h_{71}^{1/2}$ ergs s$^{-1}$ keV$^{-1}$ M$_{\odot}^{-1}$.
In fact, the KMM algorithm does not reject the single Gaussian
hypothesis ($<$40\% confidence).  If split into a bimodal distribution
the subgroups are partitioned at 0.4 $\times$ $10^{30}$~$h_{71}^{1/2}$
ergs s$^{-1}$ keV$^{-1}$ M$_{\odot}^{-1}$ which divides the sample
into 49 CC clusters and 15 NCC clusters.  The centers of the subgroups
are 0.72 $\times$ $10^{30}$~$h_{71}^{1/2}$ ergs s$^{-1}$ keV$^{-1}$
M$_{\odot}^{-1}$ and 0.33 $\times$ $10^{30}$~$h_{71}^{1/2}$ ergs
s$^{-1}$ keV$^{-1}$ M$_{\odot}^{-1}$ for the CC and NCC clusters
respectively.

\subsubsection{Central Temperature Drop~(K): $T_{0}$/$T_{\rm vir}$}
The central temperature drop ($T_{0}$/$T_{\rm vir}$) is often used as
an identifier for CC clusters \citep[e.g.][]{sanderson06,burns07}.  In
the histogram there seems to be a large peak centered at $\sim$1 and a
broad range of temperature drops that peaks around $\sim$0.4.  The KMM
algorithm rejects the single Gaussian hypothesis at $>$99\% confidence
and the LRTS suggests that $T_{0}$/$T_{\rm vir}$ has one of the most
significant bimodal distributions.  The partition between CC and NCC
clusters is $\sim$0.7, which divides the sample into 24 CC clusters
and 40 NCC clusters.  The centers of the subgroups are 0.44 and 0.97
for the CC and NCC clusters respectively.  For their sample of
simulated clusters, \citet{burns07} made a cut at $T_{0}$/$T_{\rm
  vir}$ = 0.8, which also seems to be a plausible cut for our sample.

\subsubsection{Slope of the Inner Temperature Profile~(L)}
The histogram for the slopes of the inner temperature profiles does
not seem to show any bimodality, suggesting that there is no universal
central temperature profile.  The fact that $T_{0}$/$T_{\rm vir}$
shows some evidence of bimodality hints that the lowest temperature of
the gas in the center of the clusters never cools much below $\sim$0.4
$T_{\rm vir}$.  However since the slopes show no evidence of
bimodality, the gradient of this temperature drop from $T_{\rm vir}$
to 0.4 $T_{\rm vir}$ is not universal.  This would imply that the size
of the region where the temperature drops below $T_{\rm vir}$ is not
universal (see Sect.~\ref{kTProfiles} for a more detailed discussion).
Usually, NCC clusters are thought to have no temperature drop, but the
distribution of $\log(T_0/T_{\rm vir})$ is continuous from 0 to
$\sim$$-$0.2, making it difficult to define a clean break.  We also do
not find any peak or distribution around the universal value of $-$0.4
found by \citet{sanderson06}.  The KMM algorithm rejects the single
Gaussian hypothesis at $>$99\% confidence, but the LRTS suggests the
bimodality is not as significant as in other parameters
(e.g. $T_{0}$/$T_{\rm vir}$).  The CC subgroup has a very large
variance, suggesting a broad distribution in the inner $T$-profile
slopes for CC clusters.  In fact because of the large variance in the
CC subgroup, the KMM algorithm identifies the three clusters with the
steepest centrally increasing temperature profiles as belonging to the
CC subgroup (see Fig.~\ref{Histogram}-L).  Since this is unphysical,
we take the partition value to be -0.23, which divides the sample into
15 CC clusters and 49 NCC clusters.  The centers of the subgroups are
-0.24 and -0.08 for the CC and NCC clusters respectively.

\subsubsection{Ratio of Central Temperatures in the Soft Band to the
  Hard Band~(M): [$T_{0}$ (0.5 - 2.0 keV)]/[$T_{0}$ (2.0 - 7.0 keV)]}
We also considered fitting the same region to different energy bands
in order to identify cool cores.  The idea is that if there are many
temperatures in the central region the soft band will be more
sensitive to the cool gas, whereas the harder band will be more
sensitive to the hotter gas.  We took our central annulus from the
$T$-profile and fit it in the 0.5 - 2.0 keV band and then in the 2.0 -
7.0 keV band and found the ratio.  The distribution of values does not
look bimodal at all and appears to be a Gaussian centered on
$\sim$0.7.  However, the KMM algorithm rejects the single Gaussian
hypothesis at $>$99\% confidence.  On the other hand, the LRTS
suggests the bimodality is not as significant as in other parameters
(e.g. $T_{0}$/$T_{\rm vir}$).  As with $\dot{M}_{\rm spec}$/$M_{500}$
and the slopes of the central temperature profiles, the large variance
of one of the subgroups (the NCC subgroup in this case) causes some
unphysical assignments (see Fig.~\ref{Histogram}-M).  The four cases
with the smallest ratio of [$T_{0}$ (0.5 - 2.0 keV)]/[$T_{0}$ (2.0 -
7.0 keV)] are assigned to the NCC subgroup.  Using only the clusters
with large values of [$T_{0}$ (0.5 - 2.0 keV)]/[$T_{0}$ (2.0 - 7.0
keV)] that the KMM algorithm assigns to the NCC subgroup, the CC/NCC
clusters are partitioned at $\sim$0.86.  This divides the sample into
50 CC clusters and 14 NCC clusters.  The centers of the subgroups are
0.72 and 0.87 for the CC and NCC clusters respectively.

\subsubsection{Scaled Gas Mass within 0.048$r_{500}$~(N): $M_{\rm
    gas}$/$M_{500}$}
The high density in the centers of CC clusters suggests that there
should be relatively more gas in the center compared with NCC
clusters.  The distribution, however, does not look particularly
bimodal, suggesting that size of the core does not scale with mass.
There seems to be a peak at $\sim$0.4 $\times$ 10$^{-3}$, with several
smaller peaks on either side.  The KMM algorithm does not reject the
single Gaussian hypothesis ($<$76\% confidence).  If split into two
distributions they are partitioned at $\sim$0.45 $h^{-3/2}$ $\times$
10$^{-3}$, which divides the sample into 38 CC clusters and 26 NCC
clusters.  The centers of the subgroups are 0.87 $h^{-3/2}$ $\times$
10$^{-3}$ and 0.28 $h^{-3/2}$ $\times$ 10$^{-3}$ for the CC and NCC
clusters respectively.

\subsubsection{Scaled Modified Spectral Mass Deposition Rate~(O):
  $\dot{M}_{\rm spec2}$/$M_{500}$}
The histogram for $\dot{M}_{\rm spec2}$/$M_{500}$ is similar to the
other $\dot{M}$ histograms although the upper peak is broader and not
as pronounced.  Interestingly, the scale of $\dot{M}_{\rm
  spec2}$/$M_{500}$ is similar to that of $\dot{M}_{\rm
  classical}$/$M_{500}$ about an order of magnitude above
$\dot{M}_{\rm spec}$/$M_{500}$ (see also Fig.~\ref{specCCvsdensCC} and
Fig.~\ref{specpCCvsdensCC}).  As with the other $\dot{M}$ histograms,
the clusters with $\dot{M}_{\rm spec2}$ = 0, were assigned to
$\log$[$\dot{M}_{\rm spec2}$/($M_{500}$/$10^{-14}$~$h_{71}$
yr$^{-1}$)] = -5 when applying the KMM algorithm to the $\log$ of
data.  The KMM algorithm only marginally rejects the single Gaussian
hypothesis ($\sim$90\% confidence).  The two subgroups are partitioned
at 1.5 $\times$ $10^{-14}$~$h_{71}$~yr$^{-1}$, which divides the
sample into 44 CC clusters and 20 NCC clusters.  The centers of the
subgroups are 11.7 $\times$ $10^{-14}$~$h_{71}$~yr$^{-1}$ and 0.39
$\times$ $10^{-14}$~$h_{71}$~yr$^{-1}$ for the CC and NCC clusters
respectively.

\subsubsection{Central Cooling Time~(P)}
The histogram for CCT looks similar to the histogram of $K_{0}$, which
is not surprising since they are calculated from similar quantities:
$T_{\rm ctr}$\footnote{$T_{\rm ctr}$ = $T_{48}$ (see
  Sect.~\ref{cooltimes}) for CCT and $T_{0}$ ($\approx T_{48}$) for
  $K_{0}$.} and $n_0$. Like $K_{0}$, the histogram of CCT has two
peaks with a smaller peak between the two.  The KMM algorithm rejects
the single Gaussian hypothesis at $>$99\% confidence and the LRTS
suggests the bimodality of CCT is the second most significant (after
$\dot{M}_{\rm classical}$/$M_{500}$) among the 16 parameters.  As with
$K_{0}$, adding a third subgroup increases the LRTS. For a bimodal
distribution the partition is $\sim$5~$h_{71}^{-1/2}$~Gyr, which
divides the sample into 42 CC clusters and 22 NCC clusters.  For the
tri-modal distribution the CC/NCC partition remains at
$\sim$5~$h_{71}^{-1/2}$~Gyr and the SCC/WCC partition is
$\sim$1~$h_{71}^{-1/2}$~Gyr.  There are four clusters with CCT between
5~$h_{71}^{-1/2}$~Gyr and 7.7~$h_{71}^{-1/2}$~Gyr, so that the
difference between a partition at 5~$h_{71}^{-1/2}$~Gyr and
7.7~$h_{71}^{-1/2}$~Gyr is a matter of low number statistics.  Since
7.7~$h_{71}^{-1/2}$~Gyr corresponds to a look back time for z$\sim$1,
about the time most clusters would have time to relax and form a cool
core, we decided to take the partition as 7.7~$h_{71}^{-1/2}$~Gyr
rather than 5~$h_{71}^{-1/2}$~Gyr.  Moreover we used
7.7~$h_{71}^{-1/2}$~Gyr to determine $\dot{M}_{\rm classical}$ and
cooling radius.  Using 1 and 7.7~$h_{71}^{-1/2}$~Gyr as cuts we divide
the sample into 28 SCC clusters, 18 WCC clusters and 18 NCC clusters
.The centers of the subgroups are 0.45~$h_{71}^{-1/2}$~Gyr,
1.91~$h_{71}^{-1/2}$~Gyr and 11.2~$h_{71}^{-1/2}$~Gyr for the CC, WCC
and NCC clusters respectively.

\begin{figure*}
\includegraphics[width=180.0mm,angle=0]{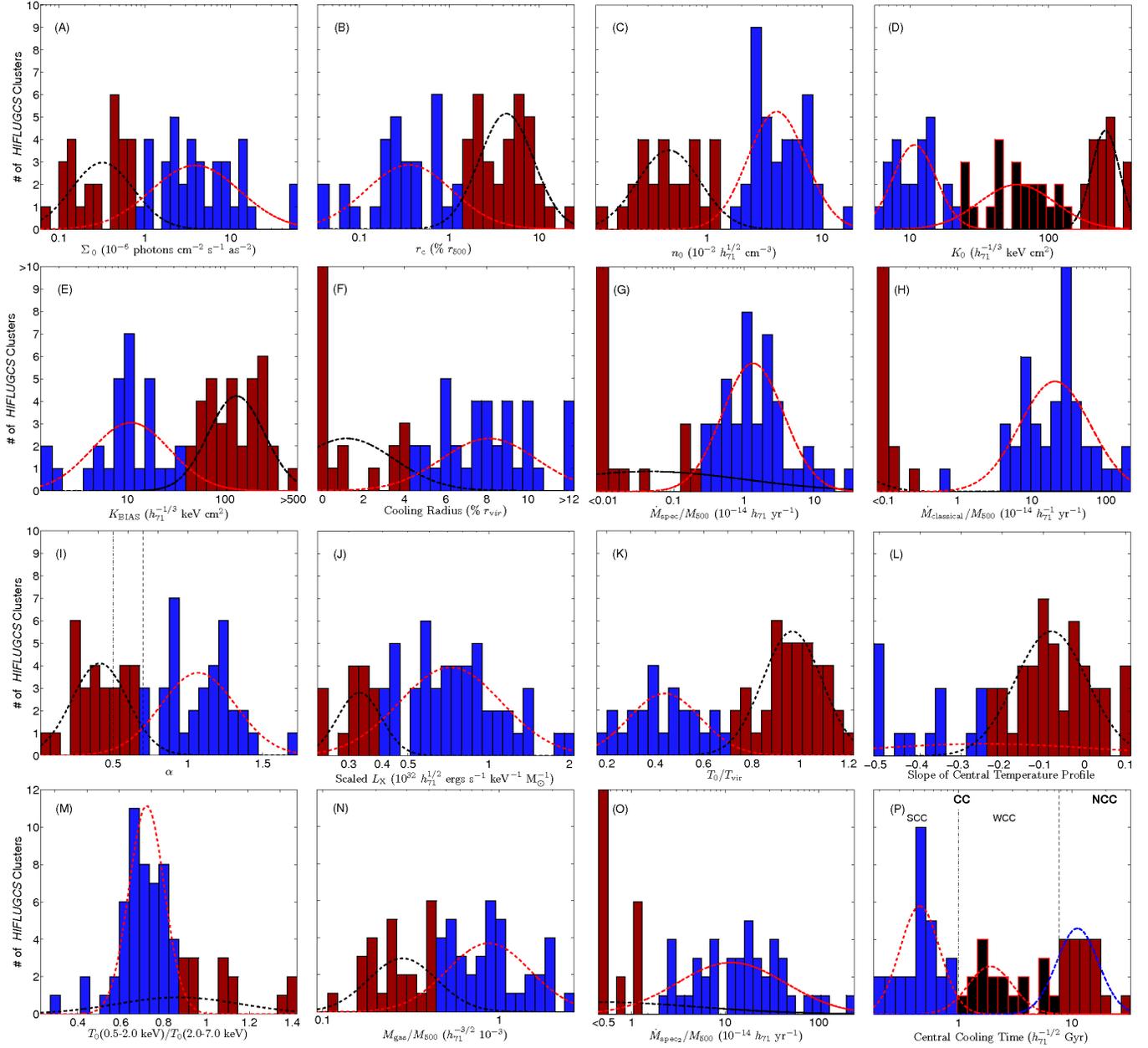}
\caption[Histograms] {Histograms of 16 parameters that may be used to
  distinguish between CC and NCC clusters.  Blue bins represent CC
  clusters, red bins indicate NCC clusters and where they appear,
  black bins represent transitional or weak CC clusters (see
  text). Row-wise left to right and starting from the top row the
  histograms are: (A) central surface brightness ($\Sigma_{0}$), (B)
  $\beta$ model core radius scaled by $r_{500}$, (C) central electron
  density ($n_{0}$), (D) central entropy ($K_{0}$), (E) biased central
  entropy ($K_{\rm BIAS}$), (F) cooling radius, (G) spectral mass
  deposition rate scaled by $M_{500}$ ($\dot{M}_{\rm
    spec}$/$M_{500}$), (H) classical mass deposition rate scaled by
  $M_{500}$ ($\dot{M}_{\rm classical}$/$M_{500}$), (I) cuspiness
  ($\alpha$), (J) bolometric X-ray luminosity of the 0 -
  0.048~$r_{500}$ region scaled by the gas mass of the region and
  virial temperature ($L_{\rm X}$/($M_{\rm gas} kT_{\rm vir}$)), (K)
  central temperature drop ($T_{0}$/$T_{\rm vir}$), (L) slope of inner
  temperature profile, (M) central temperature measured in the soft
  band divided by central temperature measured in the hard band, (N)
  gas mass within 0.048~$r_{500}$ scaled by $M_{500}$,(O) modified
  spectral mass deposition rate scaled by $M_{500}$ ($\dot{M}_{\rm
    spec2}$/$M_{500}$) and (P) central cooling time (CCT).  For the
  cuspiness histogram, the dash-dot and dashed vertical line indicates
  the CC/NCC cut and the weak CC, strong CC cut, respectively,
  suggested by \citet{vikhlinin07}. \label{Histogram}}
\end{figure*}

\begin{table*}
%\scriptsize
  \caption{Summary of the KMM algorithm results for the 16 parameters.}
\label{splits}
\centering
\begin{tabular}{l l c c c c  }
  \hline
  \hline
  Parameter    	& Break 		& $\mu_{1}$			& $\mu_{2}$ 	& $\#$ of CC  		& $\#$ of NCC  \\
  & Value$^{\dagger}$	& CC  				& NCC         	& clusters$^{\alpha}$	& clusters \\
  &			& Centroid$^{\ddagger}$		& Centroid	&			& \\
  (1)		& (2)			& (3)			 	& (4)		& (5)			& (6) \\
  \hline
  $\Sigma_{0}$ (10$^{-6}$ photons cm$^{-2}$ s$^{-1}$ arcsec$^{-2}$)							& 0.80 	& 3.88	& 0.32 		& 37	& 27 \\
  $r_{c}$/$r_{500}$ (10$^{-2}$)											& 1.3  	& 0.35	& 4.31		& 30	& 34 \\
  $n_{0}$	(10$^{-2}$ $h_{71}^{1/2}$ cm$^{-3}$)	  								& 1.5 	& 4.02	& 0.46		& 37	& 27 \\
  $K_{0}$ ($h_{71}^{-1/3}$ keV cm$^{2}$) 										& 150 (22) & 11 (59) & 257 & 46 (22) 	& 18 \\
  $K_{\rm BIAS}$ ($h_{71}^{-1/3}$ keV cm$^{2}$)									& 40	& 11	& 131		& 29	& 35 \\
  Cooling Radius (\% $r_{500}$)   										& 4.3	& 8.1	& 1.2		& 37	& 27 \\
  $\dot{M}_{\rm spec}$/$M_{500}$ ($h_{71}$ 10$^{-14}$~yr$^{-1}$)							& 0.23	& 1.3	& 0.045		& 43	& 21 \\
  $\dot{M}_{\rm classical}$/$M_{500}$ ($h_{71}^{-1}$ 10$^{-14}$~yr$^{-1}$)					& 0.5	& 20.5	& 0.0106 	& 43	& 21 \\
  $\alpha$													& 0.7$^{\beta}$ & 1.07	& 0.41	& 35	& 29 \\
  $L_{X}$/[$M_{\rm gas}$ $kT_{\rm vir}$] (10$^{30}$ $h_{71}^{1/2}$ erg s$^{-1}$ keV$^{-1}$ M$_{\odot}^{-1}$) 	& 0.4   & 0.72  & 0.33		& 49	& 15 \\
  $T_{0}$/$T_{\rm vir}$												& 0.7	& 0.97	& 0.44		& 24	& 40 \\
  Slope of Central Temperature Profile										& -0.23	& -0.24 & -0.08		& 15	& 49 \\
  $[T_{0}$ (0.5 -2.0 keV)$]$/$[T_{0}$ (2.0 - 7.0 keV)$]$								& 0.86	& 0.72	& 0.87		& 50	& 14 \\
  $M_{\rm gas}$/$M_{500}$	($h_{71}^{-3/2}$ 10$^{-3}$)								& 0.45 	& 0.87	& 0.28		& 38	& 26 \\
  $\dot{M}_{\rm spec2}$/$M_{500}$ ($h_{71}$ 10$^{-14}$~yr$^{-1}$)							& 1.5	& 11.7	& 0.39		& 44	& 20 \\
  Central Cooling Time ($h_{71}^{-1/2}$~Gyr)									& 7.7(1) & 0.45 (1.91) & 11.2 & 46 (18) & 18 \\
  \hline
  \multicolumn{6}{p{\textwidth}}{\footnotesize Columns: (1)~the parameter, (2)~the value of the parameter that separates CC and NCC clusters, (3)~the KMM determined centroid value for the CC clusters, (4)~the KMM determined centroid value for the NCC clusters, (5)~the number of CC clusters determined by the KMM algorithm and (6)~the number of NCC clusters determined by the KMM algorithm.}
\end{tabular}

\raggedright
\footnotesize
$^{\dagger}$ The number in parenthesis, if any, is the cut between SCC and WCC (or transition) clusters.\\
$^{\ddagger}$ The number in parenthesis, if any, is the centroid value for the WCC clusters.  In these cases the other number is the centroid value of the SCC clusters.\\
$^{\alpha}$ This number includes SCC and WCC clusters where appropriate.  The number in parenthesis, if any, is the number of WCC (or transition) clusters.\\
$^{\beta}$ Using the cuts suggested by \citet{vikhlinin07} of $\alpha$ = 1 (WCC and SCC) and $\alpha$ = 0.5 (CC and NCC), gives 35 SCC clusters, 10 WCC clusters and 19 NCC clusters.\\
\end{table*}

\subsection{The Defining Parameter}
\label{define_param}
The likelihood ratio test identifies $\dot{M}_{\rm
  classical}$/$M_{500}$ as having the most significant bimodality,
however we did not choose it as the best method to separate NCC and CC
clusters.  There are two reasons for rejecting it as the best method.
(1)~Clusters which have a CCT $>$ 7.7~$h_{71}^{-1/2}$~Gyr have
$\dot{M}_{\rm classical}$ = 0, making it difficult to compare
$\dot{M}_{\rm classical}$/$M_{500}$ to other parameters.  (2)~The
errors in $\dot{M}_{\rm classical}$/$M_{500}$ are quite large (see
Fig.~\ref{paramcompare}-G).  In fact the average uncertainty in
$\dot{M}_{\rm classical}$/$M_{500}$ is $\sim$60\% versus 15\% for CCT,
the parameter with the next most significant bimodality. This
uncertainty is not accounted for in either the histogram or the KMM
algorithm and therefore the significance of the bimodality and the
cluster assignments are also quite uncertain.

\begin{figure}
\includegraphics[width=90.0mm,angle=0]{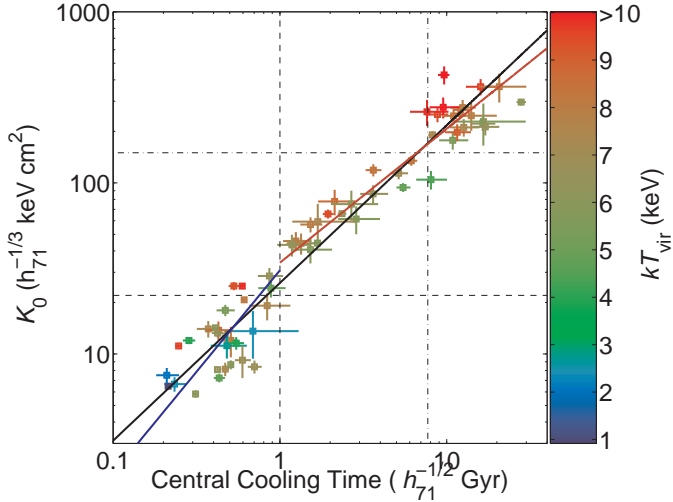}
\label{K0CCT}
\caption[$K_{0}$ versus Central Cooling Time] {This plot shows central
  entropy ($K_{0}$) versus CCT. Since both quantities are derived from
  $T_{\rm ctr}$ ($T_{48}$ $\approx$ $T_{0}$) and $n_{0}$, they are,
  not surprisingly, tightly correlated.  The dashed lines show the
  division between CC and WCC clusters.  The dash-dot lines show the
  division between WCC and NCC clusters. The black line shows the
  best fit for all clusters, the blue line shows the best-fit line for
  the SCC clusters and the red line shows the best fit to the combined
  NCC and WCCs. The data points are color coded by virial temperature
  (see scale at right). \label{Kvct}}
\end{figure}

The parameters with the next highest significance of bimodality are
CCT and $K_{0}$.  As noted earlier, $K_{0}$ and CCT are both
calculated from $n_{0}$ and $T_{\rm ctr}$ (in general $T_{48}$
$\approx$ $T_{0}$) and so it is not surprising that they have similar
distributions.  Fig.~\ref{Kvct} shows the tight correlation between
$K_{0}$ and CCT.  The dashed lines show the cut between CC and WCC
clusters for each parameter and the dot-dashed lines show the cut
between WCC and NCC clusters.  There are four clusters which are
classified as WCC clusters when using $K_{0}$ and are classified as
SCC clusters when using CCT.  It is interesting to note that these
four clusters have the lowest value of $K_{0}$ for any of the WCC
clusters (as classified by $K_{0}$) and there is also a clear break
(at $\sim$30~$h_{71}^{-1/3}$ keV cm$^{2}$) between these clusters and
the other WCC clusters (see Fig.~\ref{Kvct}).  As noted earlier this
break has also been reported by \citet{hudson07}, \citet{reiprich07}
and \citet{voit08}.  Additionally one borderline WCC cluster is
classified as an NCC cluster when using $K_{0}$ and one borderline NCC
cluster is assigned to the WCC subpopulation when classified with
$K_{0}$.  Other than these six borderline cases, all the clusters are
assigned to the same subpopluations whether CCT or $K_{0}$ is used to
classify them.

Based on Fig.~\ref{Kvct}, there is not much difference between sorting
the sample using $K_{0}$ or CCT. Assuming bremsstrahlung emission (i.e
$\Lambda(T)$ $\propto$ $T^{1/2}$) and $n_{i}$ $\approx$ $n_{e}$
$\equiv$ $n$, the dependency of the cooling time on the temperature
and density is given by $t_{\rm cool}$ $\propto$ $T^{1/2}\;n^{-1}$.
Therefore CCT is more dependent on $n$ than on $T$, while $K_{0}$ is
more affected by $T$.  Since the determination of $T$ is more
resolution dependent (i.e. it requires $\sim$100$\times$ more photons
to make a spectrum than one bin in a surface brightness profile), we
argue that $t_{\rm cool}$ is a better parameter to use.  Furthermore
it is also a more traditional metric and short cooling times are the
physical basis of the cooling flow problem.

\subsubsection{CC Fractions in Redshift and Temperature}
\label{wilcoxon}
We further investigated the subpopulations in the CCT by applying the
Wilcoxon rank-sum test \citep{wilcoxon45,mann47} to the redshift and
$T_{\rm vir}$ distributions of the CC (as well as the SCC and WCC
subsamples) and NCC clusters.  The Wilcoxon rank-sum test is a
non-parametric test that determines whether the values of a parameter
of two subpopulations derive from the same parent distribution.  For
instance, it can be used to determine if the redshifts of the CC and
NCC clusters come from the same parent distribution of redshifts.  The
results are presented in Table~\ref{zkT}.  We find that within
1~$\sigma$, the CC and NCC redshifts derive from the same redshift
population.  That is, there is no evolution of the CC/NCC fraction
within our redshift range.  Likewise, when comparing the redshift
distributions of WCC and NCC clusters and SCC and WCC clusters, the
redshifts, within 1~$\sigma$, derive from the same population.
However, for SCC and NCC clusters there is a marginally significant
difference ($\sim$1.1~$\sigma$) in the redshift distributions, with
NCC clusters having generally higher redshifts.  Na{\"i}vely, this
seems to indicate that the fraction of SCC to NCC clusters is higher
at lower redshift.  However, when examining the $T_{\rm vir}$
populations, it is clear that CC (and SCC) clusters come from a lower
temperature population than NCC clusters, with $\sim$2~$\sigma$
significance ($\sim$2.4~$\sigma$ for SCC clusters). There is even a
significant difference ($\sim$1.6~$\sigma$) between $T_{\rm vir}$ for
SCC and WCC clusters.  Only WCC and NCC clusters seem to derive from
the same population of $T_{\rm vir}$.  Since our sample is
flux-limited, low temperature clusters will only appear at low
redshifts.  Since SCC clusters come from a significantly cooler
population than NCC clusters, it makes sense that they would also have
a lower redshift distribution in a flux-limited sample.  Therefore we
conclude that within our redshift range, there is no evolution in the
CC fraction.  We also find low temperature clusters/groups are more
likely to be CC clusters than high temperature clusters, consistent
with the results of \citet{burns07} and \citet{chen07}.

\begin{table}
  \caption{Results to the Wilcoxon rank-sum test applied to redshift and $T_{\rm vir}$ for the CC, SCC, WCC and NCC subpopulations.}
\label{zkT}
\centering
\begin{tabular}{l c c c }
  \hline
  \hline
  Subsamples    	& Parameter 		& Significance	& Which subsample \\
  being 		& being 		& ($\sigma$)	& has a larger avg. \\
  compared	& used			& 		& param. value?  \\
  (1)		& (2)			& (3)		& (4)  \\
  \hline
  CC-NCC		& redshift		& 0.82          & NCC \\
  SCC-NCC		& redshift		& 1.12          & NCC \\
  SCC-WCC		& redshift		& 0.96          & WCC \\
  WCC-NCC		& redshift		& 0.16          & NCC \\
  CC-NCC		        & $T_{\rm vir}$		& 1.98          & NCC \\
  SCC-NCC		& $T_{\rm vir}$		& 2.41          & NCC \\
  SCC-WCC		& $T_{\rm vir}$		& 1.62          & WCC \\
  WCC-NCC		& $T_{\rm vir}$		& 0.81          & NCC \\
  \hline
  \multicolumn{4}{p{\columnwidth}}{\footnotesize Columns: (1)~subsamples being compared, (2)~the parameter (redshift or $T_{\rm vir}$) being used for the comparison,  (3)~the significance that the two distributions are not consistent with the null hypothesis (they are from the same parent distribution), (4)~the subpopulation that comes from the larger redshift or higher temperature distribution.}
\end{tabular}
\raggedright
\end{table}

\subsubsection{Bias on CC Fractions due to Flux-Limited Nature of the
  Sample}
\label{bias}

\begin{table}[b!]
  \caption{Results of simulations done to investigate the impact of selection 
    effects on the observed fractions of SCCs, WCCs, and NCCs.} 
  \label{simulations}
  \centering
  \begin{tabular}{|c|c|c|c|}
    \hline
    Case                    & Category & Input        & Output     \\       
                               &                & Fractions  &  Fractions \\ 
    \hline\hline                                                                                         
    Same slope         & SCC         & 0.34                  & 0.437$\pm$0.043   \\
                               & WCC        & 0.36                  & 0.281$\pm$0.041   \\
                               & NCC        & 0.30                  & 0.281$\pm$0.044   \\ 
    \hline
    Different Slopes  & SCC         & 0.310                & 0.445$\pm$0.041   \\
                               & WCC        & 0.335                 & 0.275$\pm$0.038  \\
                               & NCC        & 0.355                 & 0.280$\pm$0.041  \\
     \hline
  \end{tabular}
\end{table}

Flux-limited samples suffer from the well-known Malmquist bias, namely
that brighter objects have a higher detection rate than fainter
objects. In this section, we address how this bias may affect the
observed fractions of SCC, WCC and NCC clusters.

Strong cool-core clusters, owing to their high central densities have
enhanced central X-ray-emission. This may result in a higher chance of
their detection and serve as an explanation for observing a higher
fraction of SCC clusters in the $\hiflux$ and other flux- or
luminosity-limited samples. Since we have a complete sample, this
calculation can be done. We simulated samples of clusters which follow
the X-ray temperature function given by $\st{d} N/\st{d} V \sim
T^{-3.2}$ \citep{Ikebe2002}, in the temperature range (0.001-15)~keV
and redshift range 0.00-0.25. From the above it is clear that SCC, WCC
and NCC clusters come from the same parent redshift distribution
within $1-\sigma$ standard deviation. Hence, we assigned to the
clusters random redshifts conforming to the $N \propto D_{L}^3$ law,
where $D_{L}$ is the luminosity distance.  We calculated the
luminosities using three different $\lt$ relations as determined for
each of the three categories, the SCC, WCC, and NCC clusters,
individually (Mittal~R.~\&~Reiprich~T., in preparation).

In order to estimate the effect of imposing a flux-limit to a mixed
sample of SCCs, WCCs and NCCs on their resulting fractions, we applied
the $\hiflux$ flux-limit, $f_{\textrm x}~(0.1-2.4)$~keV$ \ge 2 \times
10^{-11}$~erg~s$^{-1}$~cm$^{-2}$, to the simulated sample. We tried
two different input sets. In the first (simplest) case, we assumed the
SCCs, WCCs and NCCs to have the same $\lt$ slope (3.33). The
normalizations were fixed to those found from the fits to the data. In
the second case, we fixed the slopes for SCCs, WCCs and NCCs to the
fitted values (Mittal~R.~\&~Reiprich~T., in preparation). We find that
in both the cases the output fractions are indeed biased. In
particular, the output SCC fraction is higher than the input
value. Thus we conclude that in reality the SCCs, WCCs and NCCs may
occur with similar fractions and due to the increased X-ray luminosity
in SCCs, their observed fraction is higher in the present sample.

% However, non-cool-core clusters, which are mainly merger systems, go
% through transient stages of enhanced X-ray luminosity and X-ray
% temperature, especially during core passage \citep{Wik2008,
%   Schindler2001}, also leading to a temporary increased likelihood
% of detection.

\subsection{Cooling Time Compared to Other Parameters}
\label{comparison}
\begin{figure*}
\includegraphics[width=180.0mm,angle=0]{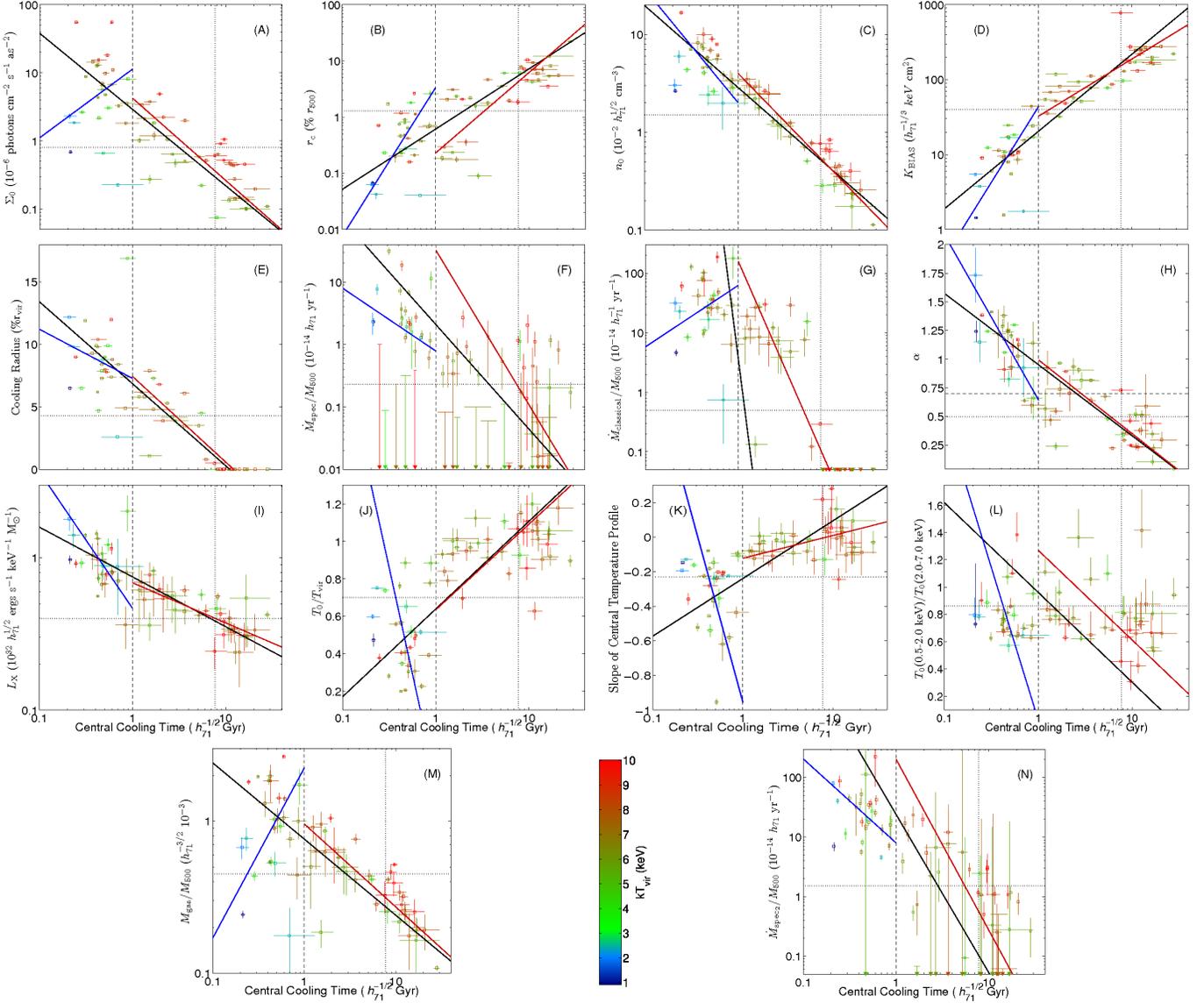}
\caption[Parameters versus Central Cooling Time] {A comparison of CCT
  with fourteen of the parameters from Fig.~\ref{Histogram}.  Row-wise
  left to right and starting from the top row the plots are: (A)
  central surface brightness $\Sigma_{0}$, (B) $\beta$ model core
  radius $r_{c}$ as a percentage of $r_{500}$, (C) central density
  $n_{0}$, (D) biased central entropy $K_{\rm BIAS}$, (E) cooling
  radius, (F) spectral mass deposition rate $\dot{M}_{\rm spec}$
  scaled by $M_{500}$, (G) classical mass deposition rate $M_{\rm
    classical}$ scaled by $M_{500}$, (H) cuspiness $\alpha$, (I)
  scaled central, bolometric, X-ray luminosity $L_{\rm X}/(M_{\rm gas}
  kT_{\rm vir})$, (J) central temperature drop $T_0$/$T_{\rm vir}$ (K)
  slope of the central temperature profile, (L) central soft band
  determined temperature divided by central hard band determined
  temperature [$T_{0}$ (0.5 - 2.0 keV)]/[$T_{0}$ (2.0 - 7.0 keV)], (M)
  central gas mass $M_{\rm gas}$ scaled by $M_{500}$ and (N) modified
  spectral mass deposition rate $\dot{M}_{\rm spec2}$ scaled by
  $M_{500}$.  The clusters are color coded by virial temperature with
  the colorscale shown in the bottom center.  The dotted lines
  represent the division between CC clusters and NCC clusters as
  determined by the KMM algorithm for that particular parameter.  The
  dashed lines represent the division between SCC clusters and WCC
  clusters.  The solid black line is the best fit to all the data, the
  blue line is the fit only to the SCC clusters (as determined by CCT)
  and the red line is the fit to the WCC and NCC clusters (as
  determined by CCT).  As noted in the text, the Fornax cluster is
  often a strong outlier.  See notes on individual parameters and
  Sect.~\ref{FC} for specifics. \label{paramcompare}}
\end{figure*}

In order to study underlying physical relations and correlate simple
observables with those requiring high quality data, we compare 14
parameters\footnote{We already compared $K_{0}$ to the CCT in
  Fig.~\ref{K0CCT}, so it is not repeated here.} from
Fig.~\ref{Histogram} to the CCT.  The relations are plotted in
Fig.~\ref{paramcompare}.  The points are color coded by virial
temperature with the color scale cropped at 10 keV.  In order to
quantitatively compare the parameters to CCT, we fit the relations to
powerlaws\footnote{Some parameters are more nearly linear in the log
  of the cooling time, in which case the parameter was exponentiated
  before fitting.}.  Since there were errors on both the parameters
and CCT, we used the bisector linear regression routine, {\it BCES}
\citep{1996ApJ...470..706A}, to fit the data.  Although some of the
parameters (e.g. $K_0$) had correlated errors with CCT, for simplicity
we assumed them to be independent.  Additionally, since the relation
between the CCT and the other parameters seemed to behave differently
for the SCC clusters, we separately fit the SCC clusters and the
non-SCC (WCC and NCC) clusters.  The black line gives the best fit for
all clusters, the blue line gives the best-fit line for the SCC
clusters and the red gives best-fit line for the non-SCC clusters.
Table~\ref{parameterfits} gives the fit values for all 14 parameters
(along with $K_{0}$) for all three lines.

\subsubsection{$\Sigma_{0}$}
It has been suggested that the central surface brightness,
$\Sigma_{0}$, is an indicator of CC strength \citep[e.g.][]{ohara06}.
We find a surprising large amount of scatter in the plot of
$\Sigma_{0}$ versus CCT, especially at short cooling times.  Since
$n_{0}$ is derived from $\Sigma_{0}$ and CCT is derived from $n_{0}$,
one would na{\"i}vely expect a tighter correlation.  Since $\Sigma$
$\propto$ $n^2$ any scatter in the $n_{0}$ versus CCT relation will be
amplified in the $\Sigma_{0}$ versus CCT relation.  Not surprisingly,
the outliers at the low end are the lower temperature objects.  Since
CCT depends both on $n_{0}$ and $T_{0}$, cooler gas will have a
shorter cooling time at the same density.  Since cooler clusters are
more likely to be CC clusters than hotter clusters (see
Sect.~\ref{define_param}), it is not surprising that there is more
scatter for the SCC clusters than for WCC and NCC clusters.  When
using $\Sigma_{0}$ to identify CC clusters, the scatter will lead to
misidentification; however, adding information about $T_{\rm vir}$ may
be used to help reduce the scatter.

\begin{figure}
\includegraphics[width=90.0mm,angle=0]{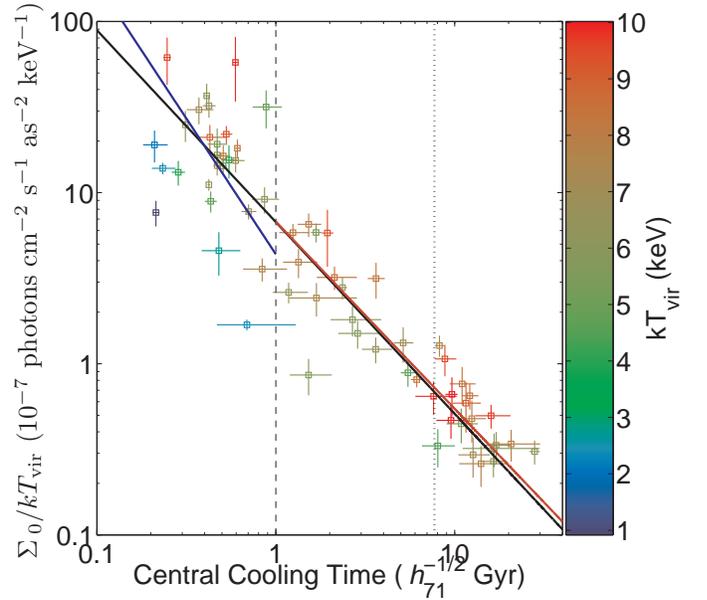}
\caption[Scaled Central Surface Brightness versus Central Cooling
Time] {The scaled central surface brightness $\Sigma_{0}$/$kT_{\rm
    vir}$ versus CCT.  The $\Sigma_{0}$ versus CCT correlation
  tightens when the $\Sigma_{0}$ is scaled by $kT_{\rm
    vir}$.  \label{sigkT_vs_CCT}}
\end{figure}

If $\Sigma_{0}$ is used to separate the CC and NCC clusters, the WCC
clusters are split into CC and NCC clusters, but not clearly on the
basis of CCT.  The WCC clusters with the longest CCTs are classified
as NCC clusters, but the WCC clusters with intermediate CCTs
($\sim$4~$h_{71}^{-1/2}$~Gyr) and short CCTs
($\sim$1-2~$h_{71}^{-1/2}$~Gyr) are sometimes categorized as CC
clusters and sometimes as NCC clusters.  Fig.~\ref{sigkT_vs_CCT} shows
how the correlation becomes tighter if the central surface brightness
is scaled by $kT_{\rm vir}$.  Cutting $\Sigma_{0}$/$kT_{\rm vir}$ at
1.5 $\times$ 10$^{-7}$ photons cm$^{-2}$ s$^{-1}$ keV$^{-1}$
successfully separates the SCC and NCC clusters, with five
misclassified WCC clusters, all but one (A2634) with relatively long
CCTs ($>$3~$h_{71}^{-1/2}$~Gyr).

\subsubsection{$r_{c} (\% r_{500}$)}
We checked to see if there was any correlation between CCT and $r_{c}$
as a percentage of $r_{500}$. We expected a correlation since CC
clusters usually require a double $\beta$ model to fit their surface
brightness profile, whereas NCC clusters do not. Therefore we expected
$r_{c}$/$r_{500}$ to be much smaller for CC clusters than NCC
clusters.  Surprisingly the correlation appears to be tighter than for
$\Sigma_{0}$, although there is still a lot of scatter.  The KMM
determined cut would classify four SCC clusters as well as most of the
WCC clusters as NCC clusters.  If we put the cut at 0.03~$r_{500}$
(just after the third peak in the histogram counting from the left),
however, we would more successfully separate the CC clusters and NCC
clusters.  On the other hand, this parameter does a poor job of
separating SCC and WCC clusters.  An interesting note is that the
coolest clusters (groups) appear to have smaller values of
$r_{c}$/$r_{500}$ compared to hotter clusters.

\subsubsection{$n_{0}$}
There is a tight correlation between $n_{0}$ and CCT.  This is not
surprising since, as noted earlier, $n_{0}$ is the strongest factor in
determining CCT.  All the SCC clusters are CC clusters when determined
with $n_{0}$ and, similarly, both methods agree on the NCC clusters as
well. The WCC clusters are split at CCT $\lesssim$
2~$h_{71}^{-1/2}$~Gyr into nine CC and nine being NCC when determined
by $n_{0}$.  It is impossible to separate SCC and WCC clusters using
$n_{0}$.  This suggests that the difference between SCC and WCC
clusters with similar values of $n_{0}$ is that SCC clusters have a
lower central temperature.  This may be due to the central temperature
drop seen in SCC clusters or simply that these SCC clusters are cooler
than their WCC counterparts.  Considering NCC and WCC clusters with
similar values of $n_{0}$, the WCC clusters must be cooler than their
NCC counterparts.

\subsubsection{$K_{\rm BIAS}$}
As with central entropy, $K_{\rm BIAS}$ seems to form a tight
correlation with CCT.  The KMM break separates the SCC clusters from
the other clusters (with one exception).  Perhaps the most interesting
aspect of $K_{\rm BIAS}$ is that the slope seems to break between the
CC and NCC clusters.  Unlike many of the other parameters this looks
truly like a broken powerlaw, as fits to both the SCC and WCC/NCC
clusters have similar values at the transition between them.  This
appears to be a good parameter for determining CC and NCC clusters,
although ideally the redshift and observation length biases would need
to be removed. The merging cluster A3266 is the outlier with
CCT$\sim$7.7~$h_{71}^{-1/2}$ and $K_{\rm BIAS}$
$\sim$900~$h_{71}^{-1/3}$ keV cm$^{2}$.

\subsubsection{Cooling Radius (\% $r_{vir})$}
Since the cooling radius is defined at $t_{\rm cool}(r_{\rm cool}) =
7.7$~$h_{71}^{-1/2}$~Gyr, the NCC clusters have no cooling radius by
definition.  These clusters were excluded when fitting the relation
between cooling radius and CCT.  There seems to be a trend of
increasing cooling radius with decreasing cooling time for the WCC
clusters.  The SCC clusters, however, do not seem to have any trend
between cooling time and size of the cooling region (as a fraction of
the virial radius).  Therefore we can conclude that the density
gradients for SCC clusters vary greatly.  The large outliers are:
S1101 at the high end with a cooling radius of $>$~0.15~$r_{500}$ and
the Fornax cluster at the low end.

\subsubsection{Scaled Spectral Mass Deposition Rate}
This plot explicitly demonstrates the danger of using $\dot{M}_{\rm
  spec}$ to divide the distribution.  The error bars that are not
visible in the histogram and not considered in the KMM-algorithm, show
how much overlap there is within errors.  There seems to be only a
very weak trend with CCT, especially if the upper limits are included.
The clusters with the highest values of $\dot{M}_{\rm
  spec}$/$M_{500}$, however, are SCC clusters.  As noted earlier and
discussed in more detail later, several of the NCC clusters have
non-negligible spectral mass deposition rates.  This fact along with
the spread in values for the SCC clusters makes it very difficult to
use $\dot{M}_{\rm spec}$ to identify CC and NCC clusters, at least
with the spectral resolution of the {\it Chandra} {\it ACIS}.

\subsubsection{Scaled Classical Mass Deposition Rate}
\label{CMDR_Comparison}
As with cooling radius, $\dot{M}_{\rm classical}$ is defined as zero
for the NCC clusters (clusters with CCT $>$ 7.7~$h_{71}^{-1/2}$~Gyr),
so these clusters were omitted when fitting the comparison of
$\dot{M}_{\rm classical}$/$M_{500}$ to CCT.  There appears to be a
weak trend of increasing $\dot{M}_{\rm classical}$/$M_{500}$ with
decreasing CCT, albeit with a great deal of scatter.  We attribute
this scatter to the differences in the gradients of the density
profiles.  That is, clusters with a very steep density gradient will
have a very short CCT but not a large value of $\dot{M}_{\rm
  classical}$, whereas clusters with relatively dense cores but
flatter density gradients will have a longer CCT but a large
$\dot{M}_{\rm classical}$.  Based on the KMM determined cut of 0.5
$\times$ 10$^{-14}$~$h_{71}^{-1}$~yr$^{-1}$ for $\dot{M}_{\rm
  classical}$/$M_{500}$, all the SCC clusters are classified as CC
clusters and all but three of the WCC clusters are classified as CC
clusters.  Two of these WCC clusters are borderline cases.

There are six WCC clusters (IIIZw54, A1650, A1651, A2142, A2244 and
A4038) that have $\dot{M}_{\rm classical}$/$M_{500}$ $>$ 15 $\times$
10$^{-14}$~$h_{71}^{-1}$~yr$^{-1}$, more consistent with SCC clusters,
although toward the lower end of the SCC clusters.  There are five SCC
clusters (A0262, MKW4, NGC4636, A3526 and A1644) that have
$\dot{M}_{\rm classical}$/$M_{500}$ $<$ 15 $\times$
10$^{-14}$~$h_{71}^{-1}$~yr$^{-1}$, more consistent with WCC clusters.
There are four (one SCC and three WCC clusters) outliers that have
very low values of $\dot{M}_{\rm classical}$/$M_{500}$ (i.e. $<$ 4
$\times$ 10$^{-14}$~$h_{71}^{-1}$~yr$^{-1}$).  They are: NGC1399,
A3266, A3667 and A2634.  It is interesting to note that all but two of
the low $\dot{M}_{\rm classical}$/$M_{500}$ SCC clusters are low
temperature ($kT < 2.5$ keV) systems, in fact they are four of the
nine coolest systems.  For the ten coolest systems, $\dot{M}_{\rm
  classical}$/$M_{500}$ is less than 45 $\times
10^{-14}~h_{71}^{-1}$~yr$^{-1}$, which is below the average for the
SCC clusters.  The WCC outliers are all merging systems suggesting
current cool-core survival, but with some disruption of the cool core.

\subsubsection{Cuspiness}
Cuspiness shows a lot of scatter when compared to CCT.  It seems to do
a good job in identifying NCC clusters; only three NCC clusters have
$\alpha > 0.5$.  All the SCC clusters have $\alpha > 0.5$, but not all
of them have $\alpha$ $> 0.75$ as suggested by \citet{vikhlinin07}.
Using the KMM determined cut of $\alpha$ = 0.7, several of the SCC and
all of the WCC would be classified as NCC clusters. Likewise, using
the cuts of \citet{vikhlinin07}, seven of the WCC clusters would be
SCC clusters, three of the SCC clusters would be WCC clusters and four
of the WCC clusters would be classified as NCC clusters.  Looking at
the fit to the data, there appears to be a shift in normalization and
slope for the SCC clusters.  That is, cuspiness increases with shorter
CCTs. However, below CCT$\sim$1~$h_{71}^{-1/2}$, the cuspiness drops
suddenly (or at least there is a large range of values). Below a CCT
of $\sim$1~$h_{71}^{-1/2}$~Gyr, the cuspiness increases more rapidly
as the CCTs get shorter.

\subsubsection{Scaled Central Luminosity}
Scaled central luminosity ($L_{X}$/[$M_{\rm gas}$ $kT_{\rm vir}$])
shows a rather flat but clear relation with CCT, with large scatter
and larger uncertainties.  Using the KMM determined cut, however, only
one SCC cluster is misclassified as an NCC cluster. Some of the WCC
clusters are classified as CC clusters and some as NCC clusters when
using the KMM determined cut in scaled central luminosity.  If the cut
is raised to 0.5 $\times$ 10$^{32}$~$h_{71}^{1/2}$ erg s$^{-1}$
keV$^{-1}$ M$_{\odot}^{-1}$, still only one SCC cluster is
misclassified, and the number of misclassified NCC clusters is reduced
to two.  Unlike, e.g. $n_{0}$, there seems to be no trend with the CCT
of the WCC and whether it is classified as CC or NCC when using scaled
central luminosity.  As with many parameters the relation seems to get
steeper for the SCC clusters.

\subsubsection{$T_{0}$/$T_{\rm vir}$}
The most interesting feature in this plot is the break in the
distribution for the SCC clusters. Although there is a lot of scatter,
there is a clear drop in central temperature for SCC clusters.
However, the value of the drop seems to be independent of CCT. There
are a few NCC clusters with central temperature drops, most notably
A2256. This may be due to a recent merger in which a core has not yet
been completely destroyed (although a possible dense core has been
disrupted).

\citet{sanderson06} suggested a similar parameter for dividing CC and
NCC clusters. Their parameter is roughly the inverse of
$T_{0}$/$T_{\rm vir}$ ($T$-ratio $\equiv$ $T_{\rm vir}$/$T_{\rm
  core}$). The major difference being that $T_{\rm core}$ is defined
as the temperature within 0.1$r_{500}$, in general much larger than
the region we used to measure $T_{0}$. They then define CC
significance, which is the difference (in units of $\sigma$) between
$T$-ratio and unity
(i.e. $\frac{T{\rm-ratio}\;-\;1}{\sigma_{T{\rm-ratio}}}$). They define
a CC cluster as any cluster with CC significance $>$3 (i.e. a
significant drop in temperature). Since they use a bright subsample of
{\it HIFLUGCS}, all 20 of their clusters are in our sample. We find
that all their CC clusters are SCC clusters in our sample. Their NCC
clusters are either WCC or NCC clusters. Likewise, employing their
method, using $T_{\rm vir}$/$T_{0}$, we find good agreement with their
classification. On the other hand we argue the values of CC
significance are very sensitive to the uncertainties in $T_{0}$ and
$T_{\rm vir}$. That is if $T$-ratio $\lesssim$ 1, then the longer the
observing time the larger CC significance becomes and the more likely
the cluster will be classified as a CC cluster. However, we concur
with their result that in general SCC clusters have a central
temperature drop and that WCC and NCC clusters have a shallow or no
central temperature drop.

\subsubsection{Slope of the Temperature Profile}
Like $T_{0}$/$T_{\rm vir}$, this plot shows a break in the
distribution at the SCC clusters.  It is clearer in this plot than in
the $T_{0}$/$T_{\rm vir}$ plot.  This is most likely because the slope
takes into account a systematic decrease in temperature as opposed to
random temperature fluctuations.  That is, for an NCC cluster with
some cool gas that has not been completely disrupted, there will be
less of a gradient than for a cluster that has a systematic decrease
in temperature in a dense, well-defined core.  For the SCC clusters
there seem to be many slopes independent of CCT, suggesting no
universal central temperature profile.  This along with
$T_{0}$/$T_{\rm vir}$ demonstrates that studies that define CC
clusters based on a temperature drop in the core are defining SCC
clusters as CC clusters and WCC and NCC clusters as NCC clusters.  The
temperature drop and profiles are discussed in more detail in
Sect.~\ref{kTProfiles}.

\subsubsection{[$T_{0}$ (0.5 - 2.0 keV)]/[$T_{0}$ (2.0 - 7.0 keV)]}
[$T_{0}$ (0.5 - 2.0 keV)]/[$T_{0}$ (2.0 - 7.0 keV)] appears to be a
very poor parameter for identifying CC clusters.  As with
$\dot{M}_{\rm spec}$, it is possible that multiple temperatures along
the line of sight (e.g. from a merging system) can produce results
that make an NCC cluster appear to be a CC cluster.  In fact the best
fit (albeit a very poor fit) shows a positive trend rather than the
expected negative trend.  Additionally, an NCC cluster (A1656) shows
the smallest fractional value.

\subsubsection{$M_{\rm gas}$/$M_{500}$}
\begin{figure}
\includegraphics[width=90.0mm,angle=0]{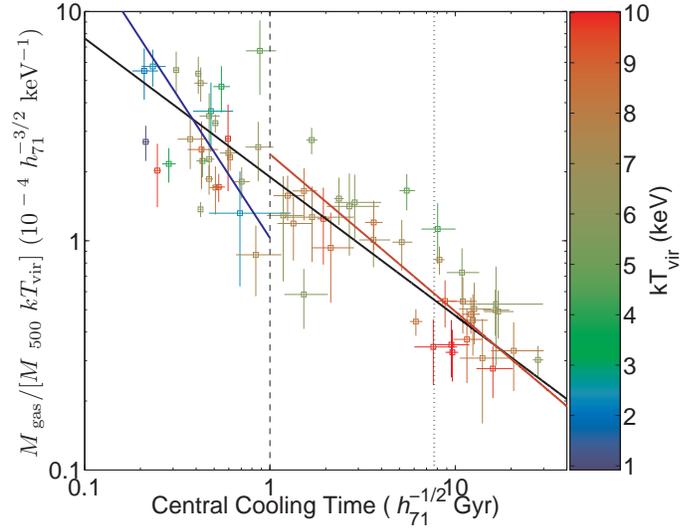}
\caption[$M_{\rm gas}$/[$M_{500}$ $kT_{\rm vir}$] {This figure shows
  that that $M_{\rm gas}$/$M_{500}$ versus CCT correlation tightens
  for SCC clusters if scaled by $kT_{\rm vir}$ This suggests that cool
  clusters cause the scatter in this relationship.  Unfortunately, but
  not surprisingly, the tight correlation between $M_{\rm
    gas}$/$M_{500}$ and CCT for WCC and NCC clusters becomes
  worse.  \label{Mgas_kT_vs_CCT}}
\end{figure}

$M_{\rm gas}$/$M_{500}$ shows a very promising trend for the WCC and
NCC clusters; however, at CCT$\lesssim$1~$h_{71}^{-1/2}$~Gyr there
appears to be no relation.  In fact, the best fit to the SCC clusters
predicts a decrease in $M_{\rm gas}$/$M_{500}$ with decreasing CCT.
The outliers appear to be cool clusters and groups, which are less
dense in the core (short cooling time comes partially from low
temperature) and possibly have steeper profiles (leading to less gas
mass within 0.048~$r_{500}$ than for rich clusters).
Fig.~\ref{Mgas_kT_vs_CCT} shows how dividing by $kT_{\rm vir}$ lessens
the scatter for the SCC clusters, explicitly showing that the large
scatter is caused by cool clusters.  Of course, since the correlation
is already tight for NCC and WCC clusters, dividing by $kT_{\rm vir}$
results in more scatter for those clusters.

\subsubsection{Scaled Modified Spectral Mass Deposition Rate}
As with $\dot{M}_{\rm spec}$/$M_{500}$, the large uncertainty in
$\dot{M}_{\rm spec2}$/$M_{500}$ shows the limitations of the
histogram.  However, $\dot{M}_{\rm spec2}$/$M_{500}$ does show a trend
with cooling time, albeit with a large scatter.  The fact that it
shows a trend is interesting since CCT and $\dot{M}_{\rm spec2}$ are
independently determined. See Sect.~\ref{CC_PROB} for a further
discussion on $\dot{M}_{\rm spec2}$.

\begin{table*}
\footnotesize
\caption{The best-fit parameters for the 15 investigated CCT versus parameter. The parameters were fit to CCT = CCT$_{0}$ (${\rm parameter})^{\Gamma}$.}
\label{parameterfits}
\centering
\begin{tabular}{l c c c c c c}
\hline
\hline
Parameter & $\Gamma$ & CCT$_{0}$       	   & $\Gamma$ & CCT$_{0}$           & $\Gamma$  & CCT$_{0}$ \\ 
          &  All     & All                 &  SCC     & SCC                 & WCC+NCC   & WCC+NCC \\
          &          & $h_{71}^{-1/2}$~Gyr &          & $h_{71}^{-1/2}$~Gyr &           & $h_{71}^{-1/2}$~Gyr \\
(1) & (2) & (3) & (4) & (5) & (6) & (7) \\
\hline
$K_{0}/(100\; h_{71}^{-1/3}\;\rm{keV}\;\rm{cm}^2)$        &  1.085$\pm$0.036 & 4.3$\pm$0.6     &  0.843$\pm$0.109 & 2.7$\pm$0.8   &  1.278$\pm$0.073 & 3.9$\pm$1.3 \\
$\Sigma_{0}/(10^{-6}$ photons cm$^{-2}$ s$^{-1}$ arcsec$^{-2})$
                                                          & -0.894$\pm$0.061 & 2.6$\pm$0.3     &  0.991$\pm$0.737 
												      & 0.090$^{+0.110}_{-0.063}$ & -0.830$\pm$0.073 & 3.3$\pm$0.4 \\
$r_{\rm c}/(r_{500}$10$^{-2})$				  &  0.932$\pm$0.060 & 1.6$\pm$0.2     &  0.378$\pm$0.068 & 0.63$\pm$0.06 &  0.698$\pm$0.085 & 2.8$\pm$0.5 \\
$n_{0}/(10^{-2}\; h_{71}^{1/2}\;\rm{cm}^{-3})$            & -1.194$\pm$0.051 & 3.5$\pm$0.2     & -0.860$\pm$0.151 & 1.8$\pm$0.4   & -1.016$\pm$0.046 & 4.1$\pm$0.2 \\
$K_{\rm BIAS}/(100\; h_{71}^{-1/3}\;\rm{keV}\;\rm{cm}^2)$ &  0.975$\pm$0.062 & 4.7$\pm$1.2     &  0.501$\pm$0.081 & 1.5$\pm$0.3   &  1.314$\pm$0.166 & 4.4$\pm$3.3 \\
exp[Cooling Radius/(10$^{-2}$ $r_{\rm vir}$)]$^\dagger$   & -0.350$\pm$0.025 & 11.1$\pm$1.1    & -0.584$\pm$1.055 
													      & 70.46$^{+653.06}_{-70.45}$ & -0.336$\pm$0.026 & 12.2$\pm$1.0 \\
$(\dot{M}_{\rm Spec}$/$M_{500})$/(10$^{-14}$ $h_{71}$~yr$^{-1}$) 
							  & -0.598$\pm$0.121 & 1.5$\pm$0.4     & -0.996$\pm$1.313 & 0.78$\pm$0.33 & -0.404$\pm$0.163 & 4.0$\pm$0.9 \\
$(\dot{M}_{\rm Classical}$/$M_{500})$/(10$^{-14}$ $h_{71}^{-1}$~yr$^{-1}$)
					  & -0.816$\pm$0.742 & 1.1$^{+2.3}_{-1.0}$ &  0.966$\pm$0.675 & 0.018$^{+0.040}_{-0.016}$ & -0.280$\pm$0.163 & 4.1$\pm$1.3 \\
$exp[\alpha$]$^{\dagger}$                                 & -3.726$\pm$0.198 &  34$\pm$6       & -1.610$\pm$0.268 & 2.8$\pm$0.9   & -3.589$\pm$0.345 & 35$\pm$7  \\
$L_{X}/(10^{32} h_{71}^{1/2}\; {\rm ergs}\; {\rm s}^{-1}\; {\rm keV}^{-1}\; {\rm M}_{\odot}^{-1}) $ 
                                                          & -3.057$\pm$0.250 & 0.41$\pm$0.07   & -1.114$\pm$0.167 & 0.43$\pm$0.04 & -3.795$\pm$0.699 & 0.24$\pm$0.14 \\
exp[$T_{0}$/$T_{\rm vir}$]$^{\dagger}$	          &  4.931$\pm$0.365 & 0.043$\pm$0.013 & -0.102$\pm$0.290 & 0.76$\pm$0.10 &  5.047$\pm$2.421 & 0.041$^{+0.099}_{-0.037}$ \\

exp[Slope of $T$-profile]$^{\dagger}$			  &  6.908$\pm$1.209 & 5.2$\pm$1.0     & -1.208$\pm$0.271 & 0.32$\pm$0.04 & 17.48$\pm$16.60  & 8.4$\pm$4.2 \\
exp[$T_{0}$(0.5-2.0 keV)/$T_{0}$(2.0-7.0 keV)]$^{\dagger}$          
                                                          & -3.469$\pm$3.389 & 28$^{+75}_{-26}$ & -1.027$\pm$0.586 & 1.0$\pm$0.5   & -3.957$\pm$2.316 & 125$^{+215}_{-64}$ \\
$M_{\rm gas}$/($10^{-3}$ $h_{71}^{-3/2}$ $M_{500}$)	  & -1.997$\pm$0.136 & 0.58$\pm$0.10   & 0.894$\pm$0.229  & 0.49$\pm$0.06 & -1.832$\pm$0.171 & 0.93$\pm$0.20 \\
$(\dot{M}_{\rm Spec2}$/$M_{500})$/(10$^{-14}$ $h_{71}$~yr$^{-1}$)
							  & -0.375$\pm$0.028 & 3.3$\pm$0.4     & -0.708$\pm$0.248 & 4.3$\pm$3.5   & -0.347$\pm$0.058 & 6.3$\pm$0.8 \\
$\Sigma_{0}/($kT$ 10^{-7}$ photons cm$^{-2}$ s$^{-1}$ keV$^{-1})$	  
							  & -0.892$\pm$0.039 & 5.5$\pm$0.4     & -0.630$\pm$0.221 & 2.5$\pm$1.6   & -0.914$\pm$0.063 & 5.7$\pm$0.4 \\
$M_{\rm gas}$/($10^{-4}$ $kT$ $M_{500}$ keV$^{-1}$)	  & -1.654$\pm$0.089 & 2.9$\pm$0.2     & -0.805$\pm$0.138 & 1.02$\pm$0.17 & -1.458$\pm$0.127 & 3.6$\pm$0.4 \\

\hline
\multicolumn{7}{p{\textwidth}}{\footnotesize Columns:  (1)~the parameter name, (2)~the fitted slope ($\Gamma$) including all clusters, (3)~the fitted normalization (CCT$_{0}$) including all clusters, (4)~$\Gamma$~for the SCC clusters, (5)~CCT$_{0}$ for the SCC clusters, (6)~$\Gamma$ for WCC and NCC clusters and (7)~CCT$_{0}$ for the WCC and NCC clusters.}
\end{tabular}
\raggedright
$^\dagger$For this parameter the fitting function was CCT = CCT$_{0}$ (exp$[{\rm parameter}]$)$^{\Gamma}$
\end{table*}

\subsection{What is the Best Diagnostic of CC Clusters?}
\label{bestproxy}
One large problem in studying the evolution of CC cluster fraction is
that resolving the cores of clusters is difficult for distant
clusters.  In sect.~\ref{comparison} we compared various parameters to
CCT.  Not surprisingly $n_{0}$ and $K_{0}$ provide the tightest
correlation to CCT.  In distant clusters $n_{0}$ is slightly easier to
determine than CCT and $K_{0}$ since it requires no spectral
information from the central region.  On the other hand it is still
dependent on one being able to resolve the very center of the cluster,
so it is not the ideal parameter to determine whether a distant
cluster is a CC or NCC cluster.

Unfortunately, when compared to $n_{0}$ and $K_{0}$, most of the other
parameters show a lot of scatter in their relation with CCT. Of those
other parameters $K_{\rm BIAS}$ appears to show the tightest
correlation.  Unlike $K_{0}$, $K_{\rm BIAS}$ is not dependent on
$n_{0}$, but on the average density in a somewhat larger region.  It
is therefore not limited by the need to resolve the central region.
There are, however, two problems with using $K_{\rm BIAS}$ as a proxy:
(1)~it requires a large enough region to fit a spectrum (which can be
physically quite large for distant clusters) and (2)~the size of the
region from which $K_{\rm BIAS}$ is determined differs between
observations (i.e. it is dependent mainly on observing time and
redshift).  As a solution to (1)~we suggest that the normalization of
the thermal model is not as difficult to constrain as the temperature
or metalicity (not as many counts are needed).  In extreme cases the
temperature and metalicity could be frozen at their overall cluster
values so that only the central normalization would need to be
constrained.  As a solution to (2)~we suggest that the region size be
defined such that it is only dependent on the density profile and is
independent of redshift, temperature and observing time.  Specifically
if two clusters have identical density profiles, then $K_{\rm BIAS}$
should be derived from physical regions of the same size.

We suggest using a constant value of the projected emission measure in
a cylinder,
\begin{equation}
\label{defnI}
I(R) \equiv 2\pi\int_{-\infty}^{\infty} \int_{0}^{R} n_{\rm e} n_{\rm H}\;r  \mathrm{d}r \mathrm{d}l,
\end{equation}
to determine region sizes that depend only on the density
profile. Here we define $n_{e}$ as the electron density and $n_{H}$ as
the proton density. One can simply use the density profile or model
for a cluster to directly determine $R$ for a predetermined value of
$I$. Alternately, for given spectral parameters and distance, the
ratio of the count rate to normalization (as defined in XSPEC for
example) is determined, so that the total observed count, $C$, can be
determined and used to define $R$.  That is, the count threshold for a
region of size $R$ that has a given value of $I$ is:
\begin{equation}
  \frac{{\mathcal C}}{{\rm cnts}} = F \left ( \frac{ t_{\rm obs}}{100\;{\rm ks}} \right )\left( \frac{\kappa}{100\;{\rm cnts}\;{\rm cm}^5\;{\rm s}^{-1}} \right) 
  \left ( \frac{200 \; h_{71}^{-1}\;{\rm Mpc}}{D_{A}(1+z)} \right )^2,
\end{equation}
where $t_{\rm obs}$ is length of the observation, $\kappa$ is the
ratio of the count rate to normalization and $F$ is a constant that
depends on the desired value of $I$,
\begin{equation}
  F \equiv \frac{10^{-11}}{16\pi} \left ( \frac{I}{h_{71}^{-2}\;{\rm Mpc}^{-2}\;{\rm cm}^{-1}} \right ).
\end{equation}
A value of $I$ = 8.6 $\times$ 10$^{65}$~$h_{71}^{-2}$ cm$^{-3}$,
corresponding to $F$ = 17\,910, produces a 10\,000 count region (in
the 0.5 - 7.0 keV band) for a metalicity of $Z$ = 0.25, a
photoelectric absorption column density of $N_{H}$ = 2 $\times$
10$^{20}$ cm$^{-2}$ and our median virial temperature (4.3 keV),
redshift (0.047) and observing time time (44 ks).  See
appendix~\ref{countscalc} for more details on these calculations.

Although this method makes the $K_{\rm BIAS}$ region size consistent
among clusters, there is an additional problem that a reasonable value
of $I$ at one redshift, will often be unreasonable for clusters at a
significantly different redshift.  For instance, for our proposed
value of $I$ = 8.6 $\times$ 10$^{65}$~$h_{71}^{-2}$ cm$^{-3}$, which
leads to a 10\,000 count region for our median cluster parameters,
would produce a 58 count region for a cluster with identical
parameters but at a redshift of $z$ = 0.5. However, if the region size
is not constant for all studies (say the chosen value of $I$ varies
depending on the study), it makes comparisons between studies
difficult.

After $K_{\rm BIAS}$, the next best parameter appears to be the scaled
central X-ray luminosity (scaled $L_{X}$).  The region from which we
extracted scaled $L_{X}$ is well-defined (0 - 0.048~$r_{500}$), but
suffers from two other problems.  The scaling depends on $M_{\rm
  gas}$, which in turn depends on $n_{0}$.  In addition, as with
$K_{\rm BIAS}$, a spectrum is needed in order to determine the
luminosity.  Of course, as with $K_{\rm BIAS}$, in extreme cases the
overall cluster temperature and metalicity can be used when getting an
estimate of the core luminosity.  The major advantage in using scaled
$L_{X}$ over $K_{\rm BIAS}$ is that it is better defined for this
study.

For distant clusters in which only a few thousand counts are
available, it is necessary to use $\Sigma_{0}$ or $\alpha$.
$\Sigma_{0}$ has the advantage that surface brightness profiles only
need $\sim$50 counts per bin, as opposed to spectra that require
thousands of counts.  Unfortunately there is the problem of
resolution.  For very distant clusters, it may not be possible to
resolve $\Sigma_{0}$ in the very central regions.  Also, generally if
$\Sigma_{0}$ is measurable then $n_{0}$ is also measurable and is a
much better proxy.  However, if $\Sigma_{0}$ is the only parameter
available, it should be divided by $kT_{\rm vir}$ before being used as
a proxy (Fig.~\ref{sigkT_vs_CCT}).  Cutting $\Sigma_{0}$/$kT_{\rm
  vir}$ at $\sim$1.5 $\times$ 10$^{-7}$ counts cm$^{-2}$ s$^{-1}$
keV$^{-1}$ does a good job of separating NCC and CC clusters.

This leaves cuspiness as the only viable candidate for distant
clusters with few counts.  Unfortunately, there is a great deal of
scatter in the $\alpha$-CCT plot.  Moreover, there seems to be a drop
in $\alpha$ at CCT $\sim$ 1~$h_{71}^{-1/2}$~Gyr, so that SCC clusters
with CCT $\sim$ 1~$h_{71}^{-1/2}$~Gyr have the same value of $\alpha$
as the WCC clusters.  Unlike $\Sigma_{0}$ there does not seem to be
any trend with $T_{\rm vir}$ and we have not found any parameter which
can be used to tighten this correlation.  Cutting $\alpha$ at 0.75
ensures for the most part that the sample above the cut will be CC
clusters, although not necessarily SCC clusters as suggested by
\citet{vikhlinin07}.  Cutting $\alpha$ at 0.5 seems to capture most of
the CC clusters above the cut, with a slight contamination from the
{\it steepest} NCC clusters.  Finally cutting at $\alpha$=1 ensures
that the clusters above the cut are SCC clusters, although there are
still several SCC clusters below the cut.  In short, it is impossible
to separate out a sample, but it is safe to say a cluster with
$\alpha>$1 is an SCC cluster and a cluster with $\alpha>$0.5 is most
likely a CC cluster.

\begin{figure}
\includegraphics[width=85.0mm,angle=0]{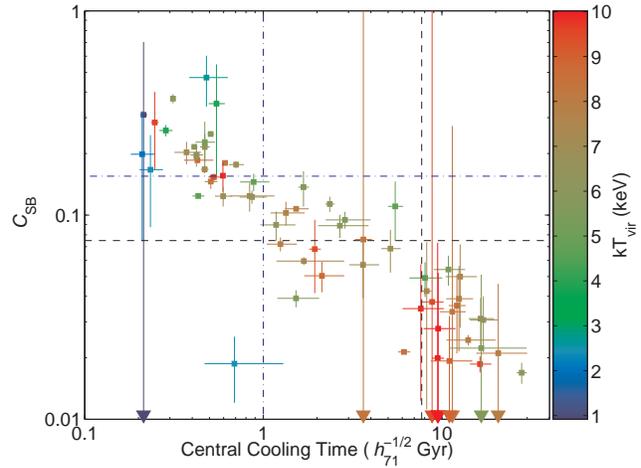}
\caption[Concentration Parameter versus Central Cooling Time] {The
  concentration parameter $C_{\rm SB}$ suggested by \citet{santos08}
  versus CCT.  These preliminary results seem very promising.  As
  discussed in the text, the outlier (Fornax cluster) is probably due
  to a problem in extrapolating the surface brightness profile. The
  horizontal blue dash-dot line and black dashed line show the
  suggested cuts of \citet{santos08} between SCC and WCC clusters
  (0.155) and WCC and NCC clusters (0.075),
  respectively.\label{CSB-CCT}}
\end{figure}

Recently, \citet{santos08} suggested a concentration parameter
($C_{\rm SB}$) as a proxy for determining whether a distant cluster is
a CC or not (see Eqn.~\ref{concen}).  Although this proxy is beyond
the scope of our analysis (the residual background must be carefully
handled to find the surface brightness at 400~kpc), we used our
$\beta$ model fits to estimate $C_{\rm SB}$.  Fig.~\ref{CSB-CCT} shows
the results of the fit, which seems very promising.  The one strong
outlier is NGC1399.  As discussed in Appendix~\ref{FC}, the surface
brightness profile of the Fornax cluster shows a flattening at
$\sim$0.04-0.05~$r_{500}$, probably due to NGC1404.  This flattening
will cause a severe overestimate of $\Sigma(r<400~{\rm kpc})$ when
extrapolating the profile out to 400~kpc.  To do a true comparison, we
would need to measure the integrated surface brightness, rather than
use models extrapolated from $\sim$0.05~$r_{500}$.

\section{Discussion}
\label{disc}

\subsection{The Cooling Flow Problem}
\label{CC_PROB}

\begin{figure}
\includegraphics[width=80.0mm,angle=0]{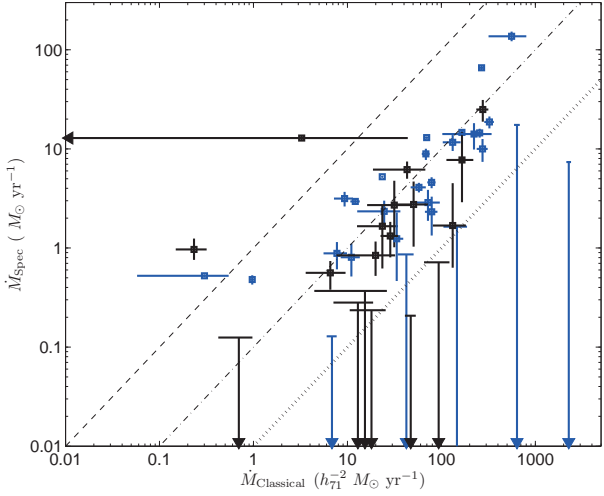}
\caption[$\dot{M}$ determined by the spectrum versus the classical
method] {Spectroscopic cooling rate $\smdr$ versus classical cooling
  rate $\mdr$ for the 46 CC clusters. The dashed-line shows parity
  between the two. The dash-dot line and the dotted line show one and
  two orders of magnitude difference respectively. In all but three
  cases, $\smdr$ $\ll$ $\dot{M}_{\rm
    classical}$. \label{specCCvsdensCC}}
\end{figure}

The {\it cooling flow problem} is the discrepancy between the
spectrally determined mass deposition rate and the classically
determined mass deposition rate. Fig.~\ref{specCCvsdensCC} clearly
demonstrates the cooling flow discrepancy.  In all but three of the 46
CC clusters, the spectral mass deposition rate is less than the
classical mass deposition rate, usually by at least one order of
magnitude (dash-dot line).

In two of these three cases, NGC1399 and A3266, the mass deposition
rates are consistent within errors.  Only in the case of A2634 is
$\smdr > \mdr$.  As noted in Sect.~\ref{CMDR_Comparison}, all three of
these clusters (A2634, NGC1399 and A3266) have anomalously low values
of $\mdr$.  The WCC clusters A2634 and A3266 appear to be involved in
mergers (although as discussed later, probably not major mergers), the
low values of $\mdr$ being then consistent with a disrupted core.
Also as noted in Sect.~\ref{define_param}, $\smdr$ can be
over-estimated in merging clusters.  Moreover in the case of A3266, it
is on the line between NCC and WCC clusters with CCT $\sim
7.7$~$h_{71}^{-1/2}$~Gyr.  As seen in Table~\ref{par}, several of the
NCC clusters have nonzero values of $\smdr$, so it is not so
surprising that A3266 would have $\smdr$ $\sim$ $\mdr$.  The case of
NGC1399 is not as clear.  In many of the plots NGC1399 is an outlier.
As discussed in Appendix~\ref{FC}, this may be due to a problem with
extrapolating the surface brightness profile.  Even if this is the
case, $\mdr$ is small ($<$ 1~$h_{71}^{-2}$ $M_{\odot}$~yr$^{-1}$) and
$\mdr$ $\approx$ $\smdr$ (see Appendix~\ref{FC} for details).  The
anomalously low $\mdr$ indicates that the density drops off quickly.
In fact the cooling radius is only 0.026 $r_{500}$, which is among the
four smallest.  The other three (in order from smallest to largest)
are A3266, A3667, A2634, all WCC merging clusters, two of which are
also outliers in this plot.  The SCC cluster with the next smallest
cooling radius is A0262 that has a cooling radius of almost twice that
of NGC1399.  Although NGC1399 shows some evidence of merging
\citep[e.g.][]{drinkwater01}, it appears to be just starting to merge
so only the outer regions have been affected.  It is also possible
that the galaxy NGC1404 has disturbed the core as it fell in.  A more
interesting scenario is that since NGC1399 is dynamically young
(Y. Schuberth, private communication), the cooling flow has just
formed so no energy has been injected yet and $\smdr \approx \mdr$.

\begin{figure}
\includegraphics[width=90.0mm,angle=0]{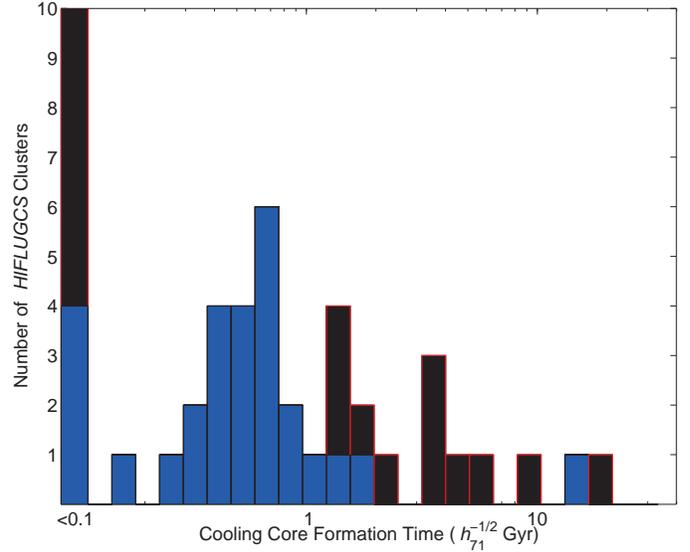}
\caption[Cool Core Formation Time] {This figure shows the histogram of
  cool core formation time (CCFT) for the 46 CC clusters. Ten of the
  clusters only have upper-limits for $\smdr$.  For there to be no
  discrepancy between $\mdr$ and $\smdr$, cooling flows would have to
  have formed very recently. \label{CCFT}}
\end{figure}

With the exception of these three discrepancies, in the other 43 CC
clusters, $\smdr$ $\ll$ $\mdr$. Assuming no energy is input to the CF
and the CF formed $\sim$7.7~$h_{71}^{-1/2}$~Gyr ago, these two
quantities should be equal. If the latter assumption is incorrect, we
can find the cooling radius such that \mbox{$\mdr$ = $\smdr$}. We
define the cooling time at this radius to be the CC formation time
(CCFT). We note that the CCFT is an upper limit to the actual time
since formation because, firstly, as the gas flows to the center, its
density increases causing it to cool more rapidly. Therefore the
actual cooling time of the gas is shorter than $t_{\rm cool}$, as
defined in Eqn.~\ref{tcooldefn}. Secondly, the spectroscopic mass
deposition model, MKCFLOW, used to fit the line emission from a
multiphase gas often provides a better fit than a single-temperature
thermal model. In other words, it may find a cool gas component even
when there is none, yielding $\smdr$ to be greater than its actual
value. Fig~\ref{CCFT} shows the histogram of CCFT for the 46 CC
clusters.  Ten of the clusters have only upper limits for $\smdr$, for
the others the distribution peaks at around 0.5~Gyr, about five to ten
times shorter than \citet{2001MNRAS.322..589A} found using {\it ASCA}
and {\it ROSAT} data and assuming intrinsic absorption.  This means in
order for the observations to match, in the absence of heating, the
CFs would have had to form very recently.

In order to check the plausibility that the cooling flow discrepancy
is due to the assumption that cooling flows have formed very recently,
we devised a rough test to estimate this likelihood. This test is
based on the hypothesis that low- and high-$z$ clusters are both drawn
from the same underlying population of clusters. Hence, the high-$z$
clusters in our sample, when evolved uninterrupted to allow their cool
cores to grow, should have mass deposition rates comparable to the
low-$z$ clusters in our sample. We show below quantitatively, for the
first time, that the likelihood of this hypothesis is very small.

% As described below, we find that the likelihood of this hypothesis
% is very small. This result is, of course, expected but we show it
% here quantitatively, for the first time.

There are two major limitations to this very simple test. Firstly, our
sample is flux-limited and hence due to selection effects we are
likely to pick clusters of increasing mass with increasing
redshift. Secondly, even though our sample spans a limited redshift
range, the cluster population changes with redshift. This is because,
in a hierarchical Universe, the more distant clusters~(high-mass) will
actually never evolve into the mostly low-mass clusters we observe
locally.
% the nearby clusters (mostly low mass) would have never looked like
% the distant clusters (high mass). In a sense we are comparing
% children today to adults of the past.
These two reasons together make a comparison of properties between
low-$z$ and high-$z$ clusters difficult. As an approximate solution to
the first problem, we scaled the mass deposition rates by $\mvir$,
even though this causes a bias against the high-$z$ clusters. This is
due to the fact that at a fixed gas density, the cooling time is an
increasing function of the temperature, ${\rm CCT} \propto T^{1/2}
\propto M^{1/3}$. Thus, if clusters are self-similar, we should expect
the cooling rate to scale with mass more slowly than $\propto
M$. Consequently, under the assumption that the cooling flows are
recent phenomena, we should not expect the high mass clusters evolved
to the lookback time of low-$z$ clusters to show higher $\mdr/\mvir$
than the lower mass ones.

\begin{figure}
  \includegraphics[width=0.35\textwidth, angle=-90]{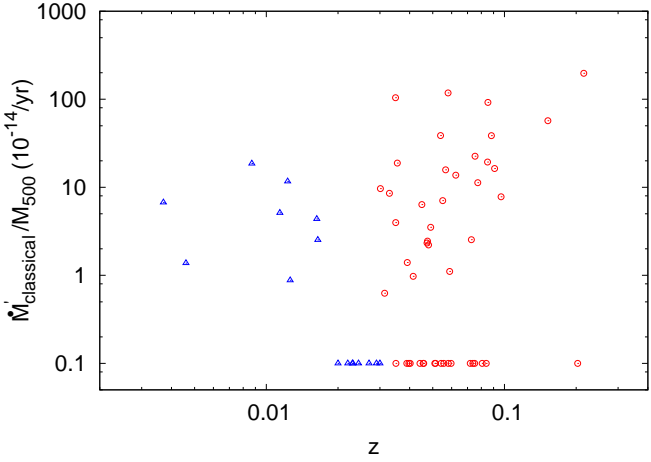}
  \caption{This figure shows the forward-evolved $\mdr$ for high-$z$
    clusters in redshift bins 2, 3 and 4~(red open circles) and
    original $\mdr$ for low-$z$ clusters in redshift bin 1~(blue open
    triangles).
%     The two samples show a marked difference in the distribution
%     especially when considering SCC clusters only~(black open
%     squares).
  }
  \label{MDR-FE}
\end{figure}

We divided our sample into four redshift bins, each with 16
clusters. Although we have a local sample (median redshift for the 16
most distant clusters is $z$ $\sim$ 0.08), most of our clusters span a
lookback time, $t_{\rm lookback}$
$\sim$1~$h_{71}^{-1/2}$~Gyr\footnote{This is the lookback time from
  $z$ = 0 to $z$ $\sim$ 0.08.} $>$ $\langle$CCFT$\rangle$. We forward
evolved each cluster in redshift bins 2, 3 and 4 by adding to its CC
formation time, $CCFT$, the difference between its lookback time,
$LBT$, and the mean lookback time of clusters in the first bin,
$\left<LBT\right>_{1}$. Hence,
\begin{equation}
  CCFT^{\prime} = CCFT + LBT - \left<LBT\right>_{1} \, ,
\end{equation}
where $CCFT^{\prime}$ is the forward-evolved CCFT of that cluster. In
order to calculate the cooling rate in these forward-evolved systems,
we determined the radius, $R^{\prime}$, at which the cooling time of
the gas equals $CCFT^{\prime}$ and calculated $\mdr$ at this radius,
$\mdr^{\prime}$. Shown in Fig.~\ref{MDR-FE} is $\mdr^{\prime}$ scaled
by $\mvir$ versus redshift. Since the low-$z$ clusters belonging to
redshift bin 1~(blue open triangles) are not evolved forward in time,
these have $\mdr^{\prime} = \mdr$. Despite the fact, as noted above,
that scaling $\mdr^{\prime}$ by $\mvir$ very likely overcompensates
the tailing-off of clusters due to selection effects, there seems to
be a marked difference in the parent low-$z$ and high-$z$~(red open
circles) samples. 

Of the 10 clusters in the highest redshift bin with $\mdr^{\prime}$
greater than zero, 8 have $\mdr^{\prime}/\mvir > 10^{-13}$~yr$^{-1}$,
while of the 8 clusters in the lowest redshift bin with
$\mdr^{\prime}$ greater than zero, only 2 have $\mdr^{\prime}/\mvir >
10^{-13}$~yr$^{-1}$. If the two subsamples are drawn from the same
distribution, the probability, $p$, of any one cluster having
$\mdr^{\prime}/\mvir > 10^{-13}$~yr$^{-1}$ should be the same for
both. The joint probability, $P(p) = P(p;m_1 \le 2)P(p; m_2 \ge 8)$,
of drawing one sample of 8 with no more than 2 clusters in this
category and a second sample of 10 with at least 8, is maximized for
$p = p_{\st max} = 0.5584$, giving $P(p_{\st max}) = P_{\st max} \le
0.00883$. Repeating this calculation for all possible pairs of draws
(always using the smaller value for the first of the pair), if the
true probability of $\mdr^{\prime}/\mvir > 10^{-13}$~yr$^{-1}$ is $q$,
the probabilities of each pair for which $P_{\st max} \le 0.00883$,
can be summed to compute the likelihood of this outcome. This is
maximized for $q=0.5$, giving 0.029 for the likelihood. Thus, the
chance that both samples are drawn from the same population is no more
than 2.9\%.

% Of the 29 high-$z$ clusters, 14 have $\mdr^{\prime}/\mvir$ greater
% than $10 \times 10^{-14}~$yr$^{-1}$, whereas of the eight low-$z$
% clusters, only two lie above this cut. Assuming the two samples come
% from the same population, the probability of reproducing the
% distribution of high-$z$ clusters given the low-$z$ clusters is
% $<0.5\%$ and, similarly, the probability of reproducing the
% distribution of low-$z$ clusters given the high-$z$ clusters is $<
% 13\%$. This difference is even more pronounced if only the SCC
% clusters are considered~(black open squares). The corresponding
% probabilites decrease to $<0.08\%$ and $<4\%$ respectively.

We also used the Kolmogorov-Smirnov~(K-S) test to distinguish between
the two samples. Even though it is well-known that the K-S method
exhibits poor sensitivity to the deviations that occur in the tails of
any given two distributions, we find the probability of the null
hypothesis, i.~e. the probability that the low-$z$ and the high-$z$
clusters are drawn from the same distribution, is only $\sim
2.4\%$. From this we can conclude that the high-$z$ clusters, after
having been evolved to match the physical state of low-$z$ clusters,
have systematically higher $\mdr$ than the low-$z$ clusters. Thus, it
is very unlikely that the discrepancy between classical and spectral
mass deposition rates in cool-core galaxy clusters can be explained
away by invoking the recent cool-core formation hypothesis.

We note there are 27 clusters with zero $\mdr^{\prime}$. Of these, 18
constitute the entire NCC cluster subsample, 6 are WCC clusters and 3
are SCC clusters. That the NCC clusters show no mass deposition rate
even after evolving them forward is not surprising. Out of the 9
WCC+SCC clusters, the five high-$z$ clusters are ones with extremely
short or zero $CCFT$ and evolving them forward still results in
$CCFT^{\prime}$ to be shorter than the cooling time at $R=0$.

% There are a number of limitations to this very simple test.  First,
% we are using a local sample allowing the very precise determination
% of the CC fraction, but making it a poor rod to measure transient
% phenomena. Still even for this limited redshift range we find a
% significant result. So, it should be possible to increase this
% significance by expanding the redshift range. Second, since our
% sample is flux limited, in a hierarchical universe the nearby
% clusters (mostly low mass) would have never looked like the distant
% clusters (high mass).  In a sense we are comparing children today to
% adults of the past.

\begin{figure}
\includegraphics[width=90.0mm,angle=0]{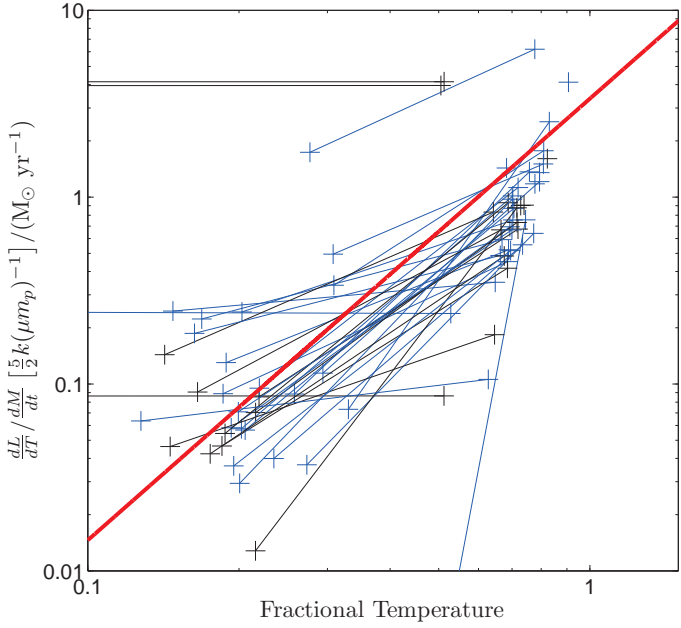}
\caption[Fraction of Cool gas]
{The fraction of differential luminosity vs $T_{0}$/$T_{\rm vir}$,
similar to Fig.~7 in \citet{peterson03}.  The weighted average value
of $\delta$ (see text) for the clusters with measurements at both data
points is $\sim$2.0 and is shown as the thick red-line. The blue
crosses and lines are SCC clusters and the black are WCC clusters.
\label{petersonplot}}
\end{figure}

Another interesting exercise is to check the emission measure
distribution of the cool gas. In terms of the power radiated per unit
temperature, \cite{peterson03} parametrized this as:
\begin{equation}
\frac{dL}{dT} = \frac{5}{2} \frac{\dot{M}k}{\mu m_{p}}(\delta + 1)
\left ( \frac{T}{T_{0}} \right )^{\delta} \,.
\label{petersoneqn}
\end{equation}
For a steady cooling flow at constant pressure, we would have
$\delta=0$, but if cooling is retarded then the amount of gas at lower
temperatures is reduced ($\delta >0$). With our limited spectral
resolution we approximate the fit in the following way.  We assume
that $\dot{M}_{\rm spec2}$ should be compared to the differential
luminosity ($\frac{dL}{dT}$) between the two fitted temperatures
$kT_{\rm high}$ (T$_{0}$ $\equiv$ $T_{\rm high}$) and $kT_{\rm low}$.
We assume that $\smdr$ should be compared to $\frac{dL}{dT}$ between
$kT_{\rm low}$ and zero.  Our argument for using $kT_{\rm low}$ and
not $kT_{\rm high}$ is that the small values of $\smdr$ (compared to
$\mdr$ and especially $\dot{M}_{\rm spec2}$) are due to the lack of
lines for the coolest gas and therefore it gives a better estimation
of the true mass deposition between $kT_{\rm low}$ and zero than
between $kT_{\rm high}$ and zero.  The fraction of observed gas is
$\dot{M}_{\rm spec2}$/$\mdr$ and $\dot{M}_{\rm spec}$/$\mdr$
respectively. Fig.~\ref{petersonplot} is similar to Fig.~7 in
\citet{peterson03}.  For each cluster with measurements at both points
(i.e. neither was an upper-limit) we did a fit to
Eqn.~\ref{petersoneqn}.  Our rough estimation of $\delta$, based on
the weighted average value of the fits, is $\sim$ 2.0 (with a very
large variance).  This value is on the steep end of the values found
by \citet{peterson03}. One may think this slight difference is due to
the fact that we are working with an unbiased sample and their sample
was picked to be SCCs (in general), so that $\dot{M}_{\rm spec}$ is
not far from $\mdr$.  However, there does not seem to be a separation
of the SCC clusters (blue) and the WCC clusters (black).  The three
outliers in the upper left are: NGC1399, A3266 and A2634, which as
mentioned earlier have $\smdr$ $\gtrsim$ $\mdr$.  This model requires
further study with higher spectral resolution instruments before any
definite conclusion can be drawn, but it is reassuring that we find
consistent results using a complete, unbiased sample of clusters that
is more than four times larger.

\subsection{Temperature Profiles}
\label{kTProfiles}

Perhaps two of the most striking plots from Fig.~\ref{paramcompare}
are subplots J and K, which show a sudden break at
$\sim$1~$h_{71}^{-1/2}$~Gyr.  In fact, we find that for the most part
only SCC clusters have a central temperature drop. All 28 SCC clusters
have a central temperature drop and of the eight non-SCC clusters with
a central temperature drop (see Table~\ref{par}), four (A1650, A2065,
A2142 and A2589) have CCT $<$ 2~$h_{71}^{-1/2}$~Gyr, putting them on
the border between SCC and WCC clusters. Additionally two (A2142 and
A3667) have well-known cold fronts.  The other three appear to be
merging clusters in which a cool core has survived: A0400
\citep{hudson06}, A0576 \citep{dupke07} and A2256
\citep{2002ApJ...565..867S}.  It is also clear from
Fig.~\ref{paramcompare}-K that the inner temperature profiles of these
eight non-SCC clusters are, for the most part, flatter than those of
the SCC clusters (slope$\gtrsim$-0.2).  We argue that based on this
fact, studies such as those of \citet{burns07} and
\citet{sanderson06}, which define CC clusters as clusters with a
central temperature drop, are basically defining CC clusters as
clusters with CCT $<$ 1~$h_{71}^{-1/2}$~Gyr (SCC clusters in our
sample).

\begin{figure}
\includegraphics[width=90.0mm,angle=0]{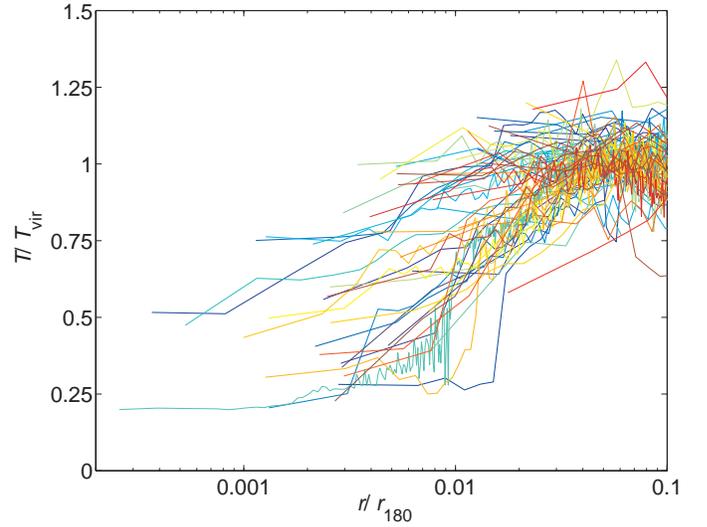}
\caption[{\it HIFLUGCS} kT Profiles] {The inner temperature profiles
  of all 64 {\it HIFLUGCS} clusters scaled by virial temperature and
  radius.  The error bars on the measurements have been omitted for
  clarity.  There appears to be a range of slopes rather than a
  bimodal distribution.  \label{kT_Profiles}}
\end{figure}

Some authors claim the existence of a universal inner temperature
profile for relaxed clusters \citep[e.g.][]{Allen2001,Sanderson2006},
while others find no such universality \citep[e.g.][]{vikhlinin05}.  A
universal inner temperature profile suggests that clusters either have
a flat central temperature profile or a drop that scales with $T_{\rm
  vir}$ and $r_{\rm vir}$.  It is clear from Fig.~\ref{Histogram}-L
that we do not see such a universal inner temperature profile.
Fig.~\ref{kT_Profiles} shows all 64 of our temperature profiles out to
0.1 $r_{\rm vir}$, scaled by $r_{\rm vir}$ and $T_{\rm vir}$.  It is
clear from this figure why we do not see a bimodal distribution in
Fig.~\ref{Histogram}-L.  There is a continuous range of slopes from a
20\% increase above $T_{\rm vir}$ (MKW8) to a decrease down to 0.2
$T_{\rm vir}$ (A3526 - Centaurus Cluster).  We note that some authors
\citep[e.g.][]{sanderson06} only use single thermal models at all
radii when constructing temperature profiles.  We, however, used
double thermal models for the inner annuli in 14 of our SCC clusters
since the addition of the second thermal component improved the fit
significantly ($>99$\% confidence according to the $F$ test).  In
these cases the lower temperature was used, causing a steeper
temperature profile than if a single thermal model had been used.

It is interesting to note that in Fig.~\ref{Histogram}-K, there does
seem to be a bimodal distribution in $T_{0}$/$T_{\rm vir}$ (albeit
with a large dispersion).  This suggests that while the radius of the
gas with $T < T_{\rm vir}$ does not scale with cluster size, the depth
of the drop does.  SCC clusters have a central temperature of
$\sim$0.4 $T_{\rm vir}$, whereas NCC and WCC clusters have $T_{0}$
$\sim$ $T_{\rm vir}$.  As we discuss in the next section, it is
unclear whether this cool gas is associated with the CF or the central
galaxy.  In the case of the former this indicates the coolest major
component of gas allowed by the feedback mechanism.  That is,
significant quantities of gas do not cool below $\sim$0.4 $T_{\rm
  vir}$.  In the case of the latter, it is consistent with observing
that the mass of the brightest cluster galaxy (BCG) scales with the
cluster mass, so that $T_{0}$/$T_{\rm vir}$ is constant.  This model
would support the simulations of \citet{burns07} who find that the CC
grows with the cluster.

\subsubsection{What Causes the Central Temperature Drop?}
An old but still ongoing controversy is whether the temperature drop
in CC clusters is related to the CF or is simply a reflection of the
potential well around the BCG.  In the case of the former, the central
temperature should be reflected by a modified cooling flow model and
should simply depend on the energy factors (energy radiated away,
along with input energy from conduction and any feedback mechanism).
In the latter case the temperature should reflect the mass of the
central galaxy and should be independent of any cooling flow model.

One check for possible multitemperature components is the need for a
double thermal model in the central annuli.  In the central annulus
(annuli) for 14 of 64 {\it HIFLUGCS} clusters (all SCC clusters), a
double thermal fit was a significantly better fit than a single
thermal model. It is possible, since these are projected temperature
profiles, this is simply due to many temperature components along the
line of sight.  Even in this case, the projected temperature
components that dominate are the ones near the cluster center where
the emission is peaked.  Whether it is the projected spectra of outer
annuli or truly the need for a double temperature fit to the central
spectra, we take this as evidence of many temperature components in
the region of the emission peak for these clusters.  Thus the change
in temperature is rapid in these regions.  To explicitly demonstrate
that a double thermal model is consistent with more than two
temperature components we ran simulations of spectra with four and
eight thermal models with temperatures equally spaced between 1 and 2
keV.  Fitting these spectra indicates that a double thermal model can,
at {\it Chandra}'s energy resolution, fully describe the plasma
emission of the underlying four- or eight- temperature components and
provides the statistically best fit to the spectrum ($\chi^2$/dof
$\sim$1)\footnote{For example, for eight thermal components, the
  average reduced $\chi^2$ for a single thermal fit was 1.46, 0.976
  for two thermal components, and 0.984 for three thermal components.
  The simulations were done for a 100ks observation with the
  background and thermal normalization taken from A1795's central
  annulus.  All thermal components were given the same input
  normalization.}.  For the eight temperature plasma the returned
temperatures were: 1.80$\pm$0.08 keV and 1.26$\pm$0.03 keV.  Therefore
we interpret annuli that require two thermal models to contain, at
minimum two thermal components, with a low temperature component $\la
T_{0}$.

\begin{figure}
\includegraphics[width=90.0mm,angle=0]{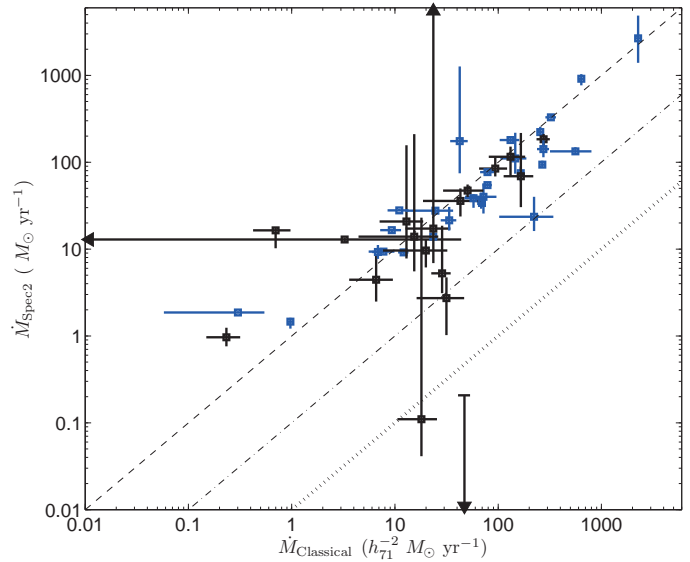}
\caption[Modified Spectral Mass Deposition Rate versus Classical Mass
Deposition Rate] {This plot is similar to Fig.~\ref{specCCvsdensCC}.
  $\dot{M}_{\rm spec2}$ is the spectrally determined mass deposition
  rate between $T_{\rm high}$ and $T_{\rm low}$, where the lower
  temperature is left free (unlike for $\dot{M}_{\rm spec}$).  The
  lines are the same as for Fig.~\ref{specCCvsdensCC}.  The mass
  deposition rates measured by both methods generally yield consistent
  results. \label{specpCCvsdensCC}}
\end{figure}

Fig.~\ref{specpCCvsdensCC} shows the modified spectral mass deposition
rate, $\dot{M}_{\rm spec2}$ versus $\dot{M}_{\rm classical}$.  The
striking thing about this plot is that the two values are very often
consistent with each other, especially for the SCC clusters.  Based on
this result, one may na\"{i}vely believe that the gas cools to $T_{\rm
  low}$ (the lowest temperature for the modified cooling flow model -
see Sect.~\ref{MDR}) at the rate predicted by the classical cooling
flow model, but does not cool at a significant rate below this
temperature.  If this is the case there are two possibilities: (1) the
gas is in thermal equilibrium throughout this range of temperatures or
(2) the gas is rapidly heated from $T_{\rm low}$ to $T_{\rm high}$,
when it reaches $T_{\rm low}$.  An alternate explanation is that with
the spectral resolution of the {\it Chandra} {\it ACIS}, we are unable
to distinguish between gas cooling from $T_{\rm high}$ to $T_{\rm
  low}$ and a rapid drop of temperature at the center.  That is, the
observation that $\dot{M}_{\rm spec2}$ $\approx$ $\dot{M}_{\rm
  classical}$ could simply be a coincidence.

\begin{figure}
\includegraphics[width=90.0mm,angle=0]{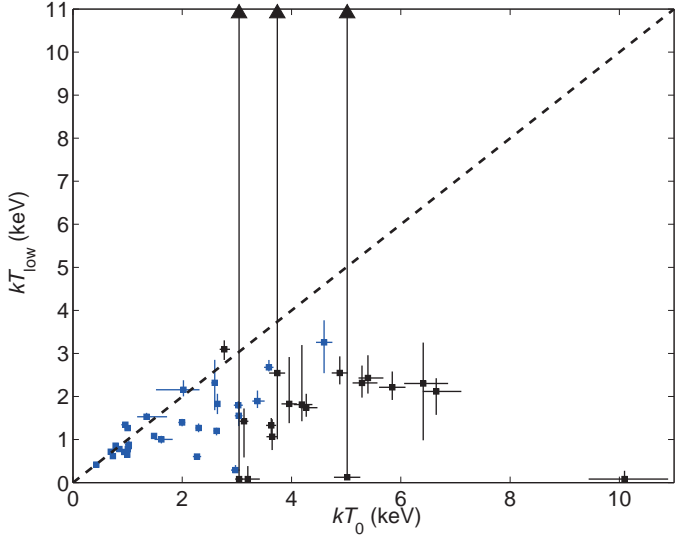}
\caption[$kT_{\rm low}$ versus $kT_{0}$] {The lower temperature in the
  modified cooling flow model ($kT_{\rm low}$) versus the cluster
  central temperature ($kT_{0}$).  Blue points are SCC clusters, while
  black points are WCC clusters.  In general $T_{\rm low}$ $<$
  $T_{0}$, which is expected if there are multitemperature components
  between $T_{\rm high}$ and $T_{\rm low}$.  The fact that the WCC
  clusters show a greater discrepancy between $T_{\rm low}$ and
  $T_{0}$ suggests that at the energy resolution of {\it Chandra's}
  {\it ACIS} it is difficult to distinguish between the modified
  cooling flow model and multitemperature components along the line of
  sight.  The three cases where $T_{\rm low}$ is unconstrained,
  $\dot{M}_{\rm spec2}$ is consistent with zero.  \label{kTCF_vs_kT0}}
\end{figure}

Investigating further, we plotted $kT_{\rm low}$ versus $kT_{0}$
(Fig.~\ref{kTCF_vs_kT0}) for the CC clusters.  The blue points are SCC
clusters and the black points are WCC clusters.  If the gas cools from
$T_{\rm high}$ to $T_{\rm low}$ at a rate consistent with a classical
cooling flow (and is somehow stopped at $T_{\rm low}$), then we would
expect $T_{\rm low}\;\lesssim\;T_{0}$.  Fig.~\ref{kTCF_vs_kT0} shows
this to be the case for SCC clusters, however no correlation is found
for WCC clusters.  As noted earlier, WCC clusters show little or no
temperature drop at the cluster center.  However, when fitted with a
modified spectral mass deposition model, there appear to be
significant quantities of gas down to $\sim$1/3 $T_{\rm vir}$.  The
simplest explanation is that $\dot{M}_{\rm spec2}$ is not the correct
model and at the energy resolution of the {\it Chandra} {\it ACIS}, it
is difficult to distinguish between multitemperature components along
the line-of-sight and a cooling flow model down to $\sim$1/3 $T_{\rm
  vir}$.  With the spectral resolution of the {\it Chandra} {\it ACIS}
we cannot draw any strong conclusions, however the results for the SCC
clusters argue for further investigation with higher spectral
resolution instruments to see if this trend of $T_{\rm low}$
$\lesssim$ $T_{0}$ persists.

% The clusters for which kT_0 < kT_low are
% 1 - A85
% 10 - NGC1399
% 12 - IIIZw54 (WCC)
% 18 - A496
% 30 - A3526
% 52 - A2199
% 64 - A4059 

\begin{figure}
\includegraphics[width=85.0mm,angle=0]{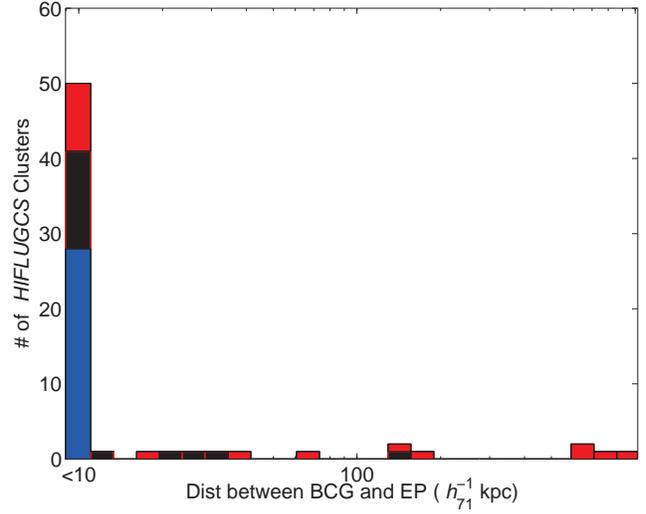}
\caption[BCG - X-ray peak Separation] {Histogram of projected
  separation of the BCG and X-ray emission peak.  The colors
  represent: blue - SCC clusters, black - WCC clusters and red - NCC
  clusters.  Most of the clusters ($\sim$78\%) have a BCG within
  12~$h_{71}^{-1}$~kpc of the X-ray peak.  100\% of the SCC clusters,
  61\% of the WCC clusters and 50\% of the NCC clusters have a BCG
  within 12~$h_{71}^{-1}$~kpc of the X-ray peak.  The uncertainty in
  the X-ray peaks is generally 4$\arcsec$, which corresponds to
  $\sim$4~$h_{71}^{-1}$~kpc for our median redshift.\label{cD_Dist}}
\end{figure}

In order to check for the correlation of the central cluster gas with
the central galaxy, we compared the position of the brightest cluster
galaxy (BCG) to the position of the emission peak. We identified the
BCGs from visual inspection of the R-band images of the Second-Epoch
Digitized Sky survey (DSS2, see e.g. {\tt
  http://archive.stsci.edu/dss}) and then extracted magnitudes and
redshifts from NED, HyperLeda\footnote{\tt http://leda.univ-lyon1.fr/}
\citep{2003A&A...412...45P} or the compilation by Andernach \& Tago
\citep[see][ for a description]{2005ASPC..329..283A}.  We measured the
position of the BCG by fitting a bidimensional Gaussian on its image,
using the NRAO program
FITSview\footnote{http://www.nrao.edu/software/fitsview; The (USA)
  National Radio Astronomy Observatory (NRAO) is operated by
  Associated Universities, Inc. and is a Facility of the (USA)
  National Science Foundation.}.  The typical uncertainty was
$\sim$0$\farcs$5.
% In the cases in which there was more than one BCG candidate
% (depending on color), we used the BCG closest to the X-ray peak.
Visual inspection of these images allowed the identification of
usually one, sometimes two or three BCG candidates.  In the cases in
which there was more than one BCG candidate, we used the candidate
closest to the X-ray peak.  Fig.~\ref{cD_Dist} shows the histogram of
distance between the X-ray peak and BCG.  The histogram is color
coded: blue for SCC clusters, black for WCC clusters and red for NCC
clusters.  The fact that all SCC clusters have a BCG at the center
seems to support the idea that the central temperature drop is related
to the presence of the galaxy.  On the other hand, 78\% of all {\it
  HIFLUGCS} clusters have a BCG within 12~$h_{71}^{-1}$~kpc of their
X-ray peak (including 61\% and 50\% of WCC and NCC clusters
respectively) and $\sim$88\% have a BCG within 50~$h_{71}^{-1}$~kpc.
There is a clear discrepancy between the number of WCC clusters (92\%)
and NCC clusters (61\%) that have a BCG at the X-ray peak ($<$
50~$h_{71}^{-1}$~kpc) and those that have cool gas at their center:
33\% and 11\% respectively.  Also as noted earlier the temperature
drop in these clusters is generally smaller than it is for SCC
clusters.

Here, we do not present this as evidence that temperature drops are or
are not related to the potential of the BCG. On the one hand, all but
one cluster (A2256) with a central temperature drop have a BCG
cospatial with the X-ray peak ($<12~h_{71}^{-1}$~kpc), suggesting that
the temperature drop is related to the BCG. On the other hand, there
are many clusters with a BCG at the center and no central temperature
drop. It appears that regardless of the mechanism that causes the
temperature of the gas in the central region to be below $T_{\rm
  vir}$, it is easier to heat the gas (to $\sim T_{\rm vir}$) in the
central region than to permanently separate it from the dark matter
potential well (i.e. the BCG). Simulations suggest that cluster
mergers can destroy cooling flows \citep[e.g.][ and references
therein]{burns07}. It is not clear though that such heating need be
done by something as energetic as a major merger. For instance, A1650
has an almost flat temperature profile and yet no evidence for a
recent major merger \citep{donahue05}.  Other examples of such
clusters are WCC clusters A2244 \citep{donahue05}, A1651 and A1060.

% It appears that regardless of the mechanism that causes the
% temperature of the gas in the central region to be below $T_{\rm
%   vir}$, it is easier to heat the gas (to $\sim T_{\rm vir}$) in the
% central region than to separate it from the BCG. This is interesting
% since simulations differ as to whether or not mergers can destroy
% cooling flows \citep[e.g.][and references therein]{burns07}. Here,
% we find that it may be easier to destroy a strong cool core than it
% is to separate the densest gas from the center of the dark matter
% potential well (i.e. the BCG). It is not clear though that such
% heating need be done by something as energetic as a major
% merger. For instance, A1650 has an almost flat temperature profile
% and yet no evidence for a recent major merger \citep{donahue05}.
% Other examples of such clusters are WCC clusters A2244
% \citep{donahue05}, A1651 and A1060.

\begin{figure}
\includegraphics[width=85.0mm,angle=0]{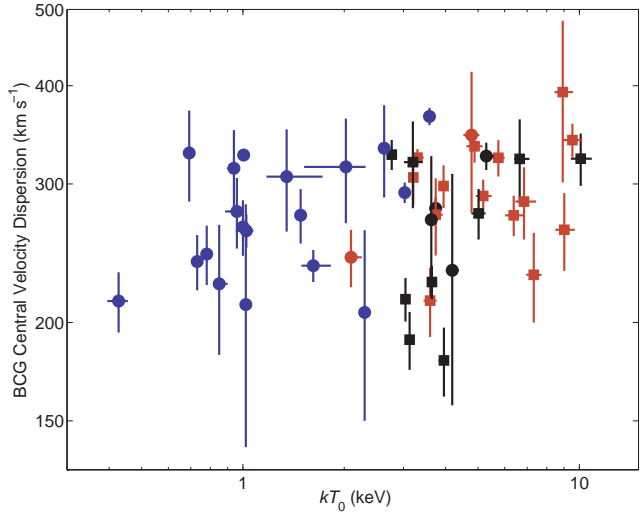}
\caption[Central BCG velocity dispersion versus $kT_{0}$] {BCG's
  central velocity dispersion versus $kT_{0}$.  Blue points are SCC
  clusters, black points are WCC clusters and red points are NCC
  clusters.  Circles signify clusters with a central temperature drop
  and squares signify clusters without a central temperature drop.
  The 16 clusters with no BCG velocity dispersion data available span
  the same range.  There appears to be no correlation between the
  BCG's central velocity dispersion and $kT_{0}$.  \label{vdisp-kT0}}
\end{figure}

To explore this further we compared the central stellar velocity
dispersion of the BCG to the central temperature.  The BCG central
stellar velocity dispersions were collected from HyperLeda, using mean
reported values and standard deviation of the values for the
uncertainty.  In a few cases, if only one measurement was reported, we
used the reported error on the measurement.  16 of our 64 {\it
  HIFLUGCS} clusters (including 8 SCC clusters) did not have any data
available. Fig.~\ref{vdisp-kT0} shows BCG central stellar velocity
dispersion versus $kT_{0}$.  The points are color coded: SCC-blue,
WCC-black and NCC-red.  Clusters with a central temperature decline
are marked with a circle and clusters without a central temperature
drop are marked with a square.  The 16 clusters with no data available
are omitted, but have the same general distribution in $kT_{0}$ as the
48 plotted clusters.  Even when looking at just the SCC clusters,
there seems to be only a weak correlation between the BCG's central
velocity dispersion and $kT_{0}$.

Assuming that our above conjecture is correct, that the gas is
relatively easy to heat without removing it from the center of the
potential, then it appears that the temperature of the central gas is
not generally determined by purely gravitational processes in the
local potential. That is, some heating mechanism heats the gas either
slightly above the local virial temperature in SCC clusters or up to
$\sim T_{\rm vir}$ in WCC and NCC clusters.  We emphasize that the
mechanisms do not have to be the same in all cases (e.g. possible AGN
outbursts or sloshing for SCC and WCC borderline cases and mergers for
NCC clusters).  This would explain the large scatter in
Fig.~\ref{vdisp-kT0}.  It would also predict that the central gas in
clusters such as A1650 will eventually return to the local virial
temperature in the center.  In such a case, in order to maintain
pressure equilibrium with the surrounding gas, the density would
increase dramatically, shortening the CCT and returning the cluster to
the SCC subsample.  It is not clear in this model which clusters would
be capable of returning to their SCC status and which ones not.
% It is not clear in this model what clusters would return to their
% SCC status and which would be incapable.
Perhaps the early major mergers of NCC clusters, as proposed by
\citet{burns07}, prevent the formation of the large central galaxy
needed to reform the CC.  Perhaps once the entropy is increased
enough, the gas cannot cool to the local virial temperature in the
time available.  Even when suppressed by magnetic fields, thermal
conduction is still likely to increase rapidly with temperature. Given
that its effect is already argued to be significant
\citep[e.g.][]{zakamska03}, it may well play a role in preventing cool
cores from being re-established after disruption by mergers in hot
clusters. We suggest that simulations of the formation of CCs
considering different central potentials could help understand which
ones can re-form and which cannot and what physical effects are
important in determining the fate of the central gas.

\subsection{BCG Separation and Mergers}
\label{bcgsepmerger}

Several works show a special relationship between the cooling activity
in cluster cores and the brightest cluster galaxies located within a
certain projected distance from the X-ray peak, typically
50~$h_{71}^{-1}$~kpc, to the X-ray peak \citep[see][ for more
details]{Mittal2009}.  Given that only eight of our 64 clusters have a
significant ($>$ 50~$h_{71}^{-1}$~kpc) projected separation between
the X-ray peak and BCG suggests that separating the BCG from the gas
is more difficult than simply heating the gas and/or disrupting the
cooling flow.  Looking more deeply, the eight clusters with BCG-X-ray
peak separations $>$50~$h_{71}^{-1}$~kpc ordered from largest to
smallest are: (1)~A3376 - $\sim$939~$h_{71}^{-1}$~kpc, (2)~A0754 -
$\sim$714~$h_{71}^{-1}$~kpc, (3)~A1367 - $\sim$666~$h_{71}^{-1}$~kpc,
(4)~A1736 - $\sim$642~$h_{71}^{-1}$~kpc, (5)~A3667 -
$\sim$155~$h_{71}^{-1}$~kpc, (6)~A2163 - $\sim$128~$h_{71}^{-1}$~kpc,
(7)~A2256 - $\sim$110~$h_{71}^{-1}$~kpc and (8)~A2255 -
$\sim$72~$h_{71}^{-1}$~kpc\footnote{Additionally the peculiar velocity
  of A2255's BCG is much larger than that of any other cluster
  (greater than A2255's velocity dispersion).}. All eight of these
clusters have been identified as merging clusters, A3667 is the only
CC cluster (and it is a borderline CC/NCC cluster) and all except
A1736 %(88\%)
have been identified as having a radio halo and/or relic(s) (see
Appendix-\ref{noic} for individual references).  The appearance of
such diffuse, Mpc scale non-thermal emission is thought to be powered
by major mergers.  Of the other 56 clusters, only eight (15\%) have
been identified as {\it possibly} containing diffuse non-thermal
emission on large scales: A1656, A3562, A85, A133, A401, A2152, A4038
and MKW8.  Of these eight, for two (A401 and A2142) the detections
seem unlikely \citep{giovannini00}, three (A85, A133 and A4038) are
small scale (a few tens of~kpc) relics associated with nearby radio
galaxies \citep[e.g][]{Slee2001} and one (MKW8) as identified as a
possible relic as seen in the VLA~Low-Frequency~Sky~Survey~(VLSS) at
74~MHz \citep{Cohen2007}. \citet{2004rcfg.proc..335K} identified three
classes of radio relics that were fundamentally different.  We argue
that in these four cases, the relics are not the large scale {\it
  Gischt} associated with mergers, but {\it AGN-relics}.  This leaves
only two (4\%) unambiguous detections of large scale diffuse radio
emission: the Coma cluster and A3562.  Based on the large discrepancy
between the number of clusters with large scale ($\sim$Mpc) radio
structure that have a large BCG-EP separation and those that do not,
we argue that a large separation of the EP and BCG is a very good
indication a major merger and therefore could be applied as a useful
method for discovering radio halos and radio relics-{\it Gischt}.

\subsection{WCC Clusters}
\label{wccclusters}
The WCC clusters are an interesting set because they seem to occupy a
transition between NCC and SCC clusters.  They are defined as having
short to moderate CCTs (1.0--7.7~$h_{71}^{-1/2}$~Gyr), and generally
have flat or shallow central temperature drops and a central entropy
that is enhanced compared to their SCC cousins.  Due to their flat or
shallow temperature profile, they are mostly classified as NCC
clusters in studies that determine CC clusters by a central
temperature drop.  Recently \citet{burns07} suggested that CC clusters
were not necessarily more relaxed than NCC clusters.  Since
\citet{burns07} define CC/NCC classification based on temperature
drop, they would classify the WCC clusters as NCC clusters.  Could
these WCC clusters be the relaxed NCCs?  \citet{donahue05} studied two
radio quiet, relaxed WCC clusters - A1650 and A2244.  They
conclude %on the other hand,
that these objects are CC clusters that have had a major AGN outburst,
raising the central entropy and temporarily disrupting the CF.

In an effort to distinguish between these two possibilities we
examined the X-ray morphology of the WCC and NCC clusters. We divided
these 36 clusters into four types: (I)~relaxed with short CCT ($<$
3~$h_{71}^{-1/2}$~Gyr), (II)~relaxed with moderate/long CCT ($>$
3~$h_{71}^{-1/2}$~Gyr), (III)~disturbed with short CCT ($<$
3~$h_{71}^{-1/2}$~Gyr), (IV)~disturbed with moderate/long CCT ($>$
3~$h_{71}^{-1/2}$~Gyr).  We classify relaxed versus disturbed by
visual inspection.  We define a relaxed cluster as having: (1)~round
or elliptical isophotes, (2)~the EP at the center of the isophotes and
(3)~little or no offset between the centers of the different levels of
isophotes (little or no sloshing).  We note, based on our
classification scheme, that all our NCC clusters will be of type II or
IV.  Our physical interpretation of the four types are: (I)~CCs that
have been disrupted, but will re-form - their CCTs are too short to be
relaxed NCCs, (II)~clusters that have not had a recent merger but have
had their CC severely disrupted (or destroyed) by a previous major
merger (III)~clusters that are merging with the core surviving
(although it may be destroyed in the future) and (IV)~merging clusters
in which there was no cool core or the core has been destroyed.

\begin{figure}
\includegraphics[width=85.0mm,angle=0]{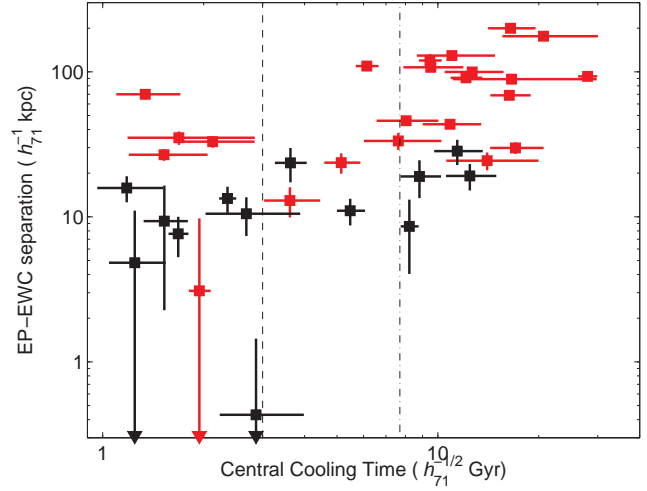}
\caption[Distance between the X-ray emission peak and emission
weighted center] {Distance between the X-ray emission peak and
  emission weighted center for the 36 WCC and NCC clusters. The black
  points are clusters that appear relaxed and the red points are
  clusters that appear disturbed.  The vertical dashed line separates
  clusters with short CCT and long CCT.  The dash-dot line divides the
  clusters between WCC and NCC.  It is clear that in general the
  separation of the EP and EWC is larger for the apparently disturbed
  clusters.  The error on the separation is 4$\arcsec$, which is the
  typical size of the smoothing kernel used when determining the
  EP.\label{EP-EWC_dist}}
\end{figure}

In order to check our categorization based on visual inspection, we
plotted the distance between the EP and the X-ray emission weighted
center (EWC) \citep[see][for details on determining the
EWC]{hudson06}.  This method gives a simple way to classify clusters,
since disturbed clusters will generally have a larger distance between
their EP and EWC than a relaxed clusters. Fig.~\ref{EP-EWC_dist} shows
the results for the 36 WCC and NCC clusters. Black points indicate the
clusters identified as being relaxed and the red points indicate the
disturbed clusters.  Fig.~\ref{EP-EWC_dist} confirms that, in general,
the disturbed clusters have a larger separation between the EP and EWC
than the relaxed clusters.

We find seven clusters of type I, six clusters of type II, five
clusters of type III~and 18 clusters of type IV.  By the definition of
\citet{burns07}, these 36 clusters are clusters that will never form a
CC. We argue, however, given the short cooling time of the clusters of
type I, it is unlikely that their CCT will never drop below
1~$h_{71}^{-1/2}$~Gyr, which in turn implies they will also have a
central temperature drop. 39\% of the WCC clusters are of this type
and are perhaps a special type of SCC cluster that experienced an
anomalous event that has temporarily disrupted the strong cool core
(e.g. raised the central temperature, entropy and cooing time).

We acknowledge that a more sophisticated method should be employed
when determining whether a cluster is relaxed or not.  However based
on preliminary results, unless WCC clusters (especially those with
short CCT) are fundamentally different from SCC clusters, some of them
will become SCC clusters.  That is, unless there is a process that
keeps their CCT $>$ 1~$h_{71}^{-1/2}$~Gyr and their
temperature-profile flat, the gas in the core will cool below $T_{\rm
  vir}$ and the CCT will drop below unity.

\section{Conclusions}
\label{concl}
We provide the most detailed systematic view into X-ray cores of
galaxy clusters to date. We find that the best method to determine
whether a cluster is a cool-core (CC) cluster is with the central
cooling time (CCT). We divide clusters into three types: strong
cool-core (SCC), weak cool-core (WCC or transition) and non-cool-core
clusters.  SCC are defined as having very short CCT ($<$
1~$h_{71}^{-1/2}$~Gyr) and are characterized by low central entropy
($\lesssim$30~$h_{71}^{-1/3}$ keV cm$^{2}$), systematic central
temperature drops (with $T_{0}$/$T_{\rm vir}$$\sim$0.4) and a
brightest cluster galaxy (BCG) at the X-ray peak.  WCC or transition
clusters are defined as having moderate CCT ( CCT between 1 -
7.7~$h_{71}^{-1/2}$~Gyr) and are characterized as having an elevated
entropy ($\gtrsim$30~$h_{71}^{-1/3}$ keV cm$^{2}$), flat or slightly
decreasing central temperature profiles and having a BCG at or near
($<$50~$h_{71}^{-1}$~kpc) the X-ray peak.  NCC clusters are defined as
having long CCT ($>$7.7~$h_{71}^{-1/2}$~Gyr), and are characterized by
large central entropies ($K_{0}$ $>$ 110~$h_{71}^{-1/3}$ keV cm$^{2}$)
and temperature profiles that generally are flat or rise towards the
center.

Based on the above classification, our main conclusions are:

\begin{enumerate}
  
\item In our flux-limited statistically complete sample, we find 72\%
  of the clusters are CC clusters with 44\% of the clusters being SCC
  clusters and 28\% being WCC clusters.
  
\item For intermediate redshift clusters (where radii as small as
  0.4\% $r_{500}$ cannot be resolved), we find that $K_{\rm BIAS}$ and
  scaled $L_{X}$ are the best proxies for CCT.  In general $K_{\rm
    BIAS}$ is better, but needs to be standardized before it is used.
  We suggest using \mbox{$I(R) \equiv 2\pi\int_{-\infty}^{\infty}
    \int_{0}^{R} n_{\rm e} n_{\rm H}\;r \mathrm{d}r \mathrm{d}l$} with
  a possible value of $I$ = 8.6 $\times$ 10$^{65}$~$h_{71}^{-2}$
  cm$^{-3}$.

\item For high redshift clusters (clusters with very few counts), we
  find that cuspiness is the most useful proxy.  However, it suffers
  from a large scatter in the relation to CCT, so precise
  categorizations can be difficult.  Preliminary results suggest that
  a concentration parameter ($\Sigma(<40$~kpc)/$\Sigma(<400$kpc)), as
  suggested by \citet{santos08}, may serve as the best proxy for
  distant clusters.

\item Dividing our representative sample into four redshift bins, we
  find it unlikely that the discrepancy between spectral mass
  deposition rate ($\dot{M}_{\rm spec}$) and classical mass deposition
  rate ($\dot{M}_{\rm classical}$) is due to the fact that CCs formed
  very recently. The probability that the lowest- and the
  highest-redshift clusters with non-zero forward-evolved mass
  deposition rates come from the same population is extremely low
  ($<3\%$). This requires the discrepancy to be explained by some
  heating method.

\item There is no evidence found in our work for a universal central
  temperature profile as claimed previously for smaller samples. This
  suggests that the radius of the cool gas is not universal. As seen
  in Table~\ref{obs}, the radius within which the temperature falls
  below $T_{\rm vir}$ differs from cluster to cluster.

\item We find the majority ($\sim$78\%) of {\it HIFLUGCS} clusters,
  including 100\% of the SCC clusters, 61\% of the WCC clusters and
  50\% of the NCC clusters, have a BCG within 12~$h_{71}^{-1}$~kpc of
  the X-ray emission peak. This number increases to 88\% for a
  distance of 50~$h_{71}^{-1}$~kpc. We find that seven out of eight
  clusters with the BCG-EP separation $>$ 50~$h_{71}^{-1}$~kpc have
  Mpc scale radio emission (halo or relic-{\it Gischt}) versus only
  two of the remaining 56 with a BCG very close to the EP.

\item There is a weak correlation between the SCC central temperature
  and the cooling of the gas as predicted by the cooling flow model.
  We also find no correlation between the SCC central temperature and
  the central velocity dispersion of the BCG.  We interpret this,
  along with the result that so many clusters have a BCG at their
  peak, as indicating that the central temperature of the gas is
  influenced by heating, which occurs differently in different
  clusters.  Therefore, the central temperature is not simply
  reflective of the central potential well or the expected cooling
  rate from $T_{\rm high}$ to $T_{\rm low}$, but requires more complex
  physics.

\item We find that $<$39\% of the WCC clusters which are relaxed also
  have relatively short cooling times ($t_{\rm cool}$ $<$
  3~$h_{71}^{-1/2}$).  We argue that these clusters are similar to SCC
  clusters but have had an event that temporarily disrupted the cool
  core, raising the core temperature, entropy and cooling time.

\end{enumerate}

Results obtained here on cluster cool-core properties and their
correlations, using a complete well-controlled sample with simple
selection criteria, can be taken as benchmark for next generation
cosmological hydrodynamical simulations aiming at describing detailed
properties of cluster cores.

\begin{acknowledgements}
  The authors wish to thank D.~A. Buote, T.~E. Clarke, M.~Markevitch
  and A.~Vikhlinin for providing proprietary data before it was
  publicly available. We would like to thank E.~Murphy for providing
  pointed radio measurements of $N_{H}$ for several clusters. We thank
  E.~Blanton, H.~B\"ohringer, Y.~Ikebe, E.~Pierpaoli, S.~Randall,
  P.~Schuecker, and G.~Sivakoff for help in the early stages of this
  work. We would like to thank R.~Smith for providing the data for the
  {\it APEC} model.  We thank the referee for a beneficial feedback
  and James Wicker for useful correspondence . T.~H.~R. and
  D.~S.~H. acknowledge support from the Deutsche
  Forschungsgemeinschaft through Emmy Noether research grant
  RE~1462/2. R.~M. acknowledges support from the Deutsche
  Forschungsgemeinschaft through the Schwerpunkt Program 1177
  (RE~1462/4). H.~A. acknowledges financial support from CONACyT under
  grants 50921-F and 81356, and partial support from the Transregional
  Collaborative Research Center TRR33 ``The Dark
  Universe''. P.~E.~J.~N. acknowledges support from NASA grant
  NAS8-03060. C.~L.~S. was supported in part by NASA Chandra grants
  GO7-8129X and AR7-8012X and NASA XMM-Newton grant NNX06AE76G.  This
  research has made use of the NASA/IPAC Extragalactic Database (NED)
  which is operated by the Jet Propulsion Laboratory, California
  Institute of Technology, under contract with the National
  Aeronautics and Space Administration. We acknowledge the usage of
  the HyperLeda database (http://leda.univ-lyon1.fr).

\end{acknowledgements}
\bibliographystyle{aa} % style aa.bst
\bibliography{BIB} % your references Yourfile.bib

%\bibliography{/raid7/dhudson/BIB/BIB} % your references Yourfile.bib

\onecolumn
\begin{appendix}
\section{Calculating Central Density for a Double $\beta$ model}
\label{DBCalc}

Starting from the definition of the {\it normalization} of the {\it
  APEC} model \citep{mewe85,mewe86,smith00,smith01b}
%\begin{equation}
%{\rm NORM} \equiv \frac{10^{-14}}{4\pi\:D_{A}\:D_{L}}\int n_{e} n_{H} dV,  
%\label{normalization}
%\end{equation}
and taking $n(r)$ = $n_{e}$ with $\zeta$ = $\frac{n_{e}}{n_{H}}$,
(calculated individually, but generally $\sim$1.2),
\begin{equation}
  \label{norm}
  {\mathcal N} \equiv \frac{10^{-14}}{4\pi\:D_{A}\:D_{L}\:\zeta}\int n(r)^2 \mathrm{d}V,  
\label{normalization2}
\end{equation}
where the terms are defined as in Eqn.~\ref{n0-beta_model}.  For a
double $\beta$ model the expression for $n(r)$ is:
\begin{equation}
  n(r) = \left[ n_{01}^{2} \left( 1 + \left( \frac{r}{r_{c_{1}}} \right)^{2} \right)^{-3\beta_{1}} +  n_{02}^{2} \left( 1 + \left( \frac{r}{r_{c_{2}}} \right)^{2} \right)^{-3\beta_{2}} \right]^{\frac{1}{2}}.
\label{dbeta-model}
\end{equation}
% so that $n_{0}$ = $\sqrt{n_{01}^2 + n_{02}^2}$.
The unabsorbed\footnote{In our case Eqn.~\ref{normalization2} already
  takes into account the absorption and any absorption in the
  subsequent calculations cancels.} surface brightness at a projected
distance, $x$ from the center over an energy range between $E_{1}$ and
$E_{2}$ is
\begin{equation}
\label{sbdefn}
\Sigma(x) = \frac{\int_{E_{1}}^{E_{2}} \Lambda_{\rm X}(T,Z,E)\mathrm{d}E}{4\pi (1+z)^{4}\:\zeta}\int_{-\infty}^{\infty} n(r)^2 \mathrm{d}l,
\label{SB1}
\end{equation}
where $r^2$ = $x^2$ + $l^2$ and $\Lambda_{\rm X}(T,Z,E)$ is the
emissivity function for a plasma of temperature $T$ and metalicity $Z$
at energy $E$.  This can be rewritten in terms of $n_{01}$ and
$n_{02}$ as:
\begin{equation}
  \Sigma(x) = \frac{\int_{E_{1}}^{E_{2}}\Lambda_{\rm X}(T,Z,E)\mathrm{d}E}{4\pi (1+z)^{4}\:\zeta}\int_{-\infty}^{\infty} n_{01}^{2} \left( 1 + \left( \frac{x^2+l^2}{x_{c_{1}}^{2}} \right) \right)^{-3\beta_{1}} \mathrm{d}l +  \frac{\int_{E_{1}}^{E_{2}}\Lambda_{\rm X}(T,Z,E)\mathrm{d}E}{4\pi (1+z)^{4}\:\zeta}\int_{-\infty}^{\infty} n_{02}^{2} \left( 1 + \left( \frac{x^2+l^2}{x_{c_{2}}^{2}} \right) \right)^{-3\beta_{2}} \mathrm{d}l.
\label{SB2}
\end{equation}
Solving the integral gives the standard expression for the double
$\beta$ model in terms of surface brightness:
\begin{equation}
  \Sigma(x) = \Sigma_{01} \left( 1 + \left( \frac{x}{x_{c_{1}}} \right)^{2} \right)^{-3\beta_{1} + \frac{1}{2}}+  \Sigma_{02} \left( 1 + \left( \frac{x}{x_{c_{2}}} \right)^{2} \right)^{-3\beta_{2} + \frac{1}{2}},
\label{SB3}
\end{equation}
where 
\begin{equation}
  \Sigma_{0i} \equiv \frac{n_{0i}^2 \int_{E_{1}}^{E_{2}} \Lambda_{\rm X}(T,Z,E)\mathrm{d}E}{4\pi (1+z)^{4}\:\zeta} \int_{-\infty}^{\infty}\left( 1 + \left( \frac{l}{x_{c_{i}}} \right)^{2} \right)^{-3\beta_{i}} \mathrm{d}l =  \frac{n_{0i}^2  x_{c_{i}} \pi^{\frac{1}{2}}}{4\pi (1+z)^{4}\:\zeta} \frac{\Gamma \left( 3\beta_{i} - \frac{1}{2} \right)}{\Gamma \left( 3 \beta_{i} \right)} \int_{E_{1}}^{E_{2}}\Lambda_{\rm X}(T,Z,E)\mathrm{d}E.
\label{SB4}
\end{equation}
Therefore,
\begin{equation}
  \frac{n_{01}^2}{n_{02}^2} =  \frac{\Sigma_{01} {\rm LI}_{2}}{\Sigma_{02} {\rm LI}_{1}} = \frac{\Sigma_{12}\:{\rm LI}_{2}}{{\rm LI}_{1}},
\label{n012}
\end{equation}
where LI$_{i}$ and $\Sigma_{12}$\footnote{Note, we have explicitly
  assumed $\Lambda_{\rm X}(T,Z,E)$ is the same for both components.
  In the case it is not, $\Sigma_{12}$ can be redefined as
  $\Lambda_{\rm X}(T_{2},Z_{2},E)\Sigma_{01}/\Lambda_{\rm
    X}(T_{1},Z_{1},E)\Sigma_{02}$ and the calculations follow
  identically.} are as defined in Eqn.~\ref{n0-dbeta_model}.  Using
this relation along with the fact that $n_{0}$ $\equiv$ $n(0)$ =
$\sqrt{n_{01}^2 + n_{02}^2}$, we find:
\begin{equation}
  n_{01}^2 =  \frac{\Sigma_{12}\:{\rm LI}_{2}}{\Sigma_{12}\:{\rm LI}_{2}+{\rm LI}_{1}} n_{0}^2,
\label{n01}
\end{equation}
and
\begin{equation}
  n_{02}^2 =  \frac{{\rm LI}_{1}}{\Sigma_{12}\:{\rm LI}_{2}+{\rm LI}_{1}} n_{0}^2.
\label{n02}
\end{equation}
Inserting these values into Eqn.~\ref{dbeta-model} to find an
expression for $n(r)$ in terms of $n_{0}$, we get
\begin{equation}
  n(r) =\frac{n_{0}}{\sqrt{\Sigma_{12}\:{\rm LI}_{2}+ {\rm LI}_{1}}} \left[ \Sigma_{12}\:{\rm LI}_{2} \left( 1 + \left( \frac{r}{r_{c_{1}}} \right)^{2} \right)^{-3\beta_{1}} + {\rm LI}_{1} \left( 1 + \left( \frac{r}{r_{c_{2}}} \right)^{2} \right)^{-3\beta_{2}} \right]^{\frac{1}{2}},  
\label{dbeta-n0}
\end{equation}
Inserting this expression of $n(r)$ into Eqn.~\ref{normalization2} and
solving for $n_{0}$, we recover Eqn.~\ref{n0-dbeta_model}.

%\end{appendix}
%\begin{appendix}

\section{$K_{\rm BIAS}$ Calculations}
\label{countscalc}
From the definition of surface brightness (Eqn.~\ref{sbdefn}), a
cluster at redshift $z$, of a region with an angular radius $x$, has
an integrated surface brightness (or Flux ${\mathcal F}$) between
energies $E_{1}$ and $E_{2}$:
\begin{equation}
  \label{intsb}
  {\mathcal F} = \frac{\int_{E_{1}}^{E_{2}} \Lambda_{\rm X}(T,Z,E)\mathrm{d}E}{2(1+z)^4} \int_{-\infty}^{\infty} \int_{0}^{x} n_{\rm e} n_{\rm H}\;x \mathrm{d}x \mathrm{d}l,
\end{equation}
where $n_{e}$ is the electron density, $n_{H}$ is the proton density,
$\Lambda_{X}(T,Z,E)$ is the emissivity function as defined in
Eqn.~\ref{sbdefn}.  To remove the redshift dependence of the projected
region size, we convert the projected region of angular radius $x$ to
a cylindrical region of physical radius $R$, such that $R$ $\approx$
$xD_{A}(z)$.  Eqn.~\ref{intsb} becomes:
\begin{equation}
  \label{intsb2}
  {\mathcal F} = \frac{\int_{E_{1}}^{E_{2}} \Lambda_{\rm X}(T,Z,E)\mathrm{d}E}{4\pi\:D_{A}\:D_{L}\:(1+z)^2} I(R),
\end{equation}
where $D_{A}$ is angular diameter distance, $D_{L}$ is the luminosity
distance and $I(R)$ is defined as in Eqn.~\ref{defnI}.  Therefore the
total counts ${\mathcal C}$ collected by a telescope for an
observation of length $t_{\rm obs}$, in an energy band from $E_{1}$ to
$E_{2}$, of a cylindrical region of physical radius $R$ is:
\begin{equation}
  \label{cnts}
  {\mathcal C} = t_{\rm obs} \frac{\int_{E_{1}}^{E_{2}} \alpha(E)\Lambda_{\rm X}(T,Z,E) A_{\rm eff}(E) \mathrm{d}E}{4\pi D_{A} D_{L}\;(1+z)^2} I(R),
\end{equation}
where $\alpha(E)$ and $A_{\rm eff}(E)$ are the absorption from
Galactic hydrogen and the effective area of the telescope at energy
$E$, respectively.  We can calculate
$\int_{E_{1}}^{E_{2}}\alpha(E)\Lambda_{\rm X}(T,Z,E)A_{\rm
  eff}(E)\mathrm{d}E$ for an absorbed thermal model using {\it XSPEC}
with an appropriate ARF and RMF. Specifically, since normalization
${\mathcal N}$ $\propto$ ${\mathcal CR}$ $\equiv$ $C/t_{\rm obs}$,
{\it XSPEC} can be used to find the constant of proportionality
$\kappa$.  From the definition of ${\mathcal N}$ (see
Eqn.~\ref{norm}):
\begin{equation}
  \int_{E_{1}}^{E_{2}}\alpha(E)\Lambda_{\rm X}(T,Z,E)A_{\rm eff}(E)\mathrm{d}E= \frac{(1+z)^2\kappa}{10^{14}},
\end{equation}
so that
\begin{equation}
\label{cnts2}
{\mathcal C} = t_{\rm obs} \frac{10^{-14}\;\kappa}{4\pi D_{A} D_{L}} I(R).
\end{equation}
Using an on-axis {\it Chandra} ARF and RMF, we determined $\kappa$
($122.3$ photons cm$^{5}$ s$^{-1}$) for an energy band from 0.5 - 7.0
keV, with $Z$ = 0.25 solar, $N_{H}$ = 2 $\times$ 10$^{20}$ cm$^{-2}$
and our median observation time (44 ks), redshift (0.047) and virial
temperature (4.3 keV).  Inserting our determined value of $\kappa$
into Eqn.~\ref{cnts2} and solving for $I$, such that $C$ = 10\,000
counts, yields $I$ = 8.6 $\times$ 10$^{65}$~$h_{71}^{-2}$ cm$^{-3}$.
Therefore using the criterion that our median observation would have
$K_{\rm BIAS}$ determined by circle with 10\,000 counts, equivalent
regions from other observations would have:
\begin{equation}
  \frac{{\mathcal C}}{{\rm cnts}} = 17\,910 \left ( \frac{ t_{\rm obs}}{100\; {\rm ks}} \right )\left( \frac{\kappa}{100\;{\rm cnts}\;{\rm cm}^5\;{\rm s}^{-1}} \right) 
  \left ( \frac{200 \;h_{71}^{-1}\;{\rm Mpc}}{D_{A}(1+z)} \right )^2.
\end{equation}

%\end{appendix}
%\begin{appendix}

\section{Notes On Individual Clusters}
\label{noic}

\subsection{A0085}  
This cluster appears to have two subclumps, one near the center and
one further to the south \citep{kempner02}.  In determining the
temperature profile and global cluster temperature the latter was
excluded.  This SCC cluster hosts a well-studied radio relic, which is
close to but not connected to the central radio galaxy
\citep[e.g][]{Slee01}.  The central region of this cluster requires a
double thermal model out to $\sim$11$\arcsec$
($\sim$12~$h_{71}^{-1}$~kpc).

\subsection{A0119}
This is possibly a merging cluster, which shows elongation towards the
northeast.  The X-ray peak of this NCC cluster, which does not
dominate the surface brightness, has a cD galaxy cospatial with it.
The cluster contains three wide-angle-tailed~(WAT) radio galaxies
which may be interacting with the ICM
\citep[e.g.][]{1999A&A...344..472F}.

\subsection{A0133}
Central regions of this cluster show an east-west elongation.  An
in-depth study with {\it XMM} and {\it Chandra} by
\cite{fujita04,fujita02} revealed an X-ray {\it tongue} extending
northwest. This SCC cluster hosts a radio relic, that is close to but
not connected to the central radio galaxy \citep[e.g][]{Slee01}. The
central region of this cluster requires a double thermal model out to
$\sim$15$\arcsec$ ($\sim$16~$h_{71}^{-1}$~kpc).

\subsection{NGC0507 Group}
The overall X-ray spectrum of this group shows a suspicious hard tail.
An additional powerlaw component was included in the overall
temperature fit. It is possible that the hard tail is due to
unresolved low mass X-ray binaries (LMXBs), although the central
region ($\sim$1$\farcm$8 = 35.7 h$_{71}^{-1}$~kpc) was removed and no
evidence of a hard excess is seen in the spectra of the central
annuli. It is possible that (given the redshift of NGC0507 z = 0.0165)
the LMXBs are only strong enough to be measured in a large region and
are insignificant compared to the group emission in the central
region.  On the other hand the powerlaw has a steep photon index
($\Gamma_{X}$ = 2.3 - 3.0) that is not consistent with LMXBs, which
usually have a photon index of $\Gamma_{X}$=1.6.  The component has a
total flux of $\sim$4 $\times$ 10$^{-12}$ ergs cm$^{-2}$ s$^{-1}$
corresponding to a luminosity of 2 $\times$ 10$^{42}$~$h_{71}^{-2}$
ergs s$^{-1}$ (over 0.4 - 10.0 keV). It is also possible that the hard
excess is related to an insufficiently subtracted particle background,
which is visible in this cool cluster. Both models (additional
particle background or powerlaw) give an identical overall
temperature. We also note that a more detailed analysis of the
residual background in outer cluster regions shows no residual
particle background. The central region of this cluster requires a
double thermal model out to $\sim$62$\arcsec$
($\sim$20.6~$h_{71}^{-1}$~kpc).

\subsection{A0262}
The spectral fits to the inner regions are poor ( $\chi^2$/dof $\sim$
1.4) even with a double thermal model.  Using non-solar abundance
ratios significantly improves the fit, but does not change the
best-fit overall temperature.  We therefore used solar ratios for
simplicity. The central region of this cluster requires a double
thermal model out to $\sim$43$\arcsec$ ($\sim$14~$h_{71}^{-1}$~kpc).

\subsection{A0400}
This cluster hosts the double radio source 3C75 within its center and
shows evidence of merging \citep{hudson06}.  As noted in
\citet{hudson06}, the hydrogen column density is higher than measured
with the radio \citep[$N_H$ = 0.85 $\times 10^{21}$
cm$^{-2}$][]{kalberla05} and therefore we left it free for all
spectral fits.  We find a hydrogen column density of $N_{H}$ =
0.98-1.22 $\times 10^{21}$ cm$^{-2}$ for our fit to the overall
cluster.

\subsection{A0399}
This cluster is near to A401 and shows evidence of interaction with
A401 \citep[e.g.][]{sakelliou04}.  The temperature profile of this
cluster peaks at the X-ray center.

\subsection{A0401}
See also A0399.  This cluster may host a radio halo
\citep{giovannini99}.  We included an early observation (before 2001),
since the later observation was offset, with the cluster center in the
corner of a CCD.  The BCG closest to the X-ray peak is
$\sim34$~$h_{71}^{-1}$~kpc away, making it one of fourteen clusters
with the BCG $>$12~$h_{71}^{-1}$~kpc from the X-ray peak.

\subsection{A3112}
Although the background flaring seen in some observations was removed,
the effect seems to have broadened a fluorescence line.  This can be
seen in the fit to the overall cluster spectrum (at $\sim$7.5 keV).
This effect seems simply to make the fit poor ($\chi^2$/dof
$\sim$1.6), but it does not affect the best-fit values whether the
line is removed or not.  \citet{takizawa03} first presented the {\it
  Chandra} data of A3112, interpreting the radio active central cD
galaxy as interacting with the ICM.  \citet{bonamente07} claim a soft
excess and hard excess in this cluster that may be related with the
central radio active BCG.  We do not see a similar effect, however we
do not separately fit the 1$\arcmin$-2$\farcm$5 annulus that
\citet{bonamente07} fit.  We do confirm that the 1$\arcmin$-2$\farcm$5
annulus is isothermal in our $kT$-profile so that their result is not
due to a temperature fluctuations in the cluster.  This SCC cluster is
one of sixteen clusters for which no data exist for the BCG central
velocity dispersion.

\subsection{NGC1399 Group/Fornax Cluster}
\label{FC}
This nearby SCC cluster has two X-ray peaks cospatial with {\it
  NGC1399} and {\it NGC1404}.  The X-ray peak is taken to be cospatial
with the BCG {\it NGC1399}.  The peak on {\it NGC1404} was removed for
spatial and spectral analysis.  Fornax appears to be an outlier in six
of the plots of parameters versus CCT in which it has an anomalously
low value for its CCT.  These parameters are: (1) $\Sigma_{0}$, (2)
$r_{c}$/$r_{500}$, (3) $K_{\rm BIAS}$, (4) cooling radius, (5)
$\dot{M}_{\rm classical}$/$M_{500}$ and (6) $M_{\rm gas}$/$M_{500}$.
Additionally it is the only SCC cluster in which $\dot{M}_{\rm
  classical}$ $\leq$ $\dot{M}_{\rm spec}$.  One possible explanation
is that {\it NGC1404} is about $0.04~r_{500}$ from the X-ray peak and
due to the extended emission around it, the surface brightness profile
severely flattens.  The extrapolated outer profile therefore
overestimates the projected gas lowering the central density (and
altering associated parameters).  To check how much this influenced
the Fornax cluster as an outlier, we fit only the central part of the
surface brightness that could be fit well to a single $\beta$-model.
This model most likely underestimates the projected gas, thereby
providing the largest possible values for central density.  In the
case of Fig.~\ref{specCCvsdensCC}, this model raises $\dot{M}_{\rm
  classical}$ to 0.75$\pm$0.04~$h_{71}^{-2}$ $M_{\odot}$ yr$^{-1}$
making it larger than $\dot{M}_{\rm spectral}$ = 0.52$\pm$0.02
$M_{\odot}$~yr$^{-1}$.  We emphasize that this result overestimates
$\dot{M}_{\rm classical}$ and in any case $\dot{M}_{\rm spec}$
$\approx$ $\dot{M}_{\rm classical}$, making it an odd SCC cluster. In
all other cases the Fornax cluster remains an outlier.  In the case of
the CCT the value falls to $\sim$0.4~$h_{71}^{-1/2}$~Gyr moving the
Fornax cluster to the left in Fig~\ref{paramcompare}.  (1)
$\Sigma_{0}$, not surprisingly, remains almost the same, $\sim$0.21
$\times$ 10$^{-6}$ photons cm$^{-2}$ s$^{-1}$ arcsec$^{-2}$.  The
shift to the left in this case makes the Fornax cluster even more of
an outlier. (2) $r_{c}/r_{500}$, likewise, remains almost identical,
$\sim$4 $\times$ 10$^{-4}$. Unlike, $\Sigma_{0}$, the shift to the
left in this case makes the Fornax cluster more similar to other SCC
clusters. (3) $K_{\rm BIAS}$ is, of course, unaffected by the density
model, however moving the Fornax cluster to the left makes it more
consistent with the other SCC clusters.  (4) The cooling radius
remains almost identical, increasing to $0.03~r_{500}$.  Additionally
moving the Fornax cluster to the left makes it even more of an
outlier. (5) $\dot{M}_{\rm classical}$/$M_{500}$ increases slightly to
$\sim$1.8 $\times$ 10$^{-14}$~$h_{71}^{-1}$~yr$^{-1}$.  Moving the
Fornax cluster to the left makes the trend of decreasing $\dot{M}_{\rm
  classical}$/$M_{500}$ with CCT worse, but it is more consistent with
other groups with short CCT.  (6) $M_{\rm gas}$/$M_{500}$, is raised
slightly, $\sim$0.2 $\times$ 10$^{-3}$.  Similar to $\dot{M}_{\rm
  classical}$/$M_{500}$, moving the Fornax cluster to the left makes
the trend with CCT worse, but makes it more consistent with the other
groups.  We emphasize that the Fornax cluster is still an outlier in
all six cases, and that these values are the other extrema, with the
true value somewhere between these and where they are plotted in
Fig~\ref{paramcompare}.  The physical interpretation is that it is
possible that Fornax is in the process of forming a cool core (i.e it
has a nascent core in the terms of \citet{burns07}) and therefore is
dynamically different from the other SCC clusters that have
well-established cores.  The central region of this cluster requires a
double thermal model out to $\sim$170$\arcsec$
($\sim$15.9~$h_{71}^{-1}$~kpc).

\subsection{2A0335+096}
This cluster, along with A0478 and NGC1550, has a significantly higher
hydrogen column density than measured at radio wavelengths
\citep[$N_H$ = 1.68 $\times 10^{21}$ cm$^{-2}$][]{kalberla05}.  We fit
all spectra with the column density free.  For the fit to the overall
spectrum we find $N_H$ = 2.35-2.47 $\times 10^{21}$ cm$^{-2}$.  This
cluster has two major galaxies near the X-ray peak, which resides
between the two of them ($\sim10$~$h_{71}^{-1}$~kpc from the closest).
Of the 16 clusters in which no information about the BCG central
velocity dispersion is available, this cluster has the shortest CCT.
The central region of this cluster requires a double thermal model out
to $\sim$38$\arcsec$ ($\sim$26~$h_{71}^{-1}$~kpc).

\subsection{IIIZw54} {\it IIIZw54} is a pair of galaxies near the
center of a poor galaxy group.  We used a 6$\arcsec$ kernel when
smoothing the image in order to determine the emission peak.  This
cluster does not have a bright core, although it appears to be quite
round and relaxed.  The brighter of two galaxies in the galaxy pair
{\it IIIZw54} (a cD galaxy) is cospatial with the X-ray peak.

\subsection{A3158}
\citet{lokas06} report A3158 as a relaxed cluster based on the
velocity dispersion of the galaxies. The X-ray emission appears to be
elliptical and there are two cDs near the cluster center, one of which
lies at the X-ray peak. This cluster definitely does not have a bright
core, with a central density of only $\sim$5 $\times$ 10$^{-3}$
cm$^{-3}$. The temperature profile peaks in the center at $\sim$5.7
keV.

\subsection{A0478}
This cluster, along with 2A0335+096 and NGC1550, has a significantly
higher hydrogen column density than measured in the radio \citep[$N_H$
= 1.64 $\times 10^{21}$ cm$^{-2}$][]{kalberla05}. We fit all spectra
with the column density free.  Our fit to the overall cluster yields a
column density of ($N_H$ = 2.89 - 2.96 $\times 10^{21}$ cm$^{-2}$),
consistent with the value found by \citet{sanderson05}.  This cluster
has the highest spectral mass deposition rate of any {\it HIFLUGCS}
cluster, making it an ideal candidate for a grating observation.
Unfortunately the {\it RGS} data from a long {\it XMM-Newton} exposure
was virtually unusable \citep{deplaa04}.  This SCC cluster is also one
of sixteen clusters in which no data for the BCG central velocity
dispersion are available.

\subsection{NGC1550 Group}
This cluster, along with 2A0335+096 and A0478, has a significantly
higher hydrogen column density than measured in the radio \citep[$N_H$
= 0.981 $\times 10^{21}$ cm$^{-2}$][]{kalberla05}.  We fit all spectra
with the column density free.  Our fit to the overall cluster yields a
column density of $N_H$ = 1.34 - 1.41 $\times 10^{21}$ cm$^{-2}$.  The
column density appears to peak towards the center of the cluster,
having a value of $N_H$ = 1.9 - 2.2 $\times 10^{21}$ cm$^{-2}$ in the
innermost annulus.

\subsection{EXO0422-086/RBS 0540}
The short observation of this SCC cluster indicates a round, centrally
peaked cluster with a moderate central temperature drop.  The BL~Lac
object EXO 0423.4-0840 at the center of this cluster was studied by
\citet{belsole05}.  This is one of sixteen clusters for which no data
about the BCG central velocity dispersion are available.

\subsection{A3266}
This cluster has a very low background scaling factor; therefore an
additional unfolded powerlaw component was included in the spectral
fits to account for any residual particle background. Reading in the
background as a corfile (i.e. a second background component with an
adjustable scaling factor), the overall best-fit temperature is found
to be consistent with our result including an unfolded powerlaw.
\citet{henriksen02,finoguenov06} presented detailed analyses of this
merging system.

\subsection{A0496}
The high abundances in the central region of this cluster are better
fit with a VAPEC model, however since this did not change the best-fit
values of temperature, solar ratios were used for simplicity.  A
double thermal model greatly improved the fits to spectra in annuli
out to 0$\farcm$3. However, the high temperature component for annuli
between 0$\farcm$18 and 0$\farcm$3 is $\gg kT_{\rm vir}$ ( $kT \sim 8$
keV).  Although this may be evidence of very hot gas near the cluster
core, the investigation is beyond the scope of this paper.  Therefore
for annuli between 0$\farcm$18 and 0$\farcm$3, we used a single
thermal fit.  \citet{2007ApJ...671..181D} studied the longest of the
three {\it Chandra} observations of this cluster in depth.  They argue
that there is a cold front at the center of this cluster, which is
caused by an off-center passage of a smaller dark matter halo.

\subsection{A3376}
This cluster was fit with an unfolded powerlaw component to account
for possible low-level flares in both observations.  This cluster
appears highly disturbed in the X-ray with a strong elongation along
the east-west direction.  \citet{bagchi06} report the existence of
double relics, one to the east of the cluster center and one to the
west.  \citet{nevalainen04} found a diffuse, hard excess with the {\it
  BeppoSAX PDS} at 2.7$\sigma$ significance.  The BCG of this cluster
is $\sim$1 Mpc from the X-ray peak, the most distant of any cluster in
the sample and one of eight clusters with a separation of
$>$50~$h_{71}^{-1}$~kpc.  There is a radio galaxy with bent jets very
close to the X-ray peak \citep{Mittal2009}. Optically it is clearly
much fainter than the BCG and is most likely an AGN that may have been
activated by the merger.  The jets are bent in the opposite direction
to the elongation of the cluster, possibly bent from ICM ram pressure.

\subsection{A3391}
The short observation of this NCC cluster shows an elliptical shaped
ICM with a BCG cospatial with the emission peak. \citet{tittley01}
discovered a filament between A3391 and the nearby cluster A3395.

\subsection{A3395s}
This cluster is very close to and may be interacting with A3395e.
A3395e was excluded from all extended analysis.  \citet{donnelly01}
claim A3395s and A3395e are near first core passage.

\subsection{A0576}
\citet{kempner04} originally presented an analysis of the {\it
  Chandra} data.  \citet{dupke07} presented a detailed analysis of the
{\it XMM-Newton} and {\it Chandra} data suggesting that it is a
line-of-sight merger. The X-ray image seems somewhat perturbed with
elliptical isophotes with alternating NW-SE shifted centers,
reminiscent of sloshing, already noted by \citet{kempner04}.  The BCG
is $\sim$24~$h_{71}^{-1}$~kpc from the X-ray peak, making it one of
fourteen clusters with the separation $>$12~$h_{71}^{-1}$~kpc.  There
is, however, a slightly fainter galaxy closer
($<$12~$h_{71}^{-1}$~kpc) to the X-ray peak that is radio active,
whereas the BCG is not.  The peculiar velocity of the BCG is one of
five clusters that is more than 50\% of the velocity dispersion.  This
WCC cluster is one of the three WCC/NCC clusters with
CCT$\gg$1~$h_{71}^{-1/2}$~Gyr (i.e. not on the border between SCC and
WCC) and a systematic temperature decrease at the cluster center.

\subsection{A0754}
This irregularly shaped cluster hosts a halo and relic
\citep{kassim01}.  \citet{henry04} presented a detailed analysis of
this merging system using the {\it XMM-Newton} observation.  Only the
pre-2001 {\it Chandra} observation is used, since it was the only one
that contained the cluster core. More recent observation have been
made but do not cover the cluster center and therefore are not useful
for core studies.  The BCG for this cluster is
$\sim$714~$h_{71}^{-1}$~kpc away from the X-ray peak, making it one of
eight clusters where this separation is $>$50~$h_{71}^{-1}$~kpc.

\subsection{A0780/Hydra-A Cluster}
This cluster is known to have a massive central AGN outburst \citep{nulsen05}.  

\subsection{A1060}
This WCC cluster is also known as the Hydra cluster.  \citet{sato07}
recently presented a {\it Suzaku} observation of this cluster.  This
cluster has two bright galaxies near the core, one of which is
cospatial with X-ray peak.  Both galaxies have a clearly visible
diffuse X-ray component \citep{yamasaki02}.

\subsection{A1367}
Due to the short exposure time and lack of a bright core, we used a
12$\arcsec$ kernel when smoothing to determine the X-ray peak.  This
is a very well-studied merging cluster. This cluster has an infalling
starburst group \citep{sun05a,cortese06} and optical evidence suggests
that this is a merging system \citep{cortese04}. The X-ray image
appears rather disturbed with several off-centered bright sources.
\citet{sun05b} studied the survival of galaxy coronae in this system.
This cluster hosts a radio relic \citep{gavazzi83}.  The BCG of this
cluster is $\sim$666~$h_{71}^{-1}$ kpc from the X-ray peak making it
one of eight clusters where this separation is
$>$50~$h_{71}^{-1}$~kpc.  It is also one of five clusters where the
BCG peculiar velocity is $>$ 50\% of the cluster velocity dispersion.

\subsection{MKW4}
A single thermal model is a poor fit to this high metalicity center.
Although a second thermal model does provide an improvement, freeing
the ratio of elements for a single thermal model provides the
best-fit.  Since freeing the abundance ratios does not change the
overall best-fit temperatures of the annuli, solar ratios with a
single thermal model were used for simplicity.

\subsection{ZwCl1215}
The short observation of this NCC cluster, shows a round cluster with
no bright central peak and an elliptical BCG located at the X-ray
emission peak.  This is one of the sixteen clusters for which no data
about the BCG central velocity dispersion are available.  The BCG of
this clusters is also $\sim$18~$h_{71}^{-1}$~kpc from the X-ray peak,
making it one of fourteen clusters where this separation is
$>$12~$h_{71}^{-1}$~kpc.

\subsection{NGC4636 Group}
This nearby group contains extended nonthermal emission in the central
region extending out $\sim$122$\arcsec$ ($\sim$9.19~$h_{71}^{-1}$
kpc). The luminosity of this emission ($L_{X}$
$\sim$10$^{40}$~$h_{71}^{-2}$ ergs s$^{-1}$) is consistent with the
expected unresolved LMXB population for NGC4636.
% Strange hard excess in $kT_{\rm total}$ for the I-Chips.  It looks
% more like residual flaring than LMXB's or something physical.
% Strange that it is identical in both observations.  $kT_{\rm total}$
% is also much higher than the temperature seen in the profile.  If
% the cosmic X-ray background (CXB) component is removed then the
% temperature drops to the profile level, but the fit is poor
% ($\chi^2$/dof $>$ 1.2 ) with clear residuals.  Is it possible there
% are two temperatures in the outer cluster or that the CXB is
% bringing down the temperature in the profile?
In addition to a powerlaw component, the central region of this
cluster requires a double thermal model out to $\sim$35$\arcsec$
($\sim$2.6~$h_{71}^{-1}$~kpc).

\subsection{A3526/Centaurus Cluster}
This is a well-studied, prototypical CC cluster, with a central
temperature drop (having the largest fractional drop, $kT_{0}$ $\sim$
0.2 $kT_{\rm vir}$) and enhanced central metalicity.  An arc-like
X-ray feature near the center has been identified as most likely being
a cold front associated with sloshing in the core
\citep{2002MNRAS.331..273S}.  The central region of this cluster
requires a double thermal model out to $\sim$72$\arcsec$
($\sim$16.5~$h_{71}^{-1}$~kpc).

\subsection{A1644}
As with A0085, this SCC cluster shows evidence of merging, with the
existence of a double X-ray peak.  \citet{reiprich04} analyzed the
{\it XMM} {\it EPIC} observation of this cluster.  They found the flux
of the northern (smaller) subclump is below the {\it HIFLUGCS} flux
limit whereas the flux of the southern (larger) subclump is above the
flux limit.  Therefore for purposes of this analysis the smaller
subclump was excluded from spatial and spectral analysis.
Additionally \citet{reiprich04} found evidence that the smaller
sub-clump was being stripped as it passes through the ICM.  This is
one of sixteen clusters in which the central velocity dispersion of
the BCG is unavailable.  The central region of this cluster requires a
double thermal model out to $\sim$32$\arcsec$ ($\sim$30~$h_{71}^{-1}$
kpc).

\subsection{A1650}
This CC cluster hosts a radio quiet cD galaxy \citep{donahue05}.
\citet{Mittal2009} find an upper-limit to the bolometric radio
luminosity of $\sim$9 $\times$ 10$^{38}$~$h_{71}^{-2}$ ergs s$^{-1}$.
The original short {\it Chandra} observation showed a flat temperature
profile \citep{donahue05}. However, the longer, mosaiced observations
show a slight temperature decrease in the central region. Due to the
elevated entropy in the core, \citet{donahue05} concluded a major AGN
outburst had disrupted the cooling flow. This cluster is one of four
clusters on the border between SCC and WCC.  Its CCT
($\sim$1.2~$h_{71}^{-1/2}$~Gyr) is slightly longer than the 1~Gyr
cutoff.  Moreover this cluster shows a central temperature decrease
typical of SCC clusters.  This is one of sixteen clusters in which the
central velocity dispersion of the BCG is unavailable.

\subsection{A1651}
As with A1650, \citet{donahue05} claim this is a radio quiet CC
cluster, however \citet{Mittal2009} detect central radio emission with
a bolometric luminosity of $\sim$10$^{40}$~$h_{71}^{-2}$ ergs
s$^{-1}$.  \citet{gonzalez00} fit the optical light out to
$\sim$670~$h_{100}^{-1}$~kpc, over one quarter of $r_{\rm vir}$.  The
X-ray structure looks quite round and shows no evidence of external
interaction.  However, the X-ray peak does not dominate as much as in
SCC clusters and there is no evidence of a central temperature drop.
This WCC is one of sixteen clusters in which the central velocity
dispersion of the BCG is unavailable.

\subsection{A1656/Coma Cluster}
This well-studied NCC cluster appears to be involved in a merger with
a group.  %There have been no observations of Coma since
% 1999\footnote{One is scheduled for 2008-21-03.}, therefore we
% used data from the period where the focal plane temperature
% was -110$^{\circ}$C.
This cluster hosts the first detected radio halo \citep{willson70}.

\subsection{NGC5044 Group}
The spectra for the inner regions of this group are not well fit by a
single thermal model ($\chi^2$/dof $>$ 2).  After trying several
different models to fit the residuals, we found that the statically
best model which is also physically motivated is a thermal model that
allows oxygen, silicon, sulfur, and iron to vary from solar ratios and
a powerlaw to account for a the clear hard tail (most likely due to
LMXBs).  We note that a double thermal model with the above elements
not constrained at solar ratios provides the statistically best-fit
(in the innermost annulus $\chi^2$/dof = 1.04 vs. $\chi^2$/dof = 1.13
for the thermal plus powerlaw model). In this model, however, the
hotter thermal model has a temperature of $kT$ = 1.4 - 3.0 which is
hotter than any gas found in the outer annuli and the measured virial
temperature ($kT_{\rm vir}$ = 1.22$^{+0.03}_{-0.04}$). Unless there is
a hot halo of gas extending from the center of the group out to
$\sim$16~$h_{71}^{-1}$~kpc, this model is unphysical. Finally, we note
that adding a powerlaw to the second thermal model does not improve
the fit and similar high temperatures are found for the hotter thermal
component as for the simple two thermal model.

\subsection{A1736}
This NCC cluster is a member of the Shapley Supercluster.  Due to the
short exposure time and lack of a bright core, the X-ray peak was
found by smoothing the image with a $\sim$10$\arcsec$ kernel.  The
X-ray morphology shows an irregular shape with no well-defined core.
A preliminary temperature map shows heating to the east and west of
the emission peak with cool gas extending to the south.  The BCG of
this cluster is $\sim$642~$h_{71}^{-1}$~kpc from the peak, making it
one of eight clusters where this separation is $>$50~$h_{71}^{-1}$
kpc.  This is the only cluster with a separation of
$>$50~$h_{71}^{-1}$~kpc that does not have any known associated radio
halo or relic.

\subsection{A3558}
This WCC cluster is located in the core of the Shapley Supercluster.
The observation was heavily flared, and even after a conservative
cleaning of the light curve there was evidence of some low-level
flaring in the back-illuminated chips.  \citet{rossetti07} presented
the {\it XMM} and {\it Chandra} analysis of this cluster, concluding
that it had a cool core that had survived a merger.  We find evidence
of a slight temperature drop in the core of this WCC cluster.

\subsection{A3562}
This WCC cluster is located in the core of the Shapley Supercluster.
The X-ray emission from this cluster appears to be elongated along the
northeast-southwest direction.  \citet{giacintucci05} report the
detection of a radio halo \citep[also see][]{venturi00} and argue for
a merger scenario between A3562 and SC 1329-313.  \citet{finoguenov04}
presented a detailed analysis of the {\it XMM} observation of this
cluster.  The BCG of this cluster is $\sim$31~$h_{71}^{-1}$~kpc from
the X-ray peak, making it one of fourteen clusters where this
separation is $>$12~$h_{71}^{-1}$~kpc.  However, the BCG is located in
a chip gap, so the separation may simply be an instrumental effect
(i.e. the peak on the BCG may not be detected due to the chip gap).
The {\it XMM} observation also shows an offset between the X-ray peak
and BCG but on a scale of only $\sim$23~$h_{71}^{-1}$~kpc
(Y.~Y.~Zhang, private communication).

% The {\it ROSAT} {\it PSPC} observation also suggests and offset
% between the X-ray peak and the BCG.

\subsection{A3571}
This WCC cluster is a member of the Shapley Supercluster.  

\subsection{A1795}
The core of this well-studied SCC cluster has a large filament seen in
X-rays and H$\alpha$ \citep{2005MNRAS.361...17C}.  Early core studies
with {\it Chandra} were done by \citet{2001MNRAS.321L..33F} and
\citet{2001ApJ...562L.153M}. \citet{2001MNRAS.321L..33F} found a CCT
of $\sim$0.4~$h_{71}^{-1/2}$~Gyr, approximately the same age as they
estimate for the filament. The difference between their measurement
for CCT and our measurement is probably due to the different values of
$r_{\rm cool}$ used to determine CCT.  In order to keep consistency
between clusters we determined the CCT at 0.004~$r_{500}$, however at
the redshift of A1795 we are able to determine the CCT at an even
smaller radius which gives a CCT $\sim$0.5~$h_{71}^{-1/2}$~Gyr,
consistent with \citet{2001MNRAS.321L..33F}.  Moreover, their
technique for determining CCT is slightly different from
ours. \citet{2001ApJ...562L.153M} found a cold front in the core of
A1795, which they attribute to sloshing gas.
\citet{2001ApJ...560..187O} studied {\it FUSE} observations and found
an upper limit for $\dot{M}_{\rm spec}$(20~kpc) $<$ 28~$h_{71}^{-1/2}$
$M_{\odot}$~yr$^{-1}$, consistent with our measurement of $\sim$15
$M_{\odot}$~yr$^{-1}$.

\subsection{A3581}
This SCC cluster is a member of the Shapley Supercluster.  It is the
only Shapley Supercluster member from the {\it HIFLUGCS} sample that
is an SCC cluster.  It is also the most distant of these clusters from
the center of the Shapely Supercluster.  The central region of this
cluster requires a double thermal model out to $\sim$40$\arcsec$
($\sim$18~$h_{71}^{-1}$~kpc).  \citet{2005MNRAS.356..237J} analyzed
the {\it Chandra} data from A3581.  They find a point source
coincident with the powerful radio source PKS 1404-267 at the cluster
center.  They find a central temperature drop to $\sim$0.4 $kT_{\rm
  vir}$ at the cluster center, similar to our measurement of $\sim$0.5
$kT_{\rm vir}$.

\subsection{MKW8}
This NCC cluster shows little substructure in the X-ray image.  The
X-ray isophotes are elliptical with the major axis along the east-west
direction.  The isophotes seem to have a common center (i.e. no
evidence of sloshing), however the X-ray peak appears to lie to the
east of the center of the isophotes.  There are two bright galaxies at
the center of the cluster.  The brighter of the two corresponds to the
X-ray peak (which unfortunately falls in a chip gap).  The second
galaxy is to the east, corresponding to the direction of the
elongation of the surface brightness. This cluster shows a possible
radio relic at 74~MHz in the VLA~Low-Frequency~Sky~Survey~(VLSS)
data. The extended radio emission is northwest of the X-ray peak and
extends southwest to northeast $\sim$165~$h_{71}^{-1}$~kpc at the
resolution of the VLSS \citep{Cohen2007}.

\subsection{RXJ1504.1-0248/RBS 1460}
RXJ1504 is the cluster with the highest redshift and X-ray luminosity
in {\it HIFLUGCS}, and shows the largest classical mass deposition
rate. \citet{bohringer05} reported the results to the {\it Chandra}
observation of this cluster.  This cluster was originally not included
in {\it HIFLUGCS} because its X-ray flux is only slightly ($<$20\%)
above the flux limit.  RXJ1504 appears only marginally extended in the
{\it ROSAT} All-Sky Survey.  Additionally the galaxy at the center of
the X-ray emission is classified as AGN
\citep{1999ApJS..123...41M} %(Machalski & Condon 1999)
and its optical spectrum shows emission lines. It was assumed that
even if there is only a small AGN contribution from the central AGN to
the total X-ray flux ($\sim$20\%), the cluster would fall below the
flux limit. However, the {\it Chandra} image reveals that there is
actually no significant point source emission at the center of this
cluster \citep{bohringer05}, which argues against any significant
contamination by AGN emission. The BCG features a compact and
flat-spectrum radio source, \citep[][]{Mittal2009}. Therefore, this
cluster is included into {\it HIFLUGCS}. This SCC cluster is one of
sixteen clusters in which the BCG's central velocity dispersion is not
available.

\subsection{A2029}
The spectra of the inner annuli fit best to non-solar metalicity
ratios, but freeing ratios does not change the best-fit temperatures,
so solar ratios were used for simplicity.  \citet{2004ApJ...616..178C}
studied the core of this cluster in detail with {\it Chandra}.

\subsection{A2052}
The central region of this cluster requires a double thermal model out
to $\sim$45$\arcsec$ ($\sim$32~$h_{71}^{-1}$~kpc).
\citet{2001ApJ...558L..15B} found prominent X-ray cavities in the
original {\it Chandra} observation.  They determined these cavities to
be cospatial with radio lobes from the central radio source.

\subsection{MKW3S/WBL~564}
This SCC cluster shows some disruption in the core and bubbles to the
south \citep{mazzotta04}.  MKW3S is one of sixteen clusters in which
data about the BCG's central velocity dispersion are not available.
This cluster is a member of the Hercules Supercluster.

\subsection{A2065}
A2065 is a member of the Corona Borealis Supercluster, in projection
close to the Hercules Supercluster but twice as distant.  This cluster
is one of four clusters on the border between SCC and WCC clusters.
Its CCT is ($\sim$1.3~$h_{71}^{-1/2}$~Gyr) is slightly longer than the
1~Gyr cutoff.  This cluster shows an inwardly decreasing temperature
profile as seen in the SCC clusters.  This cluster is one of sixteen
clusters in which the BCG's central velocity dispersion is not
available.  It is one of five clusters where the BCG peculiar velocity
is more than 50\% of cluster velocity dispersion.  This offset
suggests possible sloshing which may have disrupted the CC.  Based on
the {\it Chandra} data, \citet{chatzikos06} suggest that the cluster
is involved in an unequal mass merger and that one cool core has
survived the merger.  \citet{1994A&A...281..375F} identified a WAT
$\sim$19$\arcmin$ (1.6~$h_{71}^{-1}$ Mpc) south south-west of the
cluster center.  The jets of the WAT are bent away from the center of
the cluster.  In the NRAO VLA Sky Survey at 1.4~GHz
\citep[NVSS][]{Condon1998}, there appears to be a diffuse radio source
$\sim$91$\arcsec$ ($\sim$124~$h_{71}^{-1}$~kpc) to the southwest of
the cluster center.  It is unclear whether this source is associated
with the central radio source.

\subsection{A2063}
This WCC cluster appears to have a very regular morphology in X-rays,
with some hint of an elongation to the northeast.  The BCG resides at
the X-ray peak.  The NVSS shows three bright radio sources in a line
along an axis from southwest to northeast but only the center source
is associated with the BCG, while the other two are cospatial with two
neighbouring galaxies.  As with many WCC clusters this cluster shows a
flat central temperature profile and a raised central entropy $K_{0}$
$>$ 50~$h_{71}^{-1/3}$ keV cm$^{2}$.  This cluster is close to MKW3S.

\subsection{A2142}
This cluster has a double cold front \citep{markevitch00}.  The
separation between the BCG and the X-ray peak is
$\sim$23~$h_{71}^{-1}$~kpc for this cluster, making it one of fourteen
with this value $>$12~$h_{71}^{-1}$~kpc.  This is one of sixteen
clusters in which no data is available for the BCG's central velocity
dispersion.  It is possible this cluster hosts a radio halo, but the
evidence remains dubious \citep{giovannini00}.

\subsection{A2147}
A2147 is a member of the Hercules Supercluster. Due to the short
observing time combined with the lack of a bright core, we used a
10$\arcsec$ kernel when smoothing to determine the X-ray peak of this
NCC cluster.  There are three bright galaxies in a line near the core,
of which the northernmost (the BCG) is located at the X-ray peak.
This is one of the six NCC clusters in which $\dot{M}_{\rm spec} > 0$.
The X-ray morphology indicates that it is a merging cluster.  The
X-ray emission extends toward the south from the peak following the
line of the three bright galaxies as well as extending to the
southeast.  There is, additionally, a sharp drop in the X-ray emission
to the northwest.  We argue that the observed $\dot{M}_{\rm spec}$ is
not due to cooling, but results from multiple temperatures along the
line of sight caused by the merger.  Although the cluster has been
labeled as a CC cluster in the past \citep[e.g.][]{henriksen96},
\citet{sanderson06} found it to be an NCC cluster and likely merger
system.

\subsection{A2163}
This well-known merging cluster contains the largest known radio halo
\citep{feretti01}.  The separation between the BCG and X-ray peak is
$\sim$158~$h_{71}^{-1}$~kpc for this cluster, making it one of eight
clusters where this value is $>$50~$h_{71}^{-1}$.  Our measurement of
$kT_{\rm vir}$ ($\sim$16 keV) is higher than the value of $\sim$12 keV
found by \citet{2001ApJ...563...95M} with data from the original,
shorter {\it Chandra} observation.  However, a recent measurement by
\citet{2008arXiv0805.2207V}, using the same {\it Chandra} as we, finds
$kT_{vir}$ $\sim$15 keV, more consistent with our result.  The
difference between our result and \citet{2008arXiv0805.2207V} is
barely inconsistent within 1~$\sigma$ and is probably due to
differences in the techniques used to determine $kT_{\rm vir}$ in this
extremely hot cluster.  This is the second most distant and hottest
cluster in the {\it HIFLUGCS} sample. This is one of sixteen clusters
for which data on the BCG's central velocity dispersion are not
available, however, since the BCG is not cospatial with the X-ray peak
so this information is not important for our analysis.

\subsection{A2199}
The central region of this cluster requires a double thermal model out
to $\sim$29$\arcsec$ ($\sim$17~$h_{71}^{-1}$~kpc). 

\subsection{A2204}
Recently \citet{2008arXiv0806.2920R} determined the temperature of
this cluster out to $\sim r_{200}$ using {\it Suzaku}.  They find that
the temperature declines all the way from 0.3~$r_{200}$ to $r_{200}$,
consistent with predictions of simulations.  This is one of sixteen
clusters where data on the BCG's central velocity dispersion is not
available.

\subsection{A2244}
As with A1651, \citet{donahue05} claim it to be a radio quiet CC
cluster, but \citet{Mittal2009} detect central radio emission with a
bolometric luminosity of $\sim$7 $\times$ 10$^{39}$~$h_{71}^{-2}$ ergs
s$^{-1}$.  Although this is not particularly luminous, it is
consistent with radio activity in other CC clusters
\citep{Mittal2009}.  Due to elevated entropy in the core,
\citet{donahue05} concluded a major AGN outburst had disrupted the
cooling flow.  Like many WCC clusters, this cluster shows a flat
temperature profile.  However, we point out that the same was true of
A1650 until a deeper observation revealed a slight temperature drop in
the core.  This is one of sixteen clusters in which the central
velocity dispersion of the BCG is unavailable.

\subsection{A2256}
% Due to a short exposure time and lack of a bright core, we used a
% 10$\arcsec$ kernel when smoothing the image to determine the X-ray
% peak.  ???
This well-known merging cluster is the only one of two NCC clusters
that shows a systematic temperature decrease in the center.  The
temperature decrease is the largest of any NCC or WCC cluster.
Surprisingly, the separation between the BCG and X-ray peak is
$\sim$132~$h_{71}^{-1}$~kpc for this cluster, making it one of eight
clusters where this value is $>$ 50~$h_{71}^{-1}$~kpc.  Since this
separation is quite large, the cool gas is not associated with the
BCG.  It is most likely this gas is the remnant of a CC (perhaps from
a merging group) that has been stripped its central galaxy.  This
cluster hosts both a radio halo and relic
\citep[e.g.][]{clarke06,bridle76}

\subsection{A2255}
Due to a short exposure time and lack of a bright core, we used a
10$\arcsec$ kernel when smoothing the image to determine the X-ray
peak.  The separation between the BCG and X-ray peak is
$\sim$72~$h_{71}^{-1}$~kpc, making it one of eight clusters where this
value is $>$50~$h_{71}^{-1}$~kpc.  This cluster BCG also has by far
the largest peculiar velocity of any cluster; almost twice the
velocity dispersion of the cluster.  This cluster hosts both a radio
halo and a relic \citep[e.g.][]{feretti97}.

\subsection{A3667}
This well-known merging cluster shows a very sharp cold front
\citep{2001ApJ...549L..47V,2001ApJ...551..160V} and two radio relics
\citep[e.g][]{roettiger99}. The separation of the BCG and X-ray peak
is $\sim$155~$h_{71}^{-1}$~kpc for this clusters making it one of
eight clusters where this value is $>$50~$h_{71}^{-1}$~kpc. This is
the only WCC cluster with such a larger separation; however, A3667 is
on the border between NCC and WCC clusters, with
CCT$\sim$6~$h_{71}^{-1}$~Gyr.

\subsection{S1101/S\'{e}rsic 159-03}
\citet{kaastra01} provided a detailed analysis of the {\it XMM-Newton
  RGS} and {\it EPIC} data. Recently \citet{werner07} have claimed
discovery of a diffuse soft excess seen by {\it XMM-Newton} and {\it
  Suzaku} and suggest it is of non-thermal origin. This is one of
sixteen clusters in which the central velocity dispersion of the BCG
is unavailable.

\subsection{A2589}
This WCC shows a systematic temperature drop towards the center,
albeit rather flat ($kT_0$/$kT$ = 0.93 and $\Gamma$=-0.079).  Like
A1650, it is on the cusp between SCC and WCC clusters.
\citet{zappacosta06} studied this cluster with a radio-quiet BCG in
detail with {\it XMM-Newton}.  They find the cluster to be
exceptionally relaxed with a gravitating matter profile that fits a
NFW profile with $c_{\rm vir}$ = 6.1 $\pm$ 0.3 and $M_{\rm vir}$ = 3.3
$\pm$0.3 $\times$ 10$^{14}$ $M_{\odot}$ ($r_{\rm vir}$ = 1.74 $\pm$
0.05 Mpc).  They conclude that processes during halo formation act
against adiabatic contraction.  Additionally \citet{buote04} studied
the original short {\it Chandra} observation that suffered from
flaring.

Following a method to determine the residual CXB (similar to what is
described in Sect~\ref{kTvir}), we measured the surface brightness
profile out to 750~$h_{71}^{-1}$~kpc ($\sim$0.5 $r_{\rm vir}$).  We
fit this surface brightness profile to a double-$\beta$ model and the
temperature profile to a broken powerlaw.  The slope of the inner $kT$
profile was fixed at zero and the outer $kT$ profile fit well to a
powerlaw of slope -0.36 with a break radius of 4$\farcm$2 (204
h$_{71}^{-1}$~kpc).  Using the fit to the temperature profile and
double $\beta$-model, we find a virial\footnote{In this case $r_{\rm
    vir}$ is defined for an overdensity of 104.7, as used by
  \citet{zappacosta06}.} mass and radius of $M_{\rm vir}$ =
2.7$\pm$0.8 $\times$ 10$^{14}$~$h_{71}^{-1}$ $M_{\odot}$ and $r_{\rm
  vir}$ = 1.6 $\pm$0.2 h$_{71}^{-1}$ Mpc respectively, consistent with
the results of \citet{zappacosta06}.

\subsection{A2597}
\citet{mcnamara01} analyzed the original, short, flared observation of
A2597, noting the ghost bubbles.  \citet{morris05} found high spectral
mass deposition rates from the {\it XMM-Newton} {\it EPIC} and {\it
  RGS} consistent with $\sim$100 $M_{\odot}$~yr$^{-1}$ down to almost
0~keV, although the improvement to the spectral fits of the {\it RGS}
data from the addition of a cooling flow model is marginal.  The long
{\it Chandra} {\it ACIS} observation shows a mass deposition rate of
$\sim$150 $M_{\odot}$~yr$^{-1}$ down to $\sim$1.3 keV and dropping to
$\sim$10 $M_{\odot}$~yr$^{-1}$ down to $\sim$0 keV.

\subsection{A2634}
This WCC cluster contains the WAT source 3C465. There are a pair of
galaxies (NGC7720) located at the X-ray peak. An extended bright X-ray
halo (radius = $\sim 12\arcsec$), much brighter than the ICM emission,
is cospatial with the galaxy pair. The halo seems to be associated
with the larger southern galaxy. A2634 is the only CC cluster in the
sample with $\dot{M}_{\rm spec} > \dot{M}_{\rm classical}$.  The
temperature profile shows a sudden drop at $\sim$2$\farcm$7
($\sim100~h_{71}^{-1}$~kpc). Other than NGC7720, there is no obvious
core in A2634 and the elongation of the ICM to the southwest is
consistent with a merging cluster. Moreover the {\it inverted}
temperature profile is more common in NCC clusters than in WCC
clusters. We interpret the short cooling time and low $\dot{M}_{\rm
  classical}$ as a cool core that has either been disrupted or is in
the process of being destroyed by a merger. $\dot{M}_{\rm spec}$ may
reflect the original mass deposition rate, but probably is strongly
affected by multitemperature components along the line of sight in a
merging system.

\subsection{A2657}
This WCC cluster has a slight increase in temperature in the central
region.  The {\it Chandra} image shows a cluster similar to e.g. A1650
and A2244.  The central emission peak is clearly visibly but is not as
sharply peaked as in SCC clusters.  The overall ICM appears to be
quite round, with some sloshing features (differently centered X-ray
isophotes at different radii) in the central region.

\subsection{A4038}
The distance between the BCG and X-ray peak is
$\sim$12.4~$h_{71}^{-1}$~kpc for this cluster making it one of
fourteen clusters where this separation is $>$12~$h_{71}^{-1}$~kpc.
This cluster hosts a radio relic, close to but not connected to the
central radio galaxy \citep{slee98}.

\subsection{A4059}
The central region of this cluster requires a double thermal model out
to $\sim$22$\arcsec$ ($\sim$20~$h_{71}^{-1}$~kpc).

\end{appendix}

\label{lastpage}

%\longtab{1}{
  \begin{longtable}{c c c c c c c c}
    \caption{Observational Parameters (1)~Cluster name, (2)
      Right~Ascension and (3) Declination of the X-ray emission peak,
      (4) Number of observations used, (5) The total good time, (6)
      The fitted cluster virial temperature, (7) The fitted clustre
      metalicity and (8) The size of the region with \mbox{$T$ $<$
        $T_{\rm vir}$} (see section-\ref{kTvir}). }
    \label{obs}\\
    \hline\hline
    Cluster &    RA    &    DEC    & \# of  &  GT  & $kT_{\rm vir}$ & $Z$     & Core \\
    &  (J2000) &  (J2000)  &   Ob   &  ks  &    keV        & Solar   & \%$r_{\rm vir}$ \\
    (1)  &    (2)   &    (3)    &  (4)   &  (5) &      (6)      &   (7)   & (8)  \\
    \hline
    \endfirsthead
    \caption{continued.}\\
    \hline\hline
    Cluster &    RA    &    DEC    & \# of  &  GT  & $kT_{\rm vir}$ & $Z$     & Core \\
    &  (J2000) &  (J2000)  &   Ob   &  ks  &    keV        & Solar   & \%$r_{\rm vir}$ \\
    (1)  &    (2)   &    (3)    &  (4)   &  (5) &      (6)      &   (7)   & (8)  \\
    \hline
    \endhead
    \hline
    \endfoot
    A0085 &00$^{\rm h}41^{\rm m} 50\fs39$ & -09$^\circ 18\arcmin 11\farcs0$ & 9  & 114.7 & 6.00$_{-0.11}^{+0.11}$ & 0.38$_{-0.05}^{+0.03}$ & 6.2$_{-0.8}^{+1.3}$\\
    A0119 &00$^{\rm h}56^{\rm m} 16\fs04$ & -01$^\circ 15\arcmin 20\farcs6$ & 1  & 11.4 & 5.73$_{-0.30}^{+0.34}$ & 0.48$_{-0.11}^{+0.11}$ & 0\\
    A0133 &01$^{\rm h}02^{\rm m} 41\fs78$ & -21$^\circ 52\arcmin 56\farcs0$ & 3  & 120.5 & 3.96$_{-0.10}^{+0.08}$ & 0.41$_{-0.07}^{+0.03}$ & 4.5$_{-0.5}^{+0.6}$\\
    NGC0507 &01$^{\rm h}23^{\rm m} 39\fs82$ & +33$^\circ 15\arcmin 21\farcs5$ & 1  & 43.5 & 1.44$_{-0.10}^{+0.08}$ & 0.43$_{-0.08}^{+0.11}$ & 3.4$_{-0.2}^{+0.3}$\\
    A0262 &01$^{\rm h}52^{\rm m} 46\fs23$ & +36$^\circ 09\arcmin 14\farcs9$ & 1  & 28.0 & 2.44$_{-0.04}^{+0.03}$ & 0.60$_{-0.03}^{+0.04}$ & 2.0$_{-0.1}^{+0.1}$\\
    A0400 &02$^{\rm h}57^{\rm m} 41\fs59$ & +06$^\circ 01\arcmin 37\farcs4$ & 1  & 21.5 & 2.26$_{-0.12}^{+0.10}$ & 0.50$_{-0.06}^{+0.07}$ & 0\\
    A0399 &02$^{\rm h}57^{\rm m} 53\fs45$ & +13$^\circ 01\arcmin 52\farcs8$ & 1  & 49.0 & 6.70$_{-0.14}^{+0.14}$ & 0.30$_{-0.04}^{+0.05}$ & 0\\
    A0401 &02$^{\rm h}58^{\rm m} 56\fs66$ & +13$^\circ 34\arcmin 39\farcs8$ & 2  & 29.6 & 8.51$_{-0.22}^{+0.34}$ & 0.34$_{-0.06}^{+0.06}$ & 0\\
    A3112 &03$^{\rm h}17^{\rm m} 57\fs65$ & -44$^\circ 14\arcmin 18\farcs3$ & 5  & 90.5 & 4.73$_{-0.12}^{+0.11}$ & 0.38$_{-0.05}^{+0.05}$ & 5.2$_{-0.5}^{+0.8}$\\
    NGC1399 &03$^{\rm h}38^{\rm m} 29\fs10$ & -35$^\circ 27\arcmin 00\farcs9$ & 12  & 429.7 & 1.34$_{-0.00}^{+0.00}$ & 0.27$_{-0.00}^{+0.01}$ & 2.0$_{-0.1}^{+0.1}$\\
    2A0335$^{\dagger}$ &03$^{\rm h}38^{\rm m} 41\fs14$ & +09$^\circ 58\arcmin 01\farcs9$ & 1  & 19.7 & 3.53$_{-0.13}^{+0.10}$ & 0.54$_{-0.05}^{+0.06}$ & 5.5$_{-0.6}^{+0.4}$\\
    IIIZw54 &03$^{\rm h}41^{\rm m} 17\fs64$ & +15$^\circ 23\arcmin 37\farcs1$ & 1  & 23.3 & 2.50$_{-0.06}^{+0.05}$ & 0.37$_{-0.04}^{+0.04}$ & 0\\
    A3158 &03$^{\rm h}42^{\rm m} 52\fs27$ & -53$^\circ 37\arcmin 55\farcs5$ & 2  & 55.7 & 4.99$_{-0.07}^{+0.07}$ & 0.41$_{-0.03}^{+0.04}$ & 0\\
    A0478 &04$^{\rm h}13^{\rm m} 25\fs15$ & +10$^\circ 27\arcmin 53\farcs8$ & 8  & 101.4 & 7.34$_{-0.19}^{+0.18}$ & 0.31$_{-0.03}^{+0.03}$ & 5.2$_{-0.6}^{+0.7}$\\
    NGC1550 &04$^{\rm h}19^{\rm m} 37\fs97$ & +02$^\circ 24\arcmin 36\farcs2$ & 4  & 107.6 & 1.34$_{-0.00}^{+0.00}$ & 0.25$_{-0.01}^{+0.01}$ & 2.0$_{-0.1}^{+0.2}$\\
    EXO0422$^{\dagger}$ &04$^{\rm h}25^{\rm m} 51\fs24$ & -08$^\circ 33\arcmin 37\farcs9$ & 1  & 9.8 & 2.93$_{-0.12}^{+0.13}$ & 0.36$_{-0.07}^{+0.08}$ & 3.2$_{-1.4}^{+2.6}$\\
    A3266 &04$^{\rm h}31^{\rm m} 13\fs13$ & -61$^\circ 27\arcmin 11\farcs0$ & 1  & 29.5 & 9.45$_{-0.36}^{+0.35}$ & 0.26$_{-0.06}^{+0.06}$ & 0\\
    A0496 &04$^{\rm h}33^{\rm m} 37\fs95$ & -13$^\circ 15\arcmin 39\farcs9$ & 2  & 66.4 & 4.86$_{-0.06}^{+0.05}$ & 0.66$_{-0.03}^{+0.03}$ & 4.8$_{-0.3}^{+0.3}$\\
    A3376 &06$^{\rm h}02^{\rm m} 08\fs64$ & -39$^\circ 56\arcmin 48\farcs5$ & 2  & 64.2 & 3.80$_{-0.10}^{+0.11}$ & 0.31$_{-0.04}^{+0.04}$ & 0\\
    A3391 &06$^{\rm h}26^{\rm m} 20\fs50$ & -53$^\circ 41\arcmin 37\farcs0$ & 1  & 17.1 & 5.77$_{-0.36}^{+0.31}$ & 0.33$_{-0.09}^{+0.13}$ & 0\\
    A3395s &06$^{\rm h}26^{\rm m} 49\fs74$ & -54$^\circ 32\arcmin 33\farcs6$ & 1  & 21.0 & 4.82$_{-0.26}^{+0.26}$ & 0.09$_{-0.08}^{+0.10}$ & 0\\
    A0576 &07$^{\rm h}21^{\rm m} 30\fs26$ & +55$^\circ 45\arcmin 50\farcs6$ & 1  & 27.0 & 4.09$_{-0.10}^{+0.08}$ & 0.59$_{-0.06}^{+0.07}$ & 0\\
    A0754 &09$^{\rm h}09^{\rm m} 16\fs66$ & -09$^\circ 41\arcmin 20\farcs8$ & 1  & 44.1 & 11.13$_{-0.43}^{+0.39}$ & 0.30$_{-0.06}^{+0.05}$ & 0\\
    A0780 &09$^{\rm h}18^{\rm m} 6\fs09$ & -12$^\circ 05\arcmin 45\farcs0$ & 2  & 186.7 & 3.45$_{-0.09}^{+0.08}$ & 0.27$_{-0.03}^{+0.03}$ & 13.0$_{-1.3}^{+2.1}$\\
    A1060 &10$^{\rm h}36^{\rm m} 42\fs75$ & -27$^\circ 31\arcmin 42\farcs0$ & 1  & 31.4 & 3.16$_{-0.04}^{+0.04}$ & 0.47$_{-0.03}^{+0.03}$ & 0\\
    A1367 &11$^{\rm h}45^{\rm m} 00\fs29$ & +19$^\circ 40\arcmin 30\farcs2$ & 2  & 68.7 & 3.58$_{-0.06}^{+0.06}$ & 0.35$_{-0.02}^{+0.04}$ & 0\\
    MKW4  &12$^{\rm h}04^{\rm m} 27\fs08$ & +01$^\circ 53\arcmin 46\farcs1$ & 1  & 30.1 & 2.01$_{-0.04}^{+0.04}$ & 0.65$_{-0.04}^{+0.06}$ & 1.8$_{-0.2}^{+0.4}$\\
    ZwCl1215$^{\dagger}$ &12$^{\rm h}17^{\rm m} 41\fs71$ & +03$^\circ 39\arcmin 18\farcs4$ & 1  & 11.9 & 6.27$_{-0.29}^{+0.32}$ & 0.25$_{-0.08}^{+0.09}$ & 0\\
    NGC4636 &12$^{\rm h}42^{\rm m} 49\fs91$ & +02$^\circ 41\arcmin 12\farcs6$ & 2  & 150.3 & 0.90$_{-0.02}^{+0.02}$ & 0.20$_{-0.02}^{+0.02}$ & 2.0$_{-0.1}^{+0.1}$\\
    A3526 &12$^{\rm h}48^{\rm m} 48\fs85$ & -41$^\circ 18\arcmin 43\farcs8$ & 7  & 288.6 & 3.92$_{-0.02}^{+0.02}$ & 0.60$_{-0.01}^{+0.01}$ & 3.16$_{-0.05}^{+0.05}$\\
    A1644 &12$^{\rm h}57^{\rm m} 11\fs79$ & -17$^\circ 24\arcmin 32\farcs3$ & 2  & 70.3 & 5.09$_{-0.09}^{+0.09}$ & 0.37$_{-0.04}^{+0.05}$ & 5.6$_{-0.8}^{+0.6}$\\
    A1650 &12$^{\rm h}58^{\rm m} 41\fs48$ & -01$^\circ 45\arcmin 42\farcs7$ & 7  & 225.3 & 5.81$_{-0.07}^{+0.06}$ & 0.37$_{-0.02}^{+0.03}$ & 2.2$_{-0.6}^{+1.2}$\\
    A1651 &12$^{\rm h}59^{\rm m} 22\fs16$ & -04$^\circ 11\arcmin 49\farcs2$ & 1  & 9.3 & 6.34$_{-0.27}^{+0.27}$ & 0.47$_{-0.09}^{+0.10}$ & 0\\
    A1656 &12$^{\rm h}59^{\rm m} 35\fs73$ & +27$^\circ 57\arcmin 34\farcs9$ & 1  & 29.8 & 9.15$_{-0.16}^{+0.17}$ & 0.33$_{-0.03}^{+0.03}$ & 0\\
    NGC5044 &13$^{\rm h}15^{\rm m} 23\fs88$ & -16$^\circ 23\arcmin 06\farcs8$ & 1  & 15.8 & 1.22$_{-0.04}^{+0.03}$ & 0.25$_{-0.02}^{+0.04}$ & 5.4$_{-0.1}^{+1.3}$\\
    A1736 &13$^{\rm h}26^{\rm m} 51\fs87$ & -27$^\circ 10\arcmin 26\farcs8$ & 1  & 14.8 & 3.12$_{-0.12}^{+0.11}$ & 0.34$_{-0.06}^{+0.07}$ & 0\\
    A3558 &13$^{\rm h}27^{\rm m} 56\fs89$ & -31$^\circ 29\arcmin 43\farcs2$ & 1  & 9.9 & 4.95$_{-0.15}^{+0.13}$ & 0.27$_{-0.06}^{+0.06}$ & 5.8$_{-2.3}^{+6.6}$\\
    A3562 &13$^{\rm h}33^{\rm m} 37\fs29$ & -31$^\circ 40\arcmin 17\farcs0$ & 1  & 19.4 & 4.43$_{-0.16}^{+0.21}$ & 0.44$_{-0.07}^{+0.07}$ & 0\\
    A3571 &13$^{\rm h}47^{\rm m} 28\fs32$ & -32$^\circ 51\arcmin 57\farcs5$ & 1  & 25.9 & 7.00$_{-0.12}^{+0.13}$ & 0.45$_{-0.03}^{+0.04}$ & 0\\
    A1795 &13$^{\rm h}48^{\rm m} 52\fs58$ & +26$^\circ 35\arcmin 33\farcs1$ & 12  & 173.5 & 6.08$_{-0.07}^{+0.07}$ & 0.28$_{-0.02}^{+0.02}$ & 6.9$_{-0.5}^{+0.6}$\\
    A3581 &14$^{\rm h}07^{\rm m} 30\fs19$ & -27$^\circ 01\arcmin 10\farcs5$ & 1  & 6.2 & 1.97$_{-0.07}^{+0.07}$ & 0.53$_{-0.07}^{+0.08}$ & 3.0$_{-0.3}^{+0.4}$\\
    MKW8  &14$^{\rm h}40^{\rm m} 43\fs08$ & +03$^\circ 27\arcmin 57\farcs7$ & 1  & 23.1 & 3.00$_{-0.12}^{+0.12}$ & 0.45$_{-0.07}^{+0.08}$ & 0\\
    RXJ1504$^{\dagger}$ &15$^{\rm h}04^{\rm m} 07\fs52$ & -02$^\circ 48\arcmin 16\farcs8$ & 2  & 52.5 & 9.53$_{-1.16}^{+1.39}$ & 0.38$_{-0.22}^{+0.23}$ & 12.1$_{-3.6}^{+17.3}$\\
    A2029 &15$^{\rm h}10^{\rm m} 56\fs06$ & +05$^\circ 44\arcmin 41\farcs4$ & 3  & 107.0 & 8.26$_{-0.09}^{+0.09}$ & 0.38$_{-0.02}^{+0.02}$ & 3.7$_{-0.4}^{+0.5}$\\
    A2052 &15$^{\rm h}16^{\rm m} 43\fs51$ & +07$^\circ 01\arcmin 19\farcs8$ & 2  & 163.9 & 3.35$_{-0.02}^{+0.02}$ & 0.66$_{-0.02}^{+0.02}$ & 2.55$_{-0.07}^{+0.05}$\\
    MKW3S &15$^{\rm h}21^{\rm m} 51\fs75$ & +07$^\circ 42\arcmin 28\farcs7$ & 1  & 56.7 & 3.90$_{-0.09}^{+0.09}$ & 0.41$_{-0.04}^{+0.05}$ & 6.2$_{-1.8}^{+2.4}$\\
    A2065 &15$^{\rm h}22^{\rm m} 29\fs32$ & +27$^\circ 42\arcmin 22\farcs2$ & 1  & 49.5 & 5.40$_{-0.11}^{+0.20}$ & 0.29$_{-0.04}^{+0.05}$ & 5.1$_{-1.7}^{+4.0}$\\
    A2063 &15$^{\rm h}23^{\rm m} 05\fs11$ & +08$^\circ 36\arcmin 26\farcs9$ & 2  & 28.5 & 3.77$_{-0.06}^{+0.06}$ & 0.60$_{-0.04}^{+0.04}$ & 0 \\
    A2142 &15$^{\rm h}58^{\rm m} 20\fs65$ & +27$^\circ 13\arcmin 49\farcs2$ & 1  & 44.8 & 8.40$_{-0.76}^{+1.01}$ & 0.32$_{-0.23}^{+0.24}$ & $>$24.0 \\
    A2147 &16$^{\rm h}02^{\rm m} 16\fs78$ & +15$^\circ 58\arcmin 25\farcs6$ & 1  & 17.0 & 4.07$_{-0.12}^{+0.11}$ & 0.33$_{-0.06}^{+0.06}$ & 0 \\
    A2163 &16$^{\rm h}15^{\rm m} 46\fs69$ & -06$^\circ 09\arcmin 00\farcs3$ & 2  & 80.3 & 15.91$_{-0.81}^{+0.81}$ & 0.32$_{-0.06}^{+0.06}$ & 0 \\
    A2199 &16$^{\rm h}28^{\rm m} 38\fs32$ & +39$^\circ 33\arcmin 01\farcs2$ & 1  & 16.6 & 4.37$_{-0.07}^{+0.07}$ & 0.52$_{-0.03}^{+0.04}$ & 1.8$_{-0.6}^{+0.1}$\\
    A2204 &16$^{\rm h}32^{\rm m} 46\fs94$ & +05$^\circ 34\arcmin 31\farcs3$ & 2  & 18.6 & 8.92$_{-0.61}^{+0.72}$ & 0.28$_{-0.11}^{+0.11}$ & 5.4$_{-1.4}^{+1.8}$\\
    A2244 &17$^{\rm h}02^{\rm m} 42\fs68$ & +34$^\circ 03\arcmin 39\farcs3$ & 1  & 56.0 & 5.78$_{-0.11}^{+0.10}$ & 0.43$_{-0.03}^{+0.04}$ & 0 \\
    A2256 &17$^{\rm h}03^{\rm m} 14\fs26$ & +78$^\circ 38\arcmin 59\farcs9$ & 1  & 8.0 & 7.61$_{-0.63}^{+0.65}$ & 0.26$_{-0.17}^{+0.18}$ & 14.4$_{-4.1}^{+8.7}$\\
    A2255 &17$^{\rm h}12^{\rm m} 34\fs15$ & +64$^\circ 04\arcmin 11\farcs5$ & 1  & 39.6 & 5.81$_{-0.20}^{+0.19}$ & 0.27$_{-0.07}^{+0.06}$ & 0\\
    A3667 &20$^{\rm h}12^{\rm m} 42\fs66$ & -56$^\circ 50\arcmin 48\farcs6$ & 7  & 485.3 & 6.39$_{-0.04}^{+0.04}$ & 0.35$_{-0.01}^{+0.01}$ & 1.2$_{-0.5}^{+0.7}$\\
    S1101 &23$^{\rm h}13^{\rm m} 58\fs40$ & -42$^\circ 43\arcmin 31\farcs0$ & 1  & 7.9 & 2.57$_{-0.13}^{+0.12}$ & 0.23$_{-0.06}^{+0.06}$ & 6.4$_{-3.1}^{+4.2}$\\
    A2589 &23$^{\rm h}23^{\rm m} 57\fs40$ & +16$^\circ 46\arcmin 37\farcs9$ & 3  & 76.7 & 3.89$_{-0.05}^{+0.05}$ & 0.80$_{-0.04}^{+0.04}$ & 1.9$_{-0.6}^{+1.2}$\\
    A2597 &23$^{\rm h}25^{\rm m} 19\fs93$ & -12$^\circ 07\arcmin 27\farcs5$ & 2  & 112.0 & 4.05$_{-0.07}^{+0.07}$ & 0.38$_{-0.04}^{+0.04}$ & 5.8$_{-0.6}^{+1.0}$\\
    A2634 &23$^{\rm h}38^{\rm m} 29\fs25$ & +27$^\circ 01\arcmin 54\farcs2$ & 1  & 48.7 & 3.19$_{-0.11}^{+0.11}$ & 0.34$_{-0.05}^{+0.05}$ & 0\\
    A2657 &23$^{\rm h}44^{\rm m} 57\fs48$ & +09$^\circ 11\arcmin 31\farcs0$ & 1  & 16.1 & 3.52$_{-0.11}^{+0.12}$ & 0.38$_{-0.06}^{+0.06}$ & 0\\
    A4038 &23$^{\rm h}47^{\rm m} 43\fs18$ & -28$^\circ 08\arcmin 31\farcs2$ & 2  & 40.0 & 3.14$_{-0.04}^{+0.03}$ & 0.50$_{-0.03}^{+0.03}$ & 0\\
    A4059 &23$^{\rm h}57^{\rm m} 00\fs93$ & -34$^\circ 45\arcmin 33\farcs3$ & 2  & 109.9 & 4.22$_{-0.03}^{+0.03}$ & 0.66$_{-0.02}^{+0.03}$ & 2.4$_{-0.3}^{+0.1}$\\
    
    \hline
    \multicolumn{4}{p{7.5cm}}{$^{\dagger}$ Abbreviated Cluster Names:}  &    \multicolumn{4}{p{7.5cm}}{$^{\dagger}$ Alternative Cluster Names:} \\    
    \multicolumn{4}{p{7.5cm}}{2A0335: 2A~0335$+$096}                   &    \multicolumn{4}{p{7.5cm}}{Coma: A1656}\\                      
    \multicolumn{4}{p{7.5cm}}{EXO0422: EXO~0422$-$086}                 &    \multicolumn{4}{p{7.5cm}}{Fornax: NGC~1399}\\                    
    \multicolumn{4}{p{7.5cm}}{ZwCl1215: ZwCl~1215.1$+$0400}            &    \multicolumn{4}{p{7.5cm}}{Centaurus: A3526}\\               
    \multicolumn{4}{p{7.5cm}}{RXJ1504: RX~J1504.1$-$0248}              &    \multicolumn{4}{p{7.5cm}}{}                \\                 
    \hline\hline
  \end{longtable}
%}

%\longtab{2}{
  \setlength\LTleft{-2.5in}
  \setlength\LTright\fill
  \begin{landscape}
    \footnotesize
    \begin{longtable}{c c c c c c c c c c c c}
      \caption{Derived Parameters: (1) cluster name, (2) central
        electron density, (3) central cooling time, (4) fractional
        central temperature drop, (5) slope of the central temperature
        profile (positive indicates a declining temperature with radius,
        see text), (6) cuspiness defined as $\frac{\mathrm{d}
          \log(n)}{\mathrm{d} \log(r)}$ at $r = 0.04r_{500}$, (7) mass
        deposition rate within 0.048$r_{500}$, estimated for a classical
        cooling flow (see text), (8) mass deposition rate estimated from
        the spectrum within 0.048$r_{500}$, (9) mass deposition rate
        down to a free lower temperature ($T_{\rm low}$), estimated from
        the spectrum within 0.048$r_{500}$, (10) the lower temperature
        found for $\dot{M}_{\rm spec2}$, (11) central entropy and (12)
        cool-core classification - SCC = strong cool-core, WCC = weak
        cool-core and NCC = non-cool-core (based on the central cooling
        time, see sect.~\ref{define_param}).}
    \label{par}\\
    \hline\hline
    Cluster  & n$_{0}$  & CCT & $T_{0}/T_{\rm vir}$ & $\Gamma$ & $\alpha$ & $\dot{M}_{\rm classical}$  & $\dot{M}_{\rm spec}$ & $\dot{M}_{\rm spec2}$ & kT$_{\rm low}$ & $K_{0}$  & $CCC$ \\
    & $h_{71}^{1/2}$ 10$^{-2}$ cm$^{-3}$ & $h_{71}^{-1/2}$~Gyr & & & &
    $h_{71}^{-2}$ M$_{\odot}$/yr & M$_{\odot}$/yr & M$_{\odot}$/yr &
    keV & h$_{71}^{-1/3}$ keV cm$^2$ &
    \\
    (1) & (2) & (3) & (4) & (5) & (6) & (7) & (8) & (9) & (10) & (11) & (12) \\
    \hline
    \endfirsthead
    \caption{continued.}\\
    \hline\hline
    Cluster  & n$_{0}$  & CCT & $T_{0}/T_{\rm vir}$ & $\Gamma$ & $\alpha$ & $\dot{M}_{\rm classical}$  & $\dot{M}_{\rm spec}$ & $\dot{M}_{\rm spec2}$ & kT$_{\rm low}$ & $K_{0}$  & $CCC$ \\
    & $h_{71}^{1/2}$ 10$^{-2}$ cm$^{-3}$ & $h_{71}^{-1/2}$~Gyr & & & &
    $h_{71}^{-2}$ M$_{\odot}$/yr & M$_{\odot}$/yr & M$_{\odot}$/yr &
    keV & h$_{71}^{-1/3}$ keV cm$^2$ &
    \\
    (1) & (2) & (3) & (4) & (5) & (6) & (7) & (8) & (9) & (10) & (11) & (12) \\
    \hline
    \endhead
    \hline
    \endfoot
    A0085$^{\alpha}$ & 6.92$_{-0.52}^{+0.52}$ & 0.51$_{-0.04}^{+0.04}$ & 0.34$_{-0.07}^{+0.07}$ & -0.239$_{-0.022}^{+0.022}$ & 0.931$_{-0.092}^{+0.092}$ & 131.8$_{-28.3}^{+28.3}$ & 11.6$_{-2.1}^{+2.1}$ & 180.1$_{-14.9}^{+16.9}$ & 2.16$_{-0.25}^{+0.36}$ & 12.0$_{-2.4}^{+2.4}$ & SCC\\
    A0119 & 0.30$_{-0.08}^{+0.08}$ & 14.03$_{-3.43}^{+5.95}$ & 0.90$_{-0.06}^{+0.06}$ & -0.103$_{-0.104}^{+0.105}$ & 0.590$_{-0.278}^{+0.278}$ & - & $<$0.6 & $<$0.5 & 0.08$_{0.00}^{+79.82}$ & 246.9$_{-45.4}^{+45.4}$ & NCC\\
    A0133$^{\alpha}$ & 4.72$_{-0.25}^{+0.25}$ & 0.47$_{-0.03}^{+0.03}$ & 0.41$_{-0.04}^{+0.04}$ & -0.370$_{-0.038}^{+0.039}$ & 1.255$_{-0.003}^{+0.013}$ & 68.4$_{-6.5}^{+6.5}$ & 8.9$_{-1.2}^{+1.1}$ & 35.8$_{-6.7}^{+8.8}$ & 1.00$_{-0.15}^{+0.13}$ & 12.4$_{-1.3}^{+1.3}$ & SCC\\
    NGC0507$^{\alpha}$ & 2.43$_{-0.56}^{+0.56}$ & 0.48$_{-0.10}^{+0.15}$ & 0.65$_{-0.04}^{+0.04}$ & -0.240$_{-0.041}^{+0.039}$ & 0.828$_{-0.067}^{+0.067}$ & 7.7$_{-2.2}^{+2.2}$ & 0.9$_{-0.3}^{+0.3}$ & 9.4$_{-0.5}^{+0.7}$ & 0.71$_{-0.03}^{+0.04}$ & 11.2$_{-1.8}^{+1.8}$ & SCC \\
    A0262$^{\alpha}$ & 4.06$_{-0.28}^{+0.28}$ & 0.43$_{-0.03}^{+0.04}$ & 0.35$_{-0.01}^{+0.01}$ & -0.540$_{-0.037}^{+0.036}$ & 1.059$_{-0.003}^{+0.004}$ & 12.1$_{-1.4}^{+1.4}$ & 2.9$_{-0.2}^{+0.2}$ & 9.2$_{-0.5}^{+0.8}$ & 0.78$_{-0.04}^{+0.06}$ & 7.8$_{-3.5}^{+3.5}$ & SCC\\
    A0400 & 0.28$_{-0.05}^{+0.05}$ & 8.04$_{-1.47}^{+1.99}$ & 0.93$_{-0.06}^{+0.06}$ & -0.129$_{-0.092}^{+0.088}$ & 0.341$_{-0.034}^{+0.054}$ & - & 1.1$_{-0.5}^{+0.6}$ & 2.4$_{-1.0}^{+3.7}$ & 0.89$_{-0.81}^{+0.55}$ & 104.4$_{-12.9}^{+12.9}$ & NCC$^{\dagger}$\\
    A0399 & 0.45$_{-0.04}^{+0.04}$ & 12.13$_{-1.22}^{+1.44}$ & 1.09$_{-0.06}^{+0.06}$ & 0.009$_{-0.094}^{+0.091}$ & 0.379$_{-0.024}^{+0.022}$ & - & 1.6$_{-1.0}^{+1.5}$ & 1.6$_{-1.0}^{+5.2}$ & 0.08$_{0.00}^{+79.82}$ & 267.3$_{-19.7}^{+19.7}$ & NCC\\
    A0401$^{\gamma}$ & 0.67$_{-0.08}^{+0.08}$ & 8.81$_{-1.08}^{+1.41}$ & 1.05$_{-0.07}^{+0.07}$ & 0.028$_{-0.127}^{+0.120}$ & 0.354$_{-0.004}^{+0.048}$ & - & 4.0$_{-2.5}^{+2.7}$ & 4.7$_{-1.5}^{+4.0}$ & 0.51$_{-0.43}^{+4.06}$ & 250.9$_{-24.9}^{+24.9}$ & NCC\\
    A3112 & 8.20$_{-1.34}^{+1.34}$ & 0.37$_{-0.05}^{+0.08}$ & 0.56$_{-0.01}^{+0.01}$ & -0.226$_{-0.012}^{+0.012}$ & 1.218$_{-0.209}^{+0.209}$ & 146.3$_{-41.2}^{+41.2}$ & $<$1.6 & 110.5$_{-36.7}^{+107.8}$ & 1.83$_{-0.38}^{+0.38}$ & 14.0$_{-1.5}^{+1.5}$ & SCC\\
    NGC1399$^{\alpha}$ & 1.99$_{-0.93}^{+0.93}$ & 0.69$_{-0.22}^{+0.60}$ & 0.52$_{-0.01}^{+0.01}$ & -0.223$_{-0.008}^{+0.008}$ & 0.926$_{-0.257}^{+0.257}$ & 0.3$_{-0.2}^{+0.2}$ & 0.52$_{-0.02}^{+0.02}$ & 1.96$_{-0.04}^{+0.06}$ & 0.72$_{-0.01}^{+0.02}$ & 13.6$_{-4.2}^{+4.2}$ & SCC\\
    2A0335$^{\alpha}$ & 7.03$_{-0.17}^{+0.17}$ & 0.31$_{-0.01}^{+0.01}$ & 0.28$_{-0.01}^{+0.01}$ & -0.521$_{-0.025}^{+0.024}$ & 1.412$_{-0.004}^{+0.005}$ & 267.8$_{-12.8}^{+12.8}$ & 65.8$_{-3.7}^{+3.7}$ & 93.9$_{-7.4}^{+10.3}$ & 0.65$_{-0.08}^{+0.09}$ & 5.8$_{-0.3}^{+0.3}$ & SCC\\
    IIIZw54 & 0.51$_{-0.04}^{+0.04}$ & 5.48$_{-0.48}^{+0.59}$ & 1.11$_{-0.04}^{+0.04}$ & -0.015$_{-0.057}^{+0.057}$ & 0.481$_{-0.024}^{+0.025}$ & 18.1$_{-7.5}^{+7.5}$ & $<$0.2 & 0.1$_{-0.1}^{+22.9}$ & 3.10$_{-0.40}^{+0.33}$ & 94.0$_{-5.6}^{+5.6}$ & WCC\\
    A3158 & 0.52$_{-0.02}^{+0.02}$ & 8.22$_{-0.47}^{+0.54}$ & 1.15$_{-0.05}^{+0.05}$ & 0.044$_{-0.079}^{+0.077}$ & 0.294$_{-0.011}^{+0.011}$ & - & 0.7$_{-0.4}^{+0.9}$ & 4.0$_{-2.2}^{+141.4}$ & 1.15$_{-0.85}^{+3.05}$ & 190.8$_{-9.9}^{+9.9}$ & NCC\\
    A0478 & 10.03$_{-1.81}^{+1.81}$ & 0.43$_{-0.07}^{+0.10}$ & 0.41$_{-0.01}^{+0.01}$ & -0.281$_{-0.016}^{+0.016}$ & 0.910$_{-0.109}^{+0.109}$ & 559.3$_{-241.7}^{+241.7}$ & 136.7$_{-14.9}^{+15.8}$ & 133.4$_{-9.2}^{+15.5}$ & 0.29$_{-0.10}^{+0.19}$ & 13.8$_{-1.7}^{+1.7}$ & SCC\\
    NGC1550 & 5.99$_{-0.87}^{+0.87}$ & 0.23$_{-0.03}^{+0.04}$ & 0.75$_{-0.01}^{+0.01}$ & -0.128$_{-0.006}^{+0.006}$ & 1.148$_{-0.171}^{+0.171}$ & 9.3$_{-2.1}^{+2.1}$ & 3.2$_{-0.5}^{+0.5}$ & 16.5$_{-1.1}^{+0.6}$ & 0.78$_{-0.02}^{+0.01}$ & 6.7$_{-0.6}^{+0.6}$ & SCC\\
    EXO0422 & 5.48$_{-0.59}^{+0.59}$ & 0.47$_{-0.05}^{+0.07}$ & 0.89$_{-0.03}^{+0.03}$ & -0.147$_{-0.032}^{+0.033}$ & 1.237$_{-0.015}^{+0.015}$ & 42.5$_{-8.2}^{+8.2}$ & $<$0.9 & 175.0$_{-100.5}^{+1091.0}$ & 2.32$_{-1.02}^{+0.85}$ & 18.0$_{-1.4}^{+1.4}$ & SCC\\
    A3266 & 0.76$_{-0.18}^{+0.18}$ & 7.62$_{-1.60}^{+2.63}$ & 1.07$_{-0.08}^{+0.08}$ & 0.218$_{-0.148}^{+0.140}$ & 0.730$_{-0.023}^{+0.016}$ & 3.3$_{-3.3}^{+40.6}$ & 12.9$_{-0.8}^{+0.7}$ & 12.9$_{-0.8}^{+0.6}$ & 0.08$_{0.00}^{+0.31}$ & 260.2$_{-45.2}^{+45.2}$ & WCC\\
    A0496$^{\alpha}$ & 5.79$_{-0.27}^{+0.27}$ & 0.47$_{-0.02}^{+0.02}$ & 0.21$_{-0.01}^{+0.01}$ & -0.347$_{-0.008}^{+0.008}$ & 1.240$_{-0.081}^{+0.081}$ & 78.7$_{-8.0}^{+8.0}$ & 4.6$_{-0.6}^{+0.6}$ & 54.6$_{-4.8}^{+4.6}$ & 1.27$_{-0.07}^{+0.07}$ & 8.2$_{-0.7}^{+0.7}$ & SCC\\
    A3376$^{\beta}$ & 0.24$_{-0.03}^{+0.03}$ & 16.47$_{-2.35}^{+3.10}$ & 1.04$_{-0.04}^{+0.04}$ & -0.090$_{-0.065}^{+0.065}$ & 0.348$_{-0.036}^{+0.033}$ & - & $<$0.4 & $<$0.4 & 0.11$_{-0.03}^{+79.79}$ & 222.0$_{-21.1}^{+21.1}$ & NCC\\
    A3391 & 0.35$_{-0.05}^{+0.05}$ & 12.46$_{-1.89}^{+2.49}$ & 1.11$_{-0.08}^{+0.08}$ & 0.054$_{-0.103}^{+0.102}$ & 0.379$_{-0.040}^{+0.034}$ & - & $<$0.93 & $<$2.4 & 0.08$_{-0.00}^{+79.82}$ & 275.2$_{-29.6}^{+29.6}$ & NCC\\
    A3395s & 0.35$_{-0.05}^{+0.05}$ & 12.66$_{-2.18}^{+3.04}$ & 1.01$_{-0.07}^{+0.07}$ & 0.032$_{-0.087}^{+0.087}$ & 0.511$_{-0.050}^{+0.042}$ & - & $<$0.4 & $<$3.1 & 0.08$_{0.00}^{+79.82}$ & 211.1$_{-24.7}^{+24.7}$ & NCC\\
    A0576$^{\gamma\delta}$ & 0.90$_{-0.16}^{+0.16}$ & 3.62$_{-0.59}^{+0.84}$ & 0.91$_{-0.04}^{+0.04}$ & -0.092$_{-0.070}^{+0.070}$ & 0.816$_{-0.013}^{+0.013}$ & 12.9$_{-5.8}^{+5.8}$ & $<$0.3 & 20.8$_{-13.0}^{+135.8}$ & 2.55$_{-2.46}^{+77.35}$ & 86.2$_{-10.6}^{+10.6}$ & WCC$^{\ddagger}$\\
    A0754$^{\beta}$ & 0.64$_{-0.12}^{+0.12}$ & 9.53$_{-1.64}^{+2.37}$ & 0.86$_{-0.06}^{+0.06}$ & 0.054$_{-0.125}^{+0.120}$ & 0.462$_{-0.022}^{+0.060}$ & - & 2.0$_{-1.2}^{+2.1}$ & 42.1$_{-13.9}^{+17.9}$ & 2.91$_{-0.97}^{+1.91}$ & 276.7$_{-38.8}^{+38.8}$ & NCC\\
    A0780 & 7.73$_{-0.29}^{+0.29}$ & 0.41$_{-0.02}^{+0.02}$ & 0.76$_{-0.02}^{+0.02}$ & -0.079$_{-0.004}^{+0.004}$ & 1.289$_{-0.056}^{+0.056}$ & 165.4$_{-12.8}^{+12.8}$ & 14.7$_{-1.0}^{+1.0}$ & 74.7$_{-14.7}^{+14.5}$ & 1.20$_{-0.17}^{+0.11}$ & 14.2$_{-0.5}^{+0.5}$ & SCC\\
    A1060 & 1.15$_{-0.32}^{+0.32}$ & 2.87$_{-0.63}^{+1.11}$ & 0.99$_{-0.03}^{+0.03}$ & -0.027$_{-0.046}^{+0.045}$ & 0.817$_{-0.067}^{+0.067}$ & 6.6$_{-3.0}^{+3.0}$ & 0.6$_{-0.2}^{+0.2}$ & 4.4$_{-1.9}^{+4.0}$ & 1.43$_{-1.34}^{+0.47}$ & 61.5$_{-11.4}^{+11.4}$ & WCC\\
    A1367$^{\beta\delta}$ & 0.11$_{-0.00}^{+0.00}$ & 27.97$_{-1.77}^{+1.90}$ & 0.90$_{-0.03}^{+0.03}$ & -0.030$_{-0.065}^{+0.065}$ & 0.043$_{-0.001}^{+0.003}$ & - & 0.4$_{-0.2}^{+0.2}$ & 0.6$_{-0.2}^{+0.5}$ & 0.58$_{-0.50}^{+0.47}$ & 296.5$_{-13.2}^{+13.2}$ & NCC\\
    MKW4 & 4.36$_{-0.30}^{+0.30}$ & 0.28$_{-0.02}^{+0.03}$ & 0.74$_{-0.01}^{+0.01}$ & -0.161$_{-0.018}^{+0.018}$ & 1.242$_{-0.006}^{+0.006}$ & 6.8$_{-0.8}^{+0.8}$ & $<$0.1 & 9.3$_{-1.5}^{+1.9}$ & 1.09$_{-0.07}^{+0.09}$ & 12.0$_{-0.6}^{+0.6}$ & SCC\\
    ZwCl1215$^{\gamma}$ & 0.44$_{-0.05}^{+0.05}$ & 10.99$_{-1.61}^{+2.09}$ & 1.06$_{-0.08}^{+0.08}$ & -0.007$_{-0.124}^{+0.118}$ & 0.239$_{-0.027}^{+0.028}$ & - & 2.5$_{-1.6}^{+3.2}$ & 6.0$_{-3.8}^{+259.0}$ & 0.71$_{-0.63}^{+63.09}$ & 246.4$_{-26.4}^{+26.4}$ & NCC \\
    NGC4636$^{\alpha}$ & 2.64$_{-0.11}^{+0.11}$ & 0.21$_{-0.01}^{+0.01}$ & 0.47$_{-0.03}^{+0.03}$ & -0.147$_{-0.004}^{+0.004}$ & 1.242$_{-0.090}^{+0.090}$ & 1.0$_{-0.1}^{+0.1}$ & 0.48$_{-0.05}^{+0.05}$ & 1.5$_{-0.2}^{+0.2}$ & 0.42$_{-0.05}^{+0.02}$ & 6.5$_{-0.2}^{+0.2}$ & SCC\\
    A3526$^{\alpha}$ & 4.59$_{-0.07}^{+0.07}$ & 0.42$_{-0.01}^{+0.01}$ & 0.199$_{-0.002}^{+0.002}$ & -0.457$_{-0.002}^{+0.002}$ & 1.180$_{-0.021}^{+0.021}$ & 23.5$_{-0.5}^{+0.5}$ & 5.2$_{-0.1}^{+0.1}$ & 13.9$_{-0.3}^{+0.3}$ & 0.86$_{-0.02}^{+0.01}$ & 8.1$_{-0.2}^{+0.2}$ & SCC\\
    A1644$^{\alpha}$ & 3.37$_{-0.90}^{+0.90}$ & 0.84$_{-0.18}^{+0.32}$ & 0.39$_{-0.01}^{+0.01}$ & -0.435$_{-0.033}^{+0.033}$ & 0.599$_{-0.075}^{+0.075}$ & 24.6$_{-11.9}^{+11.9}$ & 2.3$_{-0.7}^{+0.7}$ & 27.7$_{-1.9}^{+2.0}$ & 1.40$_{-0.14}^{+0.13}$ & 19.2$_{-3.5}^{+3.5}$ & SCC \\
    A1650 & 3.49$_{-0.65}^{+0.65}$ & 1.25$_{-0.20}^{+0.29}$ & 0.84$_{-0.03}^{+0.03}$ & -0.080$_{-0.030}^{+0.029}$ & 0.925$_{-0.149}^{+0.149}$ & 93.7$_{-28.2}^{+28.2}$ & $<$0.7 & 84.5$_{-15.8}^{+28.5}$ & 2.55$_{-0.43}^{+0.62}$ & 45.7$_{-5.8}^{+5.8}$ & WCC$^{\ddagger}$ \\
    A1651 & 1.25$_{-0.10}^{+0.10}$ & 3.63$_{-0.37}^{+0.43}$ & 1.01$_{-0.07}^{+0.07}$ & 0.009$_{-0.090}^{+0.087}$ & 0.642$_{-0.029}^{+0.029}$ & 165.8$_{-52.8}^{+52.8}$ & 7.7$_{-4.8}^{+7.0}$ & 69.0$_{-38.6}^{+148.1}$ & 2.30$_{-2.11}^{+1.52}$ & 118.7$_{-9.7}^{+9.7}$ & WCC\\
    % A1656 & 0.39$_{-0.05}^{+0.05}$ & 16.33$_{-1.99}^{+2.58}$ & 1.01$_{-0.07}^{+0.07}$ & -0.055$_{-0.141}^{+0.138}$ & 0.418$_{-0.027}^{+0.027}$ & - & 2.1$_{-0.5}^{+0.6}$ & 11.8$_{-2.8}^{+2.9}$ & 1.52$_{-0.49}^{+1.04}$ & 369.1$_{-40.8}^{+40.8}$ & NCC\\
    A1656 & 0.40$_{-0.08}^{+0.08}$ & 15.97$_{-2.95}^{+4.52}$ & 0.99$_{-0.07}^{+0.07}$ & -0.031$_{-0.123}^{+0.118}$ & 0.438$_{-0.043}^{+0.040}$ & - & $<$0.3 & 11.1$_{-3.8}^{+4.8}$ & 2.29$_{-0.94}^{+1.38}$ & 364.6$_{-39.8}^{+39.8}$ & NCC \\
    NGC5044 & 3.04$_{-0.47}^{+0.47}$ & 0.21$_{-0.03}^{+0.04}$ & 0.60$_{-0.01}^{+0.01}$ & -0.194$_{-0.008}^{+0.008}$ & 1.732$_{-0.238}^{+0.238}$ & 11.0$_{-2.5}^{+2.5}$ & 0.8$_{-0.3}^{+0.3}$ & 27.9$_{-0.6}^{+1.1}$ & 0.62$_{-0.01}^{+0.01}$ & 7.5$_{-0.8}^{+0.8}$ & SCC\\
    A1736$^{\beta}$ & 0.17$_{-0.07}^{+0.07}$ & 16.59$_{-5.52}^{+13.17}$ & 1.06$_{-0.05}^{+0.05}$ & 0.094$_{-0.074}^{+0.074}$ & 0.105$_{-0.027}^{+0.058}$ & - & 0.3$_{-0.2}^{+0.5}$ & 0.5$_{-0.3}^{+61.7}$ & 0.58$_{-0.50}^{+79.32}$ & 228.0$_{-62.8}^{+62.8}$ & NCC\\
    A3558 & 2.46$_{-0.97}^{+0.97}$ & 1.69$_{-0.50}^{+1.15}$ & 1.01$_{-0.05}^{+0.05}$ & -0.083$_{-0.067}^{+0.065}$ & 0.738$_{-0.063}^{+0.063}$ & 31.5$_{-15.3}^{+15.3}$ & 2.7$_{-1.7}^{+2.1}$ & 2.7$_{-1.7}^{+2.0}$ & 0.12$_{-0.04}^{+79.78}$ & 59.4$_{-15.9}^{+15.9}$ & WCC\\
    A3562$^{\gamma}$ & 0.71$_{-0.07}^{+0.07}$ & 5.15$_{-0.57}^{+0.72}$ & 0.95$_{-0.05}^{+0.05}$ & -0.082$_{-0.073}^{+0.073}$ & 0.675$_{-0.034}^{+0.027}$ & 23.6$_{-10.9}^{+10.9}$ & 1.7$_{-1.0}^{+1.1}$ & 17.2$_{-10.3}^{+3847031.1}$ & 1.81$_{-0.62}^{+2.21}$ & 113.7$_{-8.8}^{+8.8}$ & WCC\\
    A3571 & 2.49$_{-0.60}^{+0.60}$ & 2.13$_{-0.43}^{+0.71}$ & 0.95$_{-0.05}^{+0.05}$ & -0.109$_{-0.093}^{+0.090}$ & 0.703$_{-0.157}^{+0.157}$ & 43.1$_{-24.4}^{+24.4}$ & 6.2$_{-1.2}^{+1.3}$ & 35.8$_{-12.1}^{+14.0}$ & 2.12$_{-0.87}^{+0.50}$ & 77.9$_{-13.3}^{+13.3}$ & WCC\\
    A1795 & 6.11$_{-0.19}^{+0.19}$ & 0.61$_{-0.02}^{+0.02}$ & 0.50$_{-0.01}^{+0.01}$ & -0.203$_{-0.008}^{+0.008}$ & 1.125$_{-0.028}^{+0.028}$ & 255.4$_{-15.0}^{+15.0}$ & 14.5$_{-1.5}^{+1.5}$ & 222.4$_{-20.3}^{+32.8}$ & 1.80$_{-0.13}^{+0.15}$ & 20.7$_{-0.7}^{+0.7}$ & SCC\\
    A3581$^{\alpha}$ & 2.63$_{-0.23}^{+0.23}$ & 0.55$_{-0.05}^{+0.06}$ & 0.52$_{-0.03}^{+0.03}$ & -0.353$_{-0.055}^{+0.054}$ & 1.298$_{-0.013}^{+0.013}$ & 33.5$_{-5.8}^{+5.8}$ & 1.2$_{-0.8}^{+1.0}$ & 21.5$_{-5.1}^{+9.1}$ & 0.88$_{-0.15}^{+0.15}$ & 11.6$_{-1.0}^{+1.0}$ & SCC\\
    MKW8 & 0.29$_{-0.05}^{+0.05}$ & 10.87$_{-1.87}^{+2.59}$ & 1.20$_{-0.06}^{+0.06}$ & 0.162$_{-0.076}^{+0.074}$ & 0.519$_{-0.048}^{+0.049}$ & - & $<$0.2 & 0.5$_{-0.3}^{+45.4}$ & 1.48$_{-1.40}^{+78.42}$ & 177.3$_{-21.0}^{+21.0}$ & NCC\\
    RXJ1504 & 7.92$_{-0.19}^{+0.19}$ & 0.59$_{-0.02}^{+0.02}$ & 0.48$_{-0.04}^{+0.04}$ & -0.213$_{-0.018}^{+0.018}$ & 1.171$_{-0.006}^{+0.006}$ & 2266.5$_{-129.7}^{+129.7}$ & $<$7.4 & 2671.5$_{-1278.5}^{+2201.7}$ & 3.26$_{-1.15}^{+0.82}$ & 24.9$_{-0.9}^{+0.9}$ & SCC\\
    A2029 & 9.02$_{-0.64}^{+0.64}$ & 0.53$_{-0.04}^{+0.04}$ & 0.43$_{-0.01}^{+0.01}$ & -0.202$_{-0.010}^{+0.009}$ & 1.112$_{-0.066}^{+0.066}$ & 324.2$_{-37.8}^{+37.8}$ & 18.7$_{-2.6}^{+2.5}$ & 328.9$_{-11.7}^{+21.7}$ & 2.68$_{-0.15}^{+0.27}$ & 24.9$_{-1.4}^{+1.4}$ & SCC\\
    A2052$^{\alpha}$ & 4.62$_{-0.18}^{+0.18}$ & 0.51$_{-0.02}^{+0.02}$ & 0.304$_{-0.005}^{+0.005}$ & -0.954$_{-0.018}^{+0.018}$ & 1.205$_{-0.029}^{+0.029}$ & 69.6$_{-3.8}^{+3.8}$ & 13.0$_{-0.3}^{+0.3}$ & 33.9$_{-1.1}^{+1.8}$ & 0.85$_{0.20}^{+0.25}$ & 8.7$_{-0.5}^{+0.5}$ & SCC\\
    MKW3S & 3.46$_{-0.58}^{+0.58}$ & 0.86$_{-0.13}^{+0.18}$ & 0.78$_{-0.02}^{+0.02}$ & -0.095$_{-0.015}^{+0.015}$ & 1.138$_{-0.186}^{+0.186}$ & 72.1$_{-24.2}^{+24.2}$ & 2.9$_{-1.0}^{+1.0}$ & 40.0$_{-14.4}^{+17.1}$ & 1.55$_{-0.38}^{+0.45}$ & 28.6$_{-3.2}^{+3.2}$ & SCC\\
    A2065$^{\delta}$ & 2.99$_{-0.61}^{+0.61}$ & 1.34$_{-0.24}^{+0.37}$ & 0.79$_{-0.03}^{+0.03}$ & -0.123$_{-0.047}^{+0.046}$ & 0.469$_{-0.050}^{+0.050}$ & 50.6$_{-21.3}^{+21.3}$ & 2.7$_{-1.7}^{+1.9}$ & 47.2$_{-9.7}^{+8.2}$ & 1.74$_{-0.34}^{+0.52}$ & 44.3$_{-6.2}^{+6.2}$ & WCC$^{\ddagger}$\\
    A2063 & 1.29$_{-0.06}^{+0.06}$ & 2.36$_{-0.13}^{+0.14}$ & 0.97$_{-0.03}^{+0.03}$ & -0.016$_{-0.047}^{+0.047}$ & 0.935$_{-0.069}^{+0.069}$ & 28.6$_{-6.1}^{+6.1}$ & 1.3$_{-0.5}^{+0.5}$ & 5.3$_{-2.1}^{+13.3}$ & 1.07$_{-0.50}^{+0.65}$ & 66.1$_{-2.8}^{+2.8}$ & WCC\\
    A2142$^{\gamma}$ & 2.65$_{-0.17}^{+0.17}$ & 1.94$_{-0.14}^{+0.16}$ & 0.70$_{-0.05}^{+0.05}$ & -0.149$_{-0.019}^{+0.019}$ & 0.893$_{-0.007}^{+0.006}$ & 275.7$_{-40.4}^{+40.4}$ & 25.0$_{-6.1}^{+6.2}$ & 184.1$_{-36.4}^{+26.9}$ & 2.21$_{-0.47}^{+0.59}$ & 65.8$_{-4.0}^{+4.0}$ & WCC$^{\ddagger}$ \\
    A2147 & 0.23$_{-0.03}^{+0.03}$ & 17.04$_{-2.72}^{+3.64}$ & 0.92$_{-0.05}^{+0.05}$ & -0.058$_{-0.084}^{+0.082}$ & 0.223$_{-0.025}^{+0.035}$ & - & 3.1$_{-0.8}^{+0.4}$ & 3.1$_{-0.8}^{+0.5}$ & 0.08$_{0.00}^{+0.78}$ & 212.4$_{-23.1}^{+23.1}$ & NCC \\
    A2163$^{\beta}$ & 0.84$_{-0.03}^{+0.03}$ & 9.65$_{-0.78}^{+0.73}$ & 1.11$_{-0.14}^{+0.14}$ & 0.280$_{-0.219}^{+0.202}$ & 0.251$_{-0.007}^{+0.007}$ & - & 57.3$_{-23.4}^{+27.1}$ & 87.3$_{-37.8}^{+73.0}$ & 1.43$_{-1.35}^{+3.07}$ & 427.6$_{-51.8}^{+51.8}$ & NCC \\
    A2199$^{\alpha}$ & 5.61$_{-0.58}^{+0.58}$ & 0.60$_{-0.06}^{+0.07}$ & 0.31$_{-0.06}^{+0.06}$ & -0.619$_{-0.097}^{+0.093}$ & 1.027$_{-0.101}^{+0.101}$ & 78.9$_{-12.0}^{+12.0}$ & 2.3$_{-1.0}^{+1.0}$ & 76.9$_{-5.4}^{+4.5}$ & 1.53$_{-0.08}^{+0.14}$ & 9.2$_{-2.0}^{+2.0}$ & SCC\\
    A2204$^{\delta}$ & 16.70$_{-0.65}^{+0.65}$ & 0.25$_{-0.01}^{+0.01}$ & 0.38$_{-0.02}^{+0.02}$ & -0.359$_{-0.033}^{+0.034}$ & 1.384$_{-0.007}^{+0.008}$ & 637.8$_{-44.8}^{+44.8}$ & $<$17.5 & 913.8$_{-145.6}^{+119.1}$ & 1.89$_{-0.26}^{+0.39}$ & 11.1$_{-0.5}^{+0.5}$ & SCC\\
    A2244 & 2.91$_{-0.40}^{+0.40}$ & 1.53$_{-0.20}^{+0.27}$ & 0.93$_{-0.04}^{+0.04}$ & -0.038$_{-0.071}^{+0.069}$ & 0.548$_{-0.029}^{+0.029}$ & 131.8$_{-50.7}^{+50.7}$ & 1.7$_{-1.1}^{+2.8}$ & 115.5$_{-47.3}^{+35.5}$ & 2.43$_{-0.58}^{+0.85}$ & 57.1$_{-5.8}^{+5.8}$ & WCC\\
    A2256$^{\beta}$ & 0.38$_{-0.06}^{+0.06}$ & 11.56$_{-1.81}^{+2.43}$ & 0.63$_{-0.05}^{+0.05}$ & -0.243$_{-0.061}^{+0.060}$ & 0.234$_{-0.032}^{+0.031}$ & - & 8.5$_{-2.6}^{+4.1}$ & 8.5$_{-2.5}^{+2.3}$ & 0.08$_{-0.00}^{+0.41}$ & 197.6$_{-22.4}^{+22.4}$ & NCC$^{\dagger}$\\
    A2255$^{\beta\delta}$ & 0.26$_{-0.07}^{+0.07}$ & 20.66$_{-5.09}^{+9.35}$ & 1.18$_{-0.09}^{+0.09}$ & -0.088$_{-0.134}^{+0.129}$ & 0.295$_{-0.056}^{+0.068}$ & - & 4.0$_{-2.5}^{+0.5}$ & 4.0$_{-2.5}^{+0.5}$ & 0.08$_{0.00}^{+18.87}$ & 365.0$_{-70.1}^{+70.1}$ & NCC\\
    A3667$^{\beta}$ & 0.78$_{-0.06}^{+0.06}$ & 6.14$_{-0.45}^{+0.52}$ & 0.83$_{-0.04}^{+0.04}$ & -0.133$_{-0.083}^{+0.080}$ & 0.224$_{-0.005}^{+0.005}$ & 0.7$_{-0.3}^{+0.3}$ & $<$0.1 & 16.5$_{-6.2}^{+1.2}$ & 2.31$_{-0.54}^{+0.65}$ & 134.6$_{-8.7}^{+8.7}$ & WCC$^{\ddagger}$ \\
    S1101 & 2.87$_{-0.49}^{+0.49}$ & 0.88$_{-0.14}^{+0.20}$ & 0.88$_{-0.04}^{+0.04}$ & -0.108$_{-0.034}^{+0.034}$ & 0.661$_{-0.188}^{+0.188}$ & 222.3$_{-120.0}^{+120.0}$ & 14.1$_{-4.3}^{+4.1}$ & 23.5$_{-7.4}^{+16.5}$ & 0.61$_{-0.08}^{+0.09}$ & 24.2$_{-2.9}^{+2.9}$ & SCC\\
    A2589 & 2.42$_{-0.52}^{+0.52}$ & 1.18$_{-0.22}^{+0.33}$ & 0.93$_{-0.02}^{+0.02}$ & -0.079$_{-0.042}^{+0.042}$ & 0.569$_{-0.164}^{+0.164}$ & 19.9$_{-12.3}^{+12.3}$ & 0.8$_{-0.3}^{+0.3}$ & 9.6$_{-3.5}^{+3.4}$ & 1.33$_{-0.37}^{+0.28}$ & 43.4$_{-6.3}^{+6.3}$ & WCC$^{\ddagger}$ \\
    A2597 & 7.24$_{-0.57}^{+0.57}$ & 0.42$_{-0.03}^{+0.04}$ & 0.57$_{-0.01}^{+0.01}$ & -0.181$_{-0.010}^{+0.010}$ & 1.380$_{-0.084}^{+0.084}$ & 274.2$_{-36.4}^{+36.4}$ & 10.0$_{-2.6}^{+2.6}$ & 141.8$_{-27.7}^{+36.9}$ & 1.27$_{-0.16}^{+0.17}$ & 13.2$_{-0.8}^{+0.8}$ & SCC \\
    A2634 & 2.21$_{-0.52}^{+0.52}$ & 1.52$_{-0.33}^{+0.53}$ & 1.00$_{-0.07}^{+0.07}$ & 0.006$_{-0.112}^{+0.109}$ & 0.240$_{-0.034}^{+0.034}$ & 0.2$_{-0.1}^{+0.1}$ & 1.0$_{-0.2}^{+0.3}$ & 1.0$_{-0.2}^{+0.3}$ & 0.08$_{0.00}^{+0.48}$ & 40.7$_{-6.9}^{+6.9}$ & WCC \\
    A2657 & 1.21$_{-0.36}^{+0.36}$ & 2.68$_{-0.66}^{+1.20}$ & 1.12$_{-0.05}^{+0.05}$ & 0.023$_{-0.057}^{+0.057}$ & 0.907$_{-0.028}^{+0.022}$ & 15.4$_{-10.9}^{+10.9}$ & $<$0.4 & 13.9$_{-8.4}^{+196.8}$ & 1.83$_{-0.71}^{+1.74}$ & 75.1$_{-15.0}^{+15.0}$ & WCC \\
    A4038$^{\gamma}$ & 1.79$_{-0.11}^{+0.11}$ & 1.68$_{-0.11}^{+0.12}$ & 0.97$_{-0.02}^{+0.02}$ & -0.024$_{-0.039}^{+0.039}$ & 0.966$_{-0.007}^{+0.007}$ & 47.2$_{-6.3}^{+6.3}$ & $<$0.2 & $<$0.2 & 0.08$_{0.00}^{+79.82}$ & 44.5$_{-2.1}^{+2.1}$ & WCC\\
    A4059$^{\alpha\delta}$ & 3.85$_{-0.35}^{+0.35}$ & 0.70$_{-0.06}^{+0.07}$ & 0.23$_{-0.01}^{+0.01}$ & -0.581$_{-0.028}^{+0.028}$ & 0.639$_{-0.018}^{+0.018}$ & 57.7$_{-10.6}^{+10.6}$ & 4.1$_{-0.5}^{+0.5}$ & 38.6$_{-8.7}^{+4.2}$ & 1.34$_{-0.19}^{+0.09}$ & 8.4$_{-0.7}^{+0.7}$ & SCC \\

    \hline
    \multicolumn{12}{p{22cm}}{$^{\alpha}$ One or more of the innermost annuli were fit with a double thermal model which significantly ($>99$\% confidence according to the $F$ test) improved the fit over a single thermal model.}\\
    \multicolumn{12}{p{22cm}}{$^{\beta}$ This is one of eight clusters in which the offset between the X-ray peak and BCG is quite large ($>$ 50~$h_{71}^{-1}$~kpc).}\\
    \multicolumn{12}{p{22cm}}{$^{\gamma}$ This is one of six clusters in which the offset between the X-ray peak and BCG is moderate ($>$ 12~$h_{71}^{-1}$~kpc but $<$ 50~$h_{71}^{-1}$~kpc).}\\
    \multicolumn{12}{p{22cm}}{$^{\delta}$  The BCG of this cluster is one of five that has a peculiar velocity $>$50\% of the cluster velocity dispersion.}\\
    \multicolumn{12}{p{22cm}}{$^{\dagger}$ Although this cluster is an NCC there is an appreciable amount of cool gas around the emission peak so that the temperature profile is not consistent with a non-negative slope.}\\
    \multicolumn{12}{p{22cm}}{$^{\ddagger}$ This WCC cluster has an appreciable amount of cool gas around the emission peak so that the temperature profile is not consistent with a non-negative slope.}\\
    \hline\hline    
  \end{longtable}
\end{landscape}
%}

%\input{table3}
\end{document}